%% file: arXiv.tex
\documentclass{article}

\title{Tetris is NP-hard even with $O(1)$ rows or columns}

\author{%
  Sualeh Asif%
    \thanks{Massachusetts Institute of Technology, Cambridge, MA, USA}
\and
  Michael Coulombe\footnotemark[1]
\and
  Erik D. Demaine\footnotemark[1]
\and
  Martin L. Demaine\footnotemark[1]
\and
  Adam Hesterberg\footnotemark[1]
\and
  Jayson Lynch\footnotemark[1]
\and
  Mihir Singhal\footnotemark[1]
}
\date{}

\usepackage[utf8]{inputenc}
\usepackage{gensymb}
\usepackage{pdfpages}
\usepackage[margin=1in]{geometry}
\usepackage{comment}
\usepackage{amsmath, amssymb, amsthm, import, etoolbox}
\usepackage[labelfont=bf]{caption}
\usepackage[labelfont=rm,labelformat=simple,skip=1pt,belowskip=1pt]{subcaption}

\usepackage{overpic}
\usepackage{colortbl}
\usepackage[T1]{fontenc}
\usepackage{fancybox}
\usepackage{enumitem}
\usepackage{hyperref}
\newcommand{\OO}{\texorpdfstring{\,\vcenter{\hbox{\includegraphics[scale=0.2]{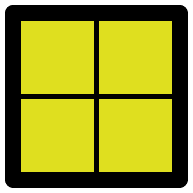}}}\,}O}
\newcommand{\TT}{\texorpdfstring{\,\vcenter{\hbox{\includegraphics[scale=0.2]{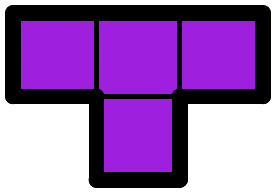}}}\,}T}
\newcommand{\LL}{\texorpdfstring{\,\vcenter{\hbox{\includegraphics[scale=0.2]{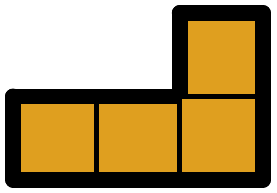}}}\,}L}
\newcommand{\JJ}{\texorpdfstring{\,\vcenter{\hbox{\includegraphics[scale=0.2]{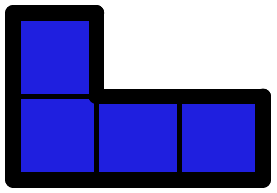}}}\,}J}
\renewcommand{\SS}{\texorpdfstring{\,\vcenter{\hbox{\includegraphics[scale=0.2]{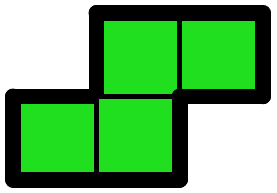}}}\,}S}
\newcommand{\ZZ}{\texorpdfstring{\,\vcenter{\hbox{\includegraphics[scale=0.2]{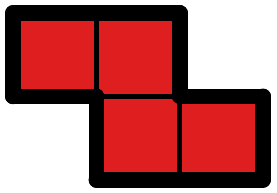}}}\,}Z}
\newcommand{\II}{\texorpdfstring{\,\vcenter{\hbox{\includegraphics[scale=0.2]{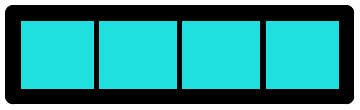}}}\,}I}

\newcommand{\mc}[1]{\mathcal{#1}}
\newcommand{\ms}[1]{\mathsf{#1}}

\newcommand{\ctet}[1]{\textsc{#1-column Tetris}}
\newcommand{\rtet}[1]{\textsc{#1-row Tetris}}
\newcommand{\ktet}[1]{\textsc{#1-tris}}
\newcommand{\cktet}[2]{\textsc{#1-column #2-tris}}
\newcommand{\rktet}[2]{\textsc{#1-row #2-tris}}
\newcommand{\tet}{\textsc{Tetris}}
\newcommand{\etet}{\textsc{Empty Tetris}}
\newcommand{\ektet}[1]{\textsc{Empty #1-tris}}
\newcommand{\cektet}[2]{\textsc{#1-column Empty #2-tris}}
\newcommand{\rektet}[2]{\textsc{#1-row Empty #2-tris}}
\newcommand{\partit}{\textsc{3-Partition}}
\newcommand{\col}{8}
\newcommand{\row}{4}

\cornersize{0.9}

\newtheorem{thm}{Theorem}[section] 
\newtheorem{lem}[thm]{Lemma}
\newtheorem{prop}[thm]{Proposition}
\theoremstyle{definition}
\newtheorem{defn}[thm]{Definition}
\newtheorem{claim}[thm]{Claim}
\newtheorem{cor}[thm]{Corollary}
\newtheorem{corq}[thm]{Corollary}

\AtEndEnvironment{corq}{\null\hfill\qedsymbol}

{\makeatletter \gdef\fps@figure{!htbp}}

\let\realbfseries=\bfseries
\def\bfseries{\realbfseries\boldmath}

{\makeatletter
 \gdef\xxxmark{%
   \expandafter\ifx\csname @mpargs\endcsname\relax % in minipage?
     \expandafter\ifx\csname @captype\endcsname\relax % in figure/caption?
       \marginpar{xxx}% not in a caption or minipage, can use marginpar
     \else
       xxx % notice trailing space
     \fi
   \else
     xxx % notice trailing space
   \fi}
 \gdef\xxx{\@ifnextchar[\xxx@lab\xxx@nolab}
 \long\gdef\xxx@lab[#1]#2{\textbf{[\xxxmark #2 ---{\sc #1}]}}
 \long\gdef\xxx@nolab#1{\textbf{[\xxxmark #1]}}
 \long\gdef\xxx@lab[#1]#2{}\long\gdef\xxx@nolab#1{}%
}

\begin{document}

\maketitle
\begin{abstract}

We prove that the classic falling-block video game \emph{Tetris}
(both survival and board clearing)
remains NP-complete even when restricted to $\col$ columns, or to $\row$ rows,
settling open problems posed over 15 years ago \cite{Tetris_IJCGA}.
Our reduction is from {\partit}, similar to the previous reduction for unrestricted board sizes, but with a better packing of buckets.
On the positive side, we prove that 2-column Tetris (and 1-row Tetris)
is polynomial.
We also prove that the generalization of Tetris to larger $k$-omino pieces
is NP-complete even when the board starts empty,
even when restricted to 3 columns or 2 rows or constant-size pieces.
Finally, we present an animated Tetris font.

\end{abstract}

\xxx{Can we generalize to $k$-tris for $k < 4$ at least (and ideally all $k$)?
  Can we mimic \cite{TotalTetris_JIP}?}

% Introduction:
\import{}{Intro.tex}

\import{tetris/}{rules.tex}

\import{tetris/}{polynomial.tex}

\import{tetris/}{nphard_col.tex}

\import{tetris/}{nphard_row.tex}

\import{tetris/}{empty.tex}

\import{tetris/}{font.tex}

\section{Open Problems} 
\label{conclusion}

The main open problems are to determine the critical threshold for the minimum
number $c^*$ of columns and the minimum number $r^*$ of rows for which
\ctet{$c^*$} and \rtet{$c^*$} are NP-complete, respectively.
We proved here that $c^* \in [3,8]$ and $r^* \in [2,4]$.
We conjecture that $r^* = 2$, i.e., that \rtet{2} is NP-complete.
These problems are open even for \cktet{$c^*$}{$O(1)$} and
\rktet{$r^*$}{$O(1)$}, i.e., allowing constant-size pieces.

Our hardness proof for \ctet{$c$} survival relies on the partial lock out rule,
which has been changed in modern versions of Tetris \cite{Tetris-guideline}.
We can avoid this assumption, and also allow constant reaction times for the
player, by adding many rows on top and using the reservoir trick from
\cite[Section 4.2]{Tetris_IJCGA}.  However, this approach works only for
$c$ odd.  Is \ctet{8} survival NP-hard without the partial lock out rule?

Modern versions of Tetris also have a ``holding'' function, where the player
can put one piece aside for later use.  Can existing results be re-established,
or existing open problems be solved, with the addition of this feature?

Many other questions posed in prior papers on Tetris still remain open.
For example, does Tetris remain hard from an empty board?
What is the complexity of Tetris with imperfect information or randomness?
Are there guaranteed loss sequences in $n$-tris for all $n$?

\section*{Acknowledgments}

This work was initiated during open problem solving in the MIT class on
Algorithmic Lower Bounds: Fun with Hardness Proofs (6.892) in Spring 2019.
We thank the other participants of that class --- in particular, Joshua Ani,
Jonathan Gabor, and Claire Tang --- for related discussions and providing an
inspiring atmosphere.
We also thank the anonymous referees for helpful comments.

\bibliographystyle{alpha}
\bibliography{references}

\end{document}

%% file: Intro.tex
\section{Introduction}
%We demonstrate that perfect information Tetris remains NP-hard even when restricted to a constant number of columns.

Tetris is among the best-selling \cite{best-selling}, and perhaps best-known,
video games ever.
Since its invention by Alexey Pajitnov 35 years ago in 1984,
%and through tumultuous legal battles for the rights,
over 80 versions have been developed on nearly every platform \cite{wikipedia}.
Perhaps most famous is the Nintendo Game Boy edition, which was bundled with
the Game Boy in the USA, resulting in 35 million copies sold
\cite{telegraph-history}.
The most recent editions ---
Tetris Effect for PS4 and PC including VR (2018) and
Tetris 99 for Nintendo Switch (2019) ---
prove Tetris's sustained popularity.
%Tetris is the subject of two documentaries \cite{wikipedia}

In standard Tetris, tetromino pieces ($\OO, \JJ, \LL, \SS, \ZZ, \TT, \II$)
are chosen at (pseudo)random and fall from the top of a 10-wide 20-tall
rectangular playfield.
While 10 is the typical width of most Tetris implementations, the height
varies between 16 and 24, and some editions change the width to anywhere
between 6 and 20 \cite{playfield}.
The player can rotate each piece by $\pm 90^\circ$ and/or slide it left/right
as it falls down, until the piece ``hits'' another piece beneath it and
the piece freezes.
If any rows are completely full, they get removed (shifting higher rows down),
and then the next piece starts falling from the top.

To make this game easier to analyze from a computational complexity perspective,
the \emph{perfect-information {\tet} problem} \cite{Tetris_IJCGA} asks,
given an initial board state of filled cells and a sequence of pieces that will
arrive, whether the pieces can be played in sequence to either
survive (not go above the top row) or clear the entire board.
(See Section~\ref{Rules of Tetris} for precise game rules.)
This problem was proved
NP-hard for arbitrary board sizes in 2002 \cite{Tetris_IJCGA}, and more
recently for the generalization to $k$-omino pieces for various~$k$
\cite{TotalTetris_JIP}.

\paragraph{Our results.}

In this paper, we analyze the following special cases of \tet;
refer to Table~\ref{table:results}.

\begin{enumerate}
\parindent=1.5em
\parskip=0em

\item \ctet{$c$}, where the playfield has exactly $c$ columns
  (and an arbitrary number of rows).
  The original Tetris paper \cite{Tetris_IJCGA} asked specifically about the
  complexity of \ctet{$c$} for $c=O(1)$,
  motivated by standard Tetris where $c=10$.

  In Section~\ref{sec:8-column},
  we prove that it is NP-complete to survive or clear the board in \ctet{$c$}
  for any $c \geq \col$.
  This result includes the width of most Tetris variants,
  including the already small Tetris Jr.\ ($c=8$),
  but excludes one variant, Tetris Wristwatch ($c=6$)
  \cite{playfield}.
  As an extra feature, this result immediately implies NP-hardness of {\tet}
  where the player can make only a bounded number of moves
  between each unit piece fall (``bounded reaction speed'').

  Complementarily, in Section~\ref{sec:2-column},
  we prove that \ctet{$c$} can be solved in polynomial time
  for $c \leq 2$.  The case $c=2$ was claimed without proof in the conclusion of
  \cite{Tetris_IJCGA}; we provide the first written proof, and generalize to
  \cktet{2}{$O(n)$}, by reducing to nondeterministic pushdown automata.
  The critical hardness threshold for $c$ is thus between $3$ and $8$.

\item \rtet{$r$}, where the playfield has exactly $r$ rows
  (and an arbitrary number of columns).
  The original Tetris paper \cite{Tetris_IJCGA} also asked about the
  complexity of \rtet{$r$} for $r=O(1)$.

  In Section~\ref{sec:4-row},
  we prove that it is NP-complete to survive or clear the board in \rtet{$r$}
  for any $r \geq \row$.

  Complementarily, we observe the trivial result that \rtet{$r$} can be solved
  in polynomial time for $r = 1$.
  The critical hardness threshold for $r$ is thus between $2$ and $4$. 
%   \xxx{This said 2 to 8 before, should be 4 right?}

\end{enumerate}

Both the $O(1)$-row and $O(1)$-column NP-hardness results are based on more
efficient packings of the ``buckets'' in the original reduction from {\partit}
\cite{Tetris_IJCGA}.  While they share the main idea with the original proof,
they require substantial care in how they provide a corridor that can reach
all of the buckets without allowing unintended solutions.
In particular, we prove NP-hardness of Tetris survival for the first time
with even-width boards (e.g., $c=8$ columns); the previous ``reservoir''
approach \cite[Section 4.2]{Tetris_IJCGA} required an odd number of columns.

\begin{enumerate}[resume]

\item \etet, where the playfield starts empty instead of having a specified
  configuration.
  The original Tetris paper \cite{Tetris_IJCGA} highlighted the complexity of
  this variant as a ``major open question'', as all existing Tetris hardness
  proofs (including those in this paper) rely on a high-complexity initial
  configuration.

  In Section~\ref{sec:empty-start},
  we solve this problem for the generalization of {\tet} to $k$-omino pieces,
  denoted \ktet{$k$},
  as implemented in the video games \emph{Pentris} and \emph{ntris},
  and previously analyzed from a complexity perspective \cite{TotalTetris_JIP}.
  Specifically, we prove the following results:

  \begin{enumerate}
  \item \cektet{8}{$(\leq 65)$} is NP-hard.
    %(assuming lines clear and lines descend before the game checks whether
    % the player lost by extending above the topmost row).
  \item \cektet{3}{$O(n)$} is NP-hard.
    This result is tight against our polynomial-time algorithm for
    \cktet{2}{$O(n)$} mentioned above.
  \item \rektet{2}{$O(n)$} is NP-hard.
  %\item If we also relax the definition to allow the game to start in a position that could not be reached in normal Tetris play, then the problem is (strongly) NP-hard even with only 2 columns.
  \end{enumerate}
\end{enumerate} 

\definecolor{header}{rgb}{0.29,0,0.51}
\definecolor{gray}{rgb}{0.85,0.85,0.85}
\definecolor{hard}{rgb}{1,0.85,0.85}
\definecolor{open}{rgb}{1,1,0.85}
\definecolor{easy}{rgb}{0.85,0.85,1}
\def\header#1{\textcolor{white}{\textbf{#1}}}

\begin{table}
  \centering
  %\small
  %\setlength\tabcolsep{0.9\tabcolsep}
  \begin{tabular}{ccccll}
    \rowcolor{header}
    \header{Rows} & \header{Columns} & \header{Empty?} & \header{Piece Sizes} & \header{Complexity} & \header{Reference} \\
    \rowcolor{hard}
    1      & $O(n)$ & no  & $O(n)$ & strongly NP-hard & Proposition~\ref{prop:1 row big} \\
    \rowcolor{easy}
    1      & $O(n)$ & yes & $O(n)$ & linear & Proposition~\ref{prop:1 row empty} \\
    \rowcolor{easy}
    1      & $O(n)$ & no  & $k$    & linear & Proposition~\ref{prop:1 col 1 row} \\
    \rowcolor{hard}
    2      & $O(n)$ & yes & $O(n)$ & strongly NP-hard & Theorem~\ref{thm:2-row empty} \\
    \rowcolor{open}
    3      & $O(n)$ & no  & 4      & OPEN & \\
    \rowcolor{hard}
    4      & $O(n)$ & no  & 4      & strongly NP-hard & Theorem~\ref{thm:4 row} \\
    \rowcolor{easy}
    $O(n)$ & 1      & no  & $O(n)$ & linear & Proposition~\ref{prop:1 col 1 row} \\
    \rowcolor{easy}
    $O(n)$ & 2      & no  & $O(n)$ & polynomial & Theorem~\ref{thm:2 row} \\
    \rowcolor{hard}
    $O(n)$ & 3      & yes & $O(n)$ & strongly NP-hard & Theorem~\ref{thm:3 columns} \\
    \rowcolor{open}
    $O(n)$ & 3--7   & no  & 4      & OPEN & \\
    \rowcolor{hard}
    $O(n)$ & 8+     & no  & 4      & strongly NP-hard & Theorem~\ref{thm:8 col} \\
    \rowcolor{hard}
    $O(n)$ & 8      & yes &$\leq65$& strongly NP-hard & Theorem~\ref{thm:65} \\
    \rowcolor{gray}
  \end{tabular}
  \caption{Summary of our results.  Each row of the table specifies the
    supported board size (numbers of rows and columns), whether the board
    starts empty or from an adversarial position (empty is stronger for
    hardness, while nonempty is stronger for algorithms),
    the allowed piece polyomino sizes (4 for Tetris),
    the complexity result for this case (red for hardness, blue for algorithm,
    and yellow for open), and where we prove the result.}
  \label{table:results}
\end{table}
 
In Section~\ref{sec:font}, we present a Tetris font where each
letter is made from exactly one copy of each tetromino piece.
The font has several variants, including a puzzle font and an animated font,
demonstrated in a companion web app.%
\footnote{\url{http://erikdemaine.org/fonts/tetris/}}

%% file: tetris/rules.tex
\section{Rules of Tetris}
\label{Rules of Tetris}

We give a brief summary of the rules of Tetris and its generalization
\ktet{$k$}, referring the reader to \cite[Section~2]{Tetris_IJCGA} and
\cite[Section~2]{TotalTetris_JIP} respectively for a complete description of
the rules.
There are in fact many real-world variations of the rules, eventually
formalized by The Tetris Company into a modern rule set \cite{Tetris-guideline}.
The rules we give here are consistent with some, but not all, implementations
of Tetris, as detailed below.

Tetris consists of a rectangular \emph{board} or \emph{playfield}
\cite{playfield}, which is rectangular in shape.
% and, in this paper, has at most $O(1)$ columns.
Each cell is \emph{filled} or \textit{unfilled}.
In the initial state and after each move, no row is completely filled.

In {\tet}, there are seven tetromino piece types (distinguished by reflection),
labeled with letters that resemble their shape:
$\OO (\ms O), \JJ (\ms J), \LL (\ms L), \SS (\ms S), \ZZ (\ms Z),
\TT (\ms T), \II (\ms I)$.
% See Figure~\ref{fig:blocks}.
In \ktet{$k$}, the piece types consist of all $k$-ominoes, i.e.,
all connected shapes made by $k$ unit squares joined along edges.
We also define \ktet{$\leq k$}, where the piece types are all polyominoes
made from $\leq k$ unit squares.
In a game instance, $n$ pieces arrive in a fixed order $p_1, p_2, \dots, p_n$.
Each piece $p_i$ falls, starting above the top row of the board, and the player
can rotate by $\pm 90^\circ$, and translate left and/or right,
as the piece drops down one unit at a time.
When the piece tries to drop but would collide with another piece,
then it stops moving (``locks down'').
If a locked-down piece extends above the top row of the board,
then the player immediately loses the game.
This rule is called \emph{partial lock out} \cite{topout}, and applies to
older versions of Tetris (e.g., Atari and NES), though modern Tetris rules
\cite{Tetris-guideline}
end the game only when a locked-down piece is entirely above the board;
see Section~\ref{conclusion}.
Finally, if any row is now entirely filled, then that row gets removed,
and all rows above shift down, creating one new empty row at the top.
%If the piece extends above the top row of the board,
%then the player loses the game.\footnote{Several variations on loss conditions are used in different Tetris games including: clearing the row before checking for out of bounds pieces, or checking if a piece is entirely (rather than partially) out of bounds. Our choice is consistent with the classic Atari and NES versions of the game.\cite{lockout}}

%For specific rules regarding rotations and piece movement, we again refer the reader to \cite{Tetris_IJCGA}.

%\begin{figure}
%\centering
%    \includegraphics[width=0.8\textwidth]{tetris/figures/Tdrawings_Tpieces.pdf}
%\caption{Tetris blocks $\ms O, \ms J, \ms L, \ms Z, \ms S, \ms I, \ms T$, from left to right}
%\label{fig:blocks}
%\end{figure}%TODO add diagram; cite figure number

\label{reasonable rotation}
Like \cite[Section~2]{Tetris_IJCGA}, we allow any model of piece rotation that satisfies
two ``reasonable'' restrictions.  First, a piece cannot ``jump'' from one
connected component of the unfilled space to another.
Second, any piece that is not $1 \times k$ (in particular, not $\II$)
cannot ``squeeze through'' a single-cell choke point.
Precisely, if cell $\langle i, j \rangle$ is unfilled and
either $\langle i \pm 1, j \rangle$ are both filled
or $\langle i, j \pm 1 \rangle$ are both filled,
and the filling of cell $\langle i, j \rangle$ would partition the
unfilled space into two connected components, then a piece that is not
$1 \times k$ cannot jump between these two connected components.
This model is a slight strengthening of the one in \cite{Tetris_IJCGA},
which only considered the $\langle i, j \pm 1 \rangle$ case because it
only had to consider vertical passage; in our $O(1)$-row proof, we need to
also consider horizontal passage.
Notably, this rotation model includes the Classic Rotation System
where each piece rotates about a center interior to the piece,
and the operation fails if that rotation would overlap a filled square
\cite{Tetris-guideline};
and it includes the more-complex Super Rotation System (SRS)
that is now standard to Tetris \cite{SRS,Tetris-guideline},
which adds a series of possible translation checks that attempt to avoid
collisions via ``wall kicks''.

The \ktet{$(\leq) k$} problem is the following decision problem:
given a starting configuration of filled cells, and the sequence
$p_1, p_2, \dots, p_n$ of $(\leq) k$-omino pieces that will arrive,
can the player maneuver the pieces to avoid pieces freezing above the
top row, and optionally, reach a state where the entire board is unfilled?
{\tet} is the special case \ktet{4}.
%By \cite[Lemma~2.1]{Tetris_IJCGA}, {\tet} is in NP,
%and thus so are the special cases considered in this paper:
These problems are trivially in NP: a certificate is the sequence of
player moves (a linear number of translations and/or rotations) between each
unit piece drop.
As a result, so are the following special cases considered in this paper:

\begin{enumerate}
\item \cktet{$c$}{$(\leq) k$}, where the board has $c$ columns.
\item \rktet{$r$}{$(\leq) k$}, where the board has $r$ rows.
\item \ektet{$(\leq) k$}, where the board's initial configuration is entirely unfilled.
\end{enumerate}

%The problem we consider here is the \textit{\col-column Tetris problem} (the \textbf{\tet} problem), which is the same as the {\tet} problem, except that the number of columns in the board is equal to \col. We will show that the {\tet} problem is NP-complete.

%% file: tetris/polynomial.tex
\section{2-column Tetris is polynomial}
\label{sec:2-column}

For completeness, we start with an easy result about one row or column:

\begin{prop} \label{prop:1 col 1 row}
  \cktet{1}{$\leq k$} and \rktet{1}{$k$} are solvable in linear time.
  %and \erktet{1}{$\leq k$} is solvable in polynomial time
  %(assuming $k$ is polynomial).
\end{prop}

\begin{proof}
  In \cktet{1}{$\leq k$}, only $1 \times j$ pieces (e.g., $\II$) are valid.
  (Any other piece does not fit in the board.)
  Every such piece immediately fills any rows it occupies, so immediately
  disappears.  Thus every sequence of pieces that fit in the board
  is a trivial win for the player.

  In \rktet{1}{$k$}, any piece that is not $1 \times k$
  results in an immediate loss for the player.
  If there are only $1 \times k$ pieces in the sequence,
  then we can follow a greedy strategy: place each piece in the leftmost
  position where it fits.  If there is any way to clear the initial row
  configuration, then this algorithm will produce one.
  If there is any way to fill a then-empty row (i.e., $k$ divides the board
  width), then we claim that this algorithm will produce one.  
  Furthermore, if the row cannot be cleared, then we claim that this algorithm
  will make the most moves possible before getting stuck.
  These claims follow from a simple greedy argument: take any strategy,
  sort its piece placements between line clears from left to right,
  and if a piece placement is not maximally left, shift it so.
  Thus, this algorithm will play for as long as is possible.
\end{proof}

%\xxx[Michael]{Someone besides me double check this (looks fine to jaysonl)}
%\xxx{Written in terms of tetromino pieces, but it should generalize to any set of $O(1)$-sized pieces}
%\xxx{TODO: figures}
%\xxx{TODO: find best source for $A_{PDA} \in P$}

\begin{thm}\label{thm:2 row}
  \cktet{2}{$O(n)$} is solvable in polynomial time.
\end{thm}

\begin{proof}
We reduce to $A_{PDA}$, the acceptance problem for nondeterministic pushdown automata, which is known to be in P by reduction to the acceptance problem for CNF Context-Free Grammars \cite{sipser2006} and the CYK algorithm that solves it \cite{younger1967recognition}.
Given a set $\Sigma$ of $\leq k$-omino pieces, a piece sequence $p$ of length $n$ over $\Sigma$, the initial board configuration $B$, and board height $h$, we output $(M,p)$, where $M$ is a PDA that recognizes piece sequences from $\Sigma$ that permit staying under $h$ rows starting from board $B$, and optionally clearing the board.

The constructed pushdown automaton $M$ represents the board state in its stack, with the topmost occupied row at the top. Its stack alphabet $\Gamma = \{\blacksquare\square, \square\blacksquare\}$ represents the two possible configurations of a 2-column row (as $\blacksquare\blacksquare$ is invalid and $\square\square$ is invalid below the topmost occupied row). Each piece $p_i \in \Sigma$ is described by a string
over $\{\blacksquare\square, \square\blacksquare, \blacksquare\blacksquare\}$, 
%with the pattern $(\blacksquare\square \mid \square\blacksquare \mid \blacksquare\blacksquare)^*$,
as wider pieces cannot fit within two columns.

When a piece $p_i$ comes in, it would suffice for $M$ to pop and observe the top $|p_i| \leq k$ rows to determine how each placement choice changes the board state. Because a move can either delete or add up to $k$ occupied rows at the top of the board, at most $2k$ new rows need to be pushed to realize the outcome of a chosen placement. This locality of updates may not seem true for $1 \times j$ pieces with pattern $(\square\blacksquare)^*$ like the $\II$ tetromino, because they can descend arbitrarily far below the topmost row; but observe that each row it passes through and clears must be identical, either all $\blacksquare\square$ or all $\square\blacksquare$, so the resulting board state would be the same as if the top $j$ rows were cleared instead.

However, while this implementation would be sufficient, it will not be efficient for large $k$: there are $2^k$ possible sequences of $k$ rows, so if $M$ always popped $k$ rows into its finite state space, we would have exponential blowup. We can fix this issue by noticing that there are only two possible row patterns that $M$ needs to handle, $(\blacksquare\square)^* (\square\blacksquare)?$ and its mirror image $(\square\blacksquare)^* (\blacksquare\square)?$, as (by the reasonable rotation assumption) no piece can pass through two unequal rows or affect any rows below them. This reduces the number of possibilities to $O(k)$, keeping $M$'s state space small.
Figures~\ref{fig:2col-z-t}, \ref{fig:2col-j}, and \ref{fig:2col-o-i} show
how to handle the standard tetromino pieces.

\begin{figure}
    \centering
    \begin{subfigure}{0.22\textwidth}
    \centering
    \includegraphics[scale=0.48]{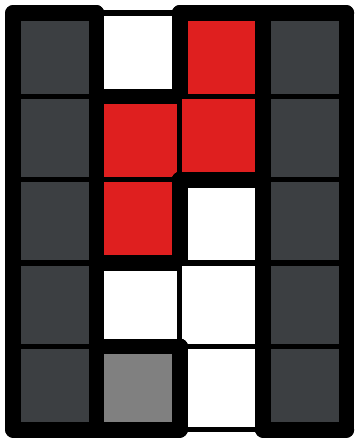}
    \includegraphics[scale=0.48]{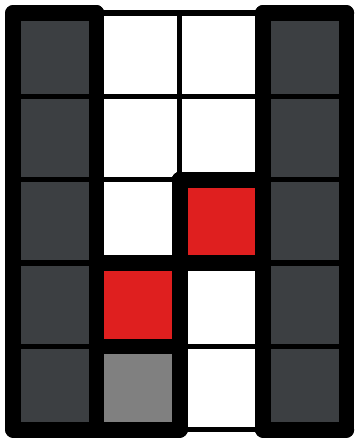}
    \caption{$\protect\ZZ$ onto $\blacksquare\square$}
    \end{subfigure}
    \hspace{0.015\textwidth}
    \begin{subfigure}{0.22\textwidth}
    \centering
    \includegraphics[scale=0.48]{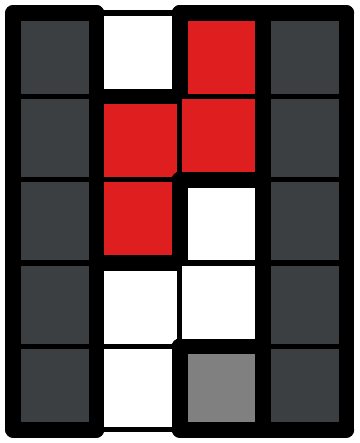}
    \includegraphics[scale=0.48]{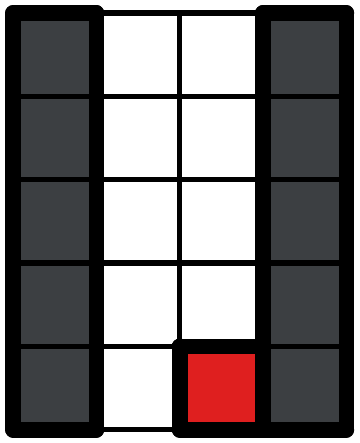}
    \caption{$\protect\ZZ$ onto $\square\blacksquare$}
    \end{subfigure}
    \hspace{0.015\textwidth}
    \begin{subfigure}{0.22\textwidth}
    \centering
    \includegraphics[scale=0.48]{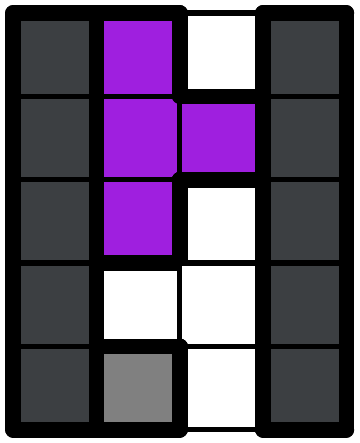}
    \includegraphics[scale=0.48]{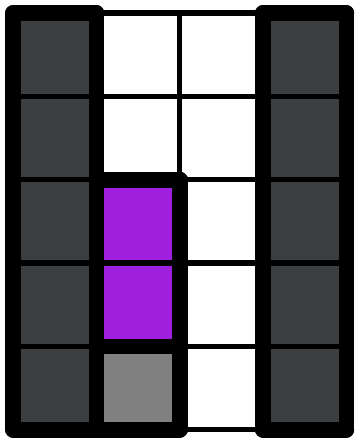}
    \caption{$\protect\TT$ onto $\blacksquare\square$}
    \end{subfigure}
    \hspace{0.015\textwidth}
    \begin{subfigure}{0.22\textwidth}
    \centering
    \includegraphics[scale=0.48]{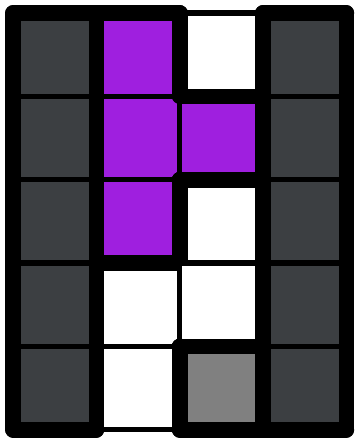}
    \includegraphics[scale=0.48]{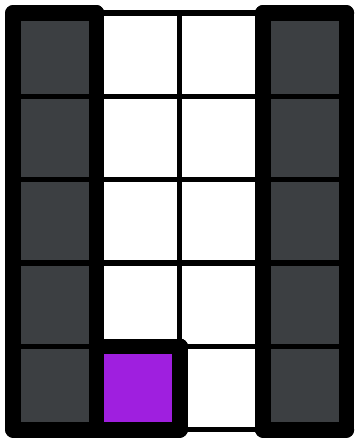}
    \caption{$\protect\TT$ onto $\square\blacksquare$}
    \end{subfigure}
    \caption{Possible 2-column outcomes for placing $\protect\ZZ$ or $\protect\TT$ pieces. $\protect\SS$ and flipped $\protect\TT$ cases are symmetric.}
    \label{fig:2col-z-t}
\end{figure}

\begin{figure}
    \centering
    %\begin{subfigure}{0.24\textwidth}
    %\centering
    %\includegraphics[scale=0.5]{tetris/figures/2col/2columns_J_up_1.pdf}
    %\includegraphics[scale=0.5]{tetris/figures/2col/2columns_J_up_2.pdf}
    %\caption{J:Add tower of two}
    %\end{subfigure}
    \begin{subfigure}{0.22\textwidth}
    \centering
    \includegraphics[scale=0.48]{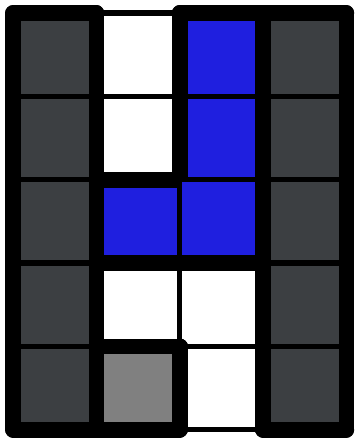}
    \includegraphics[scale=0.48]{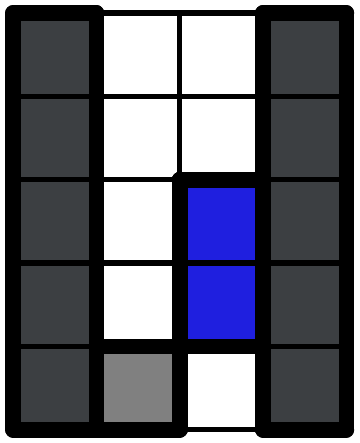}
    \caption{$\protect\JJ$ onto $\blacksquare\square$ (or $\square\blacksquare$)}
    \end{subfigure}
    \hspace{0.015\textwidth}
    \begin{subfigure}{0.22\textwidth}
    \centering
    \includegraphics[scale=0.48]{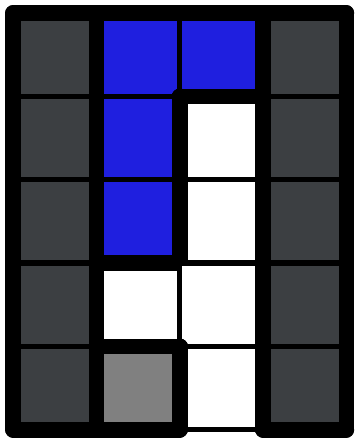}
    \includegraphics[scale=0.48]{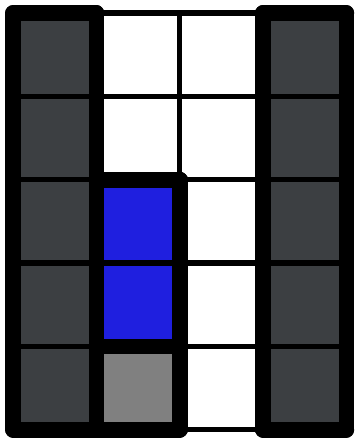}
    \caption{$\protect\JJ$ flipped onto $\blacksquare\square$}
    \end{subfigure}
    \hspace{0.015\textwidth}
    \begin{subfigure}{0.22\textwidth}
    \centering
    \includegraphics[scale=0.48]{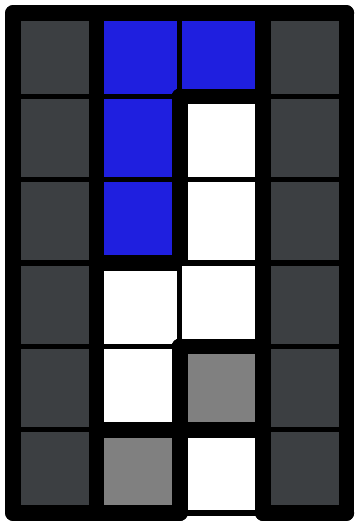}
    \includegraphics[scale=0.48]{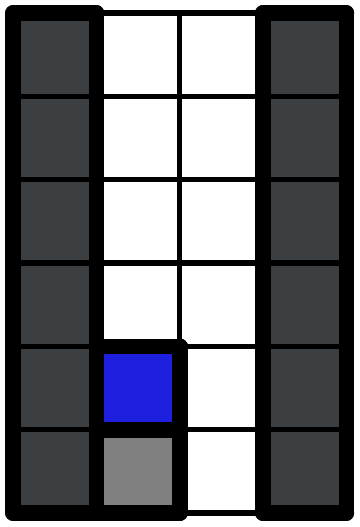}
    \caption{\raggedright $\protect\JJ$ flipped onto $\square\blacksquare, \blacksquare\square$}
    \end{subfigure}
    \hspace{0.015\textwidth}
    \begin{subfigure}{0.22\textwidth}
    \centering
    \includegraphics[scale=0.48]{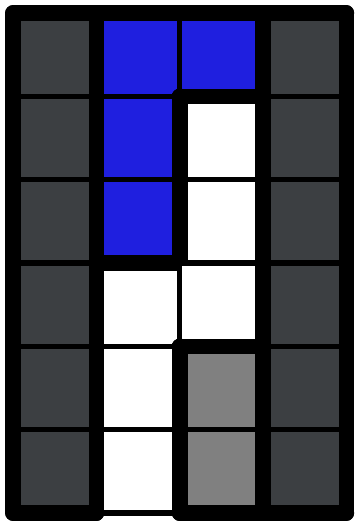}
    \includegraphics[scale=0.48]{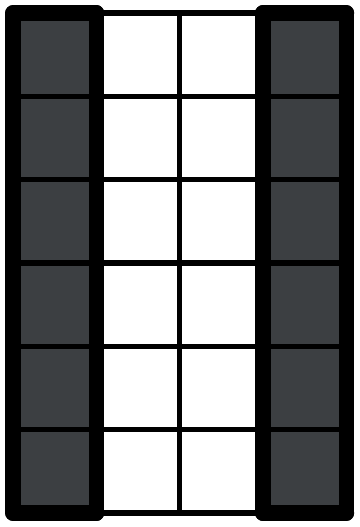}
    \caption{\raggedright $\protect\JJ$ flipped onto $\square\blacksquare, \square\blacksquare$}
    \end{subfigure}
    \caption{Possible 2-column outcomes for placing $\protect\JJ$ pieces. $\protect\LL$ cases are symmetric.}
    \label{fig:2col-j}
\end{figure}

\begin{figure}
    \centering
    \begin{subfigure}{0.24\textwidth}
    \centering
    \includegraphics[scale=0.48]{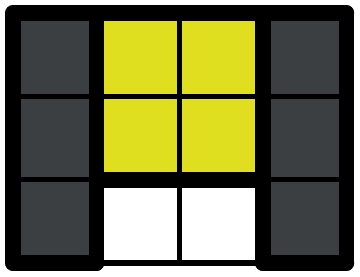}
    \includegraphics[scale=0.48]{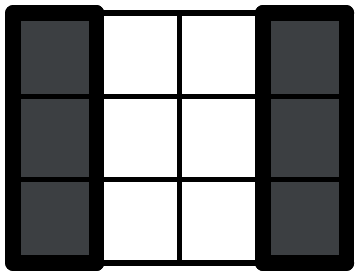}
    \caption{$\protect\OO$ on any board.}
    \end{subfigure}
    \hspace{0.015\textwidth}
    \begin{subfigure}{0.24\textwidth}
    \centering
    \includegraphics[scale=0.48]{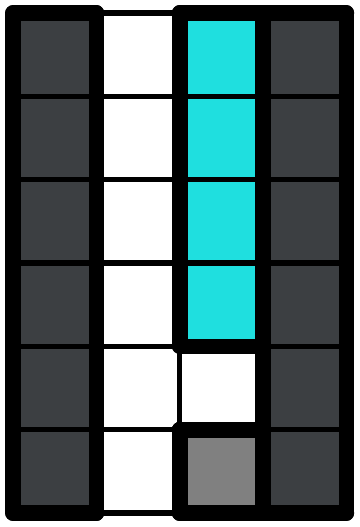}
    \includegraphics[scale=0.48]{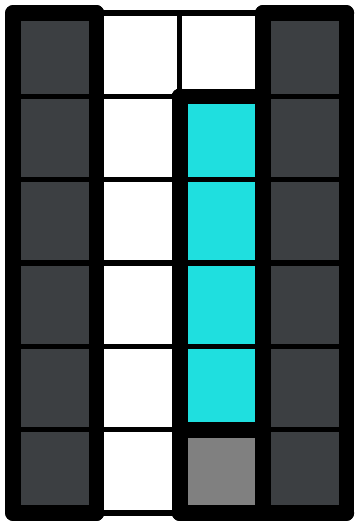}
    \caption{$\protect\II$ on top of $\square\blacksquare$}
    \end{subfigure}
    \hspace{0.015\textwidth}
    \begin{subfigure}{0.28\textwidth}
    \centering
    \includegraphics[scale=0.48]{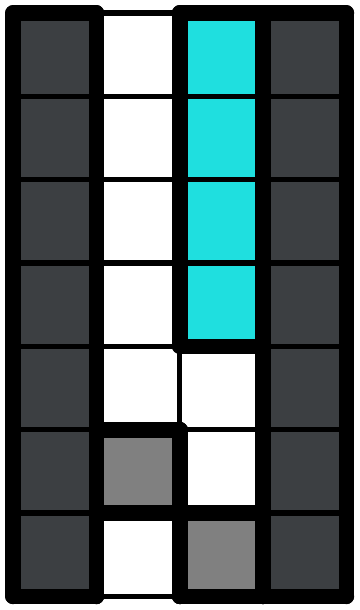}
    \includegraphics[scale=0.48]{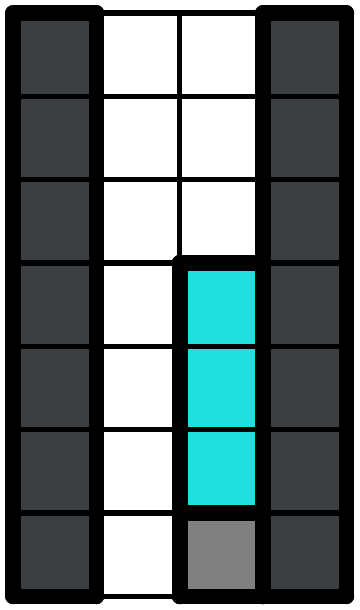}
    \caption{$\protect\II$ into $(\blacksquare\square)^1(\square\blacksquare)$}
    \end{subfigure}

    \bigskip

    \begin{subfigure}{0.24\textwidth}
    \centering
    \includegraphics[scale=0.48]{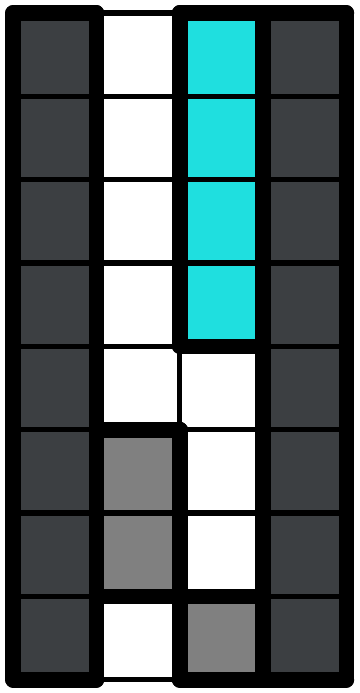}
    \includegraphics[scale=0.48]{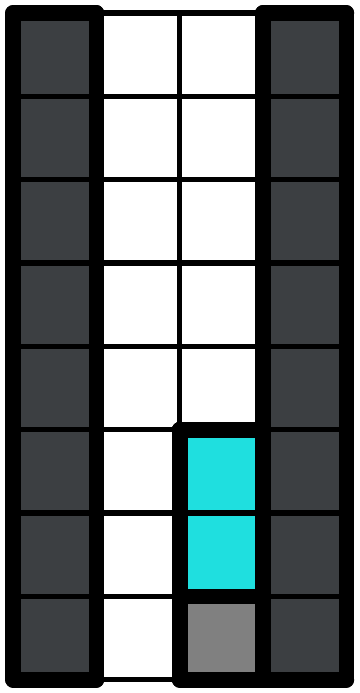}
    \caption{$\protect\II$ into $(\blacksquare\square)^2(\square\blacksquare)$}
    \end{subfigure}
    \hspace{0.015\textwidth}
    \begin{subfigure}{0.24\textwidth}
    \centering
    \includegraphics[scale=0.48]{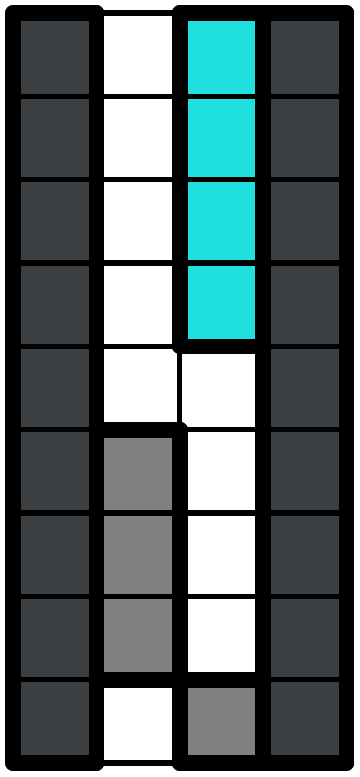}
    \includegraphics[scale=0.48]{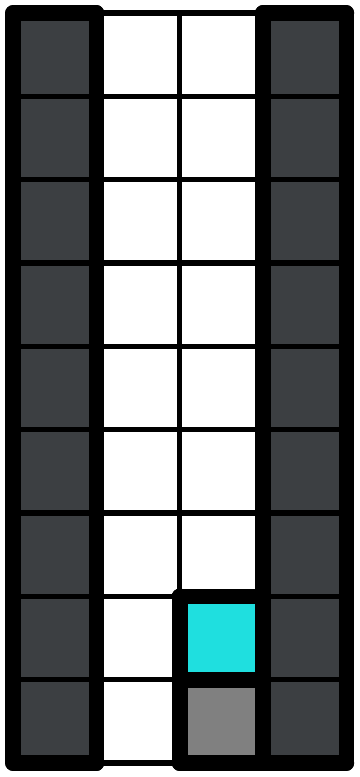}
    \caption{$\protect\II$ into $(\blacksquare\square)^3(\square\blacksquare)$}
    \end{subfigure}
    \hspace{0.015\textwidth}
    %\begin{subfigure}{0.22\textwidth}
    %\centering
    %\includegraphics[scale=0.48]{tetris/figures/2col/2columns_I_9.pdf}
    %\includegraphics[scale=0.48]{tetris/figures/2col/2columns_I_10.pdf}
    %\caption{$\protect\II$ into $(\blacksquare\square)^4(\blacksquare\square)$}
    %\end{subfigure}
    %\hspace{0.015\textwidth}
    \begin{subfigure}{0.28\textwidth}
    \centering
    \includegraphics[scale=0.48]{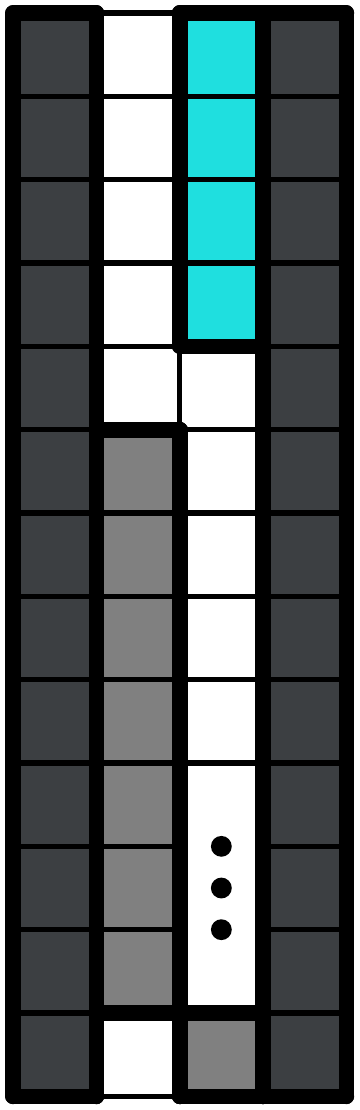}
    \includegraphics[scale=0.48]{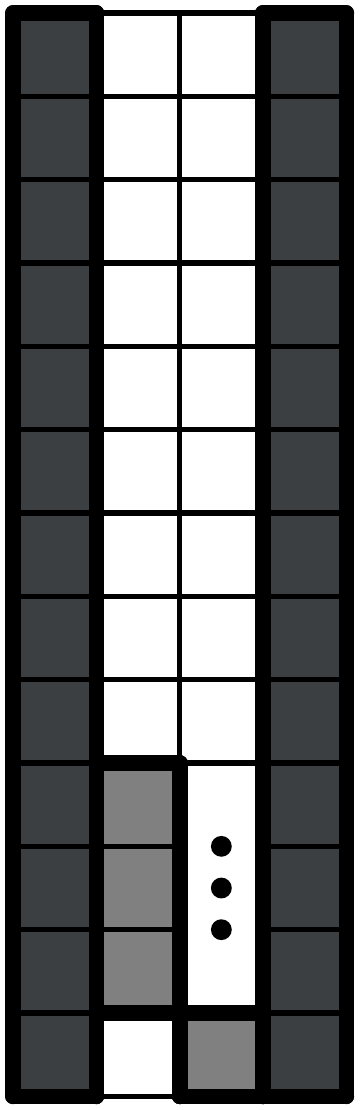}
    \caption{$\protect\II$ into $(\blacksquare\square)^4 (\blacksquare\square)^* (\square\blacksquare)$}
    \end{subfigure}
    \caption{Possible 2-column outcomes for placing $\protect\OO$ and $\protect\II$ pieces. Mirrored cases are symmetric.}
    \label{fig:2col-o-i}
\end{figure}

The execution of $M$ starts by pushing $B$ onto the stack before reading any input, initializing the board. For each piece $p_i$, $M$ nondeterministically chooses which orientation to place $p_i$ on the board. As it runs, $M$ enforces the maximum row constraint by keeping a stack height counter in its state, which is incremented and decremented appropriately as $M$ pushes and pops rows. If $M$ places a piece over the height limit $h$, then $M$ enters an inescapable rejecting state on that branch, whereas all other states (or optionally only those with an empty stack) are accepting states. With these finite-state implementable rules, $M$ performs a correct simulation which recognizes winning games, therefore $(\Sigma,p,B,h)$ is in \cktet{2}{$\leq k$} if and only if $(M,p)$ is in $A_{PDA}$.

To bound the running time,
note that %we can bound $h$ to be polynomially large by assuming $|B| \leq h \leq |B| + k n$, as if $|B| > h$ then the game is lost at the start, and if
we may assume $h \leq |B| + k n$,
as otherwise it would be impossible to reach the top row and lose.
%These bounds allow us to show that this reduction takes polynomial time.
Because producing $M$ is dominated by the computation of its transition function --- enumerating and simulating the outcomes of the $O(|\Sigma| \times h \times k)$ scenarios of placing every piece in each orientation at all legal heights with every pattern of the top $k$ rows --- the size of $M$ and the time required to produce it is polynomial in the input size.
\end{proof}

%% file: tetris/nphard_col.tex
\section{\col-column Tetris is NP-hard}
\label{sec:8-column}

In this section, we prove the following theorem:

%\begin{thm} \label{maintet}
%{\tet} is NP-hard.
%\end{thm}
%
%The original Tetris paper \cite{Tetris_IJCGA} asked specifically about the complexity of \ctet{$c$} for $c=O(1)$, motivated by real-world Tetris where $c=10$.
%Our main result is the following:
%In the (perfect-information) {\tet} problem \cite{Tetris_IJCGA}, we are given an initial board state of filled cells and a sequence of pieces that will arrive, and the goal is to place the pieces in sequence to either survive (not go above the top row) or clear the entire board. This problem was proved NP-hard for arbitrary board sizes in 2002 \cite{Tetris_IJCGA}, and more recently for other polyomino pieces \cite{TotalTetris_JIP}. The variant we consider here is the \emph{$c$-column Tetris problem} (abbreviated \emph{\ctet{$c$}}), which is the {\tet} problem restricted to boards with exactly $c$ columns.

\begin{thm} \label{thm:8 col}
It is NP-complete to survive or clear the board in \ctet{$c$} for any $c \geq 8$.
\end{thm}

%Membership in NP follows from the same result for general {\tet} \cite[Lemma~2.1]{Tetris_IJCGA}.
Like \cite{Tetris_IJCGA}, we reduce from the strongly NP-hard {\partit} problem.
%given a multiset of nonnegative integers $\{a_1, a_2, \dots, a_{3s}\}$ and a nonnegative integer $T$ satisfying the constraints $\sum_{i=1}^{3s} a_i = sT$ and $\frac T4 < a_i < \frac T2$ for all $1 \le i \le 3s$,
%determine whether $\{a_1, a_2, \dots, a_{3s}\}$ can be partitioned into $s$ (disjoint) triples, each of which sum to exactly $T$.
%\xxx{This definition is repeated on the next page. Consolidate?}
%

\begin{defn}
The \textbf{\partit} problem is defined as follows:

\textbf{Input:} A set of nonnegative integers $\{a_1, a_2, \dots, a_{3s}\}$ and a nonnegative integer $T$ satisfying the constraints $\sum_{i=1}^{3s} a_i = sT$ and $\frac T4 < a_i < \frac T2$ for all $1 \le i \le 3s$.

\textbf{Output:} Whether $\{a_1, a_2, \dots, a_{3s}\}$ can be partitioned into $s$ (disjoint) sets of size 3, each of which sum to exactly $T$.
\end{defn}

For the reduction, we exhibit a mapping from {\partit} instances to {\tet} instances so that the following is satisfied:

\begin{lem}[\ctet{$c$} $\iff {\partit}$] For a ``yes'' instance of {\partit}, there is a way to drop the pieces that clears the entire board without triggering a loss. Conversely, if the board can be cleared, then the {\partit} instance has a solution.
\end{lem}

\begin{proof}[Proof sketch]
The initial board, illustrated in Figure~\ref{fig:init}
(where filled cells are grey and the rest of the cells are unfilled), has
%$8$ columns and 
%$8sT + 24s + 17$
$12sT+48s+17$
rows. The reduction is polynomial size.
%Do we want to say that dots in cells represent a continuation of the pattern? I'm not sure it is entirely obvious.

\begin{figure}
    \centering
    \begin{subfigure}[t]{0.1\textwidth}
      \begin{overpic}[width=\textwidth]{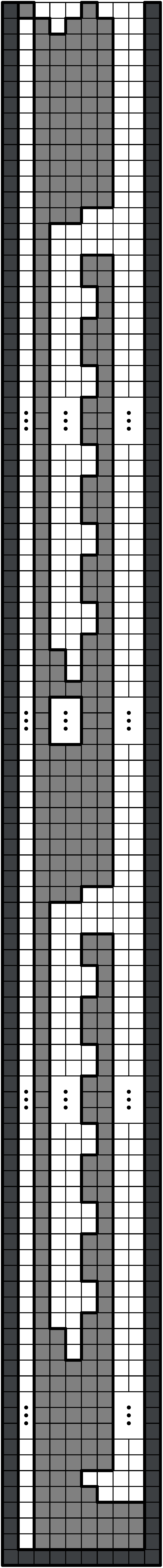}
        \footnotesize\bf
        \color{purple}
        \put(0.85,50){\rotatebox{90}{\hbox to 0pt{\hss alley\hss}}}
        \put(3.7,65){\rotatebox{90}{\hbox to 0pt{\hss bucket\hss}}}
        \put(3.7,20){\rotatebox{90}{\hbox to 0pt{\hss bucket\hss}}}
        \put(7.7,40){\rotatebox{90}{\hbox to 0pt{\hss corridor\hss}}}
      \end{overpic}
      \caption{}
      \label{fig:init}
    \end{subfigure}
    \hspace{0.015\textwidth}
    \begin{subfigure}[t]{0.1\textwidth}
        \includegraphics[width=\textwidth]{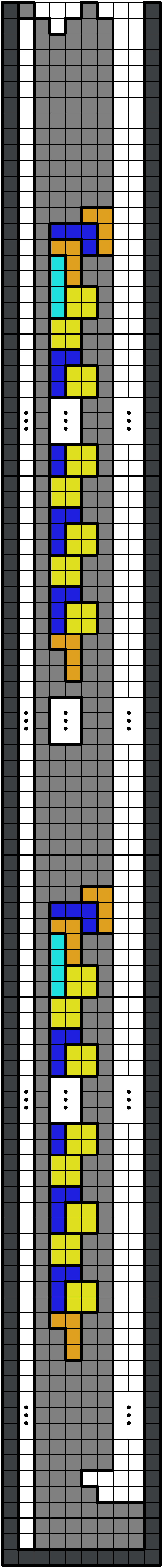}
        \caption{}
    \end{subfigure}
    \hspace{0.015\textwidth}
    \begin{subfigure}[t]{0.1\textwidth}
        \includegraphics[width=\textwidth]{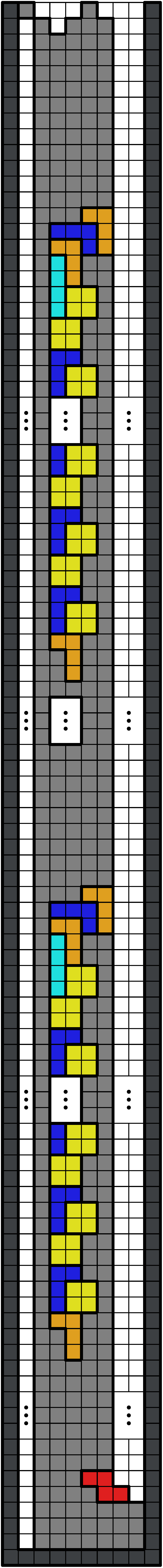}
        \caption{}
    \end{subfigure}
    \hspace{0.015\textwidth}
    \begin{subfigure}[t]{0.1\textwidth}
        \includegraphics[width=\textwidth]{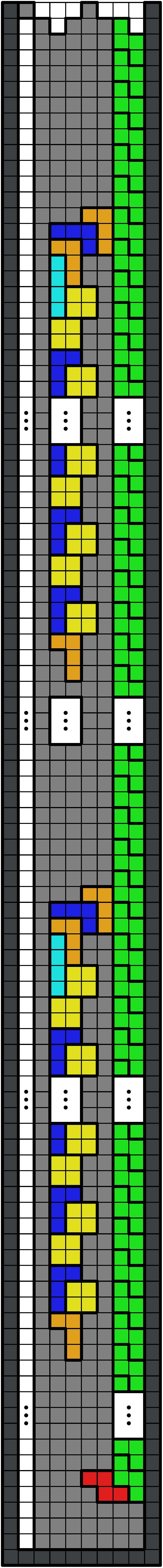}
        \caption{}
        \label{fig:corridorFill}
    \end{subfigure}
    \hspace{0.015\textwidth}
    \begin{subfigure}[t]{0.1\textwidth}
        \includegraphics[width=\textwidth]{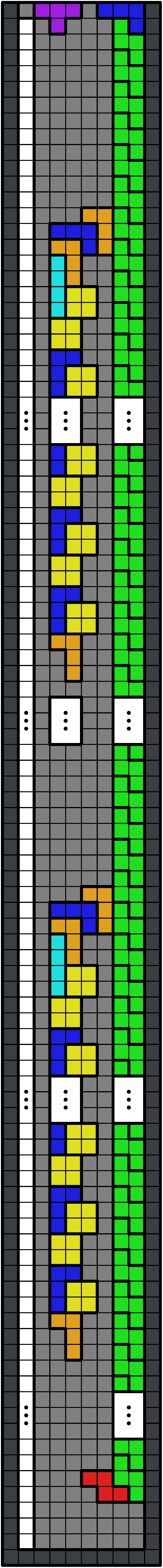}
        \caption{}
    \end{subfigure}
    \hspace{0.015\textwidth}
    \begin{subfigure}[t]{0.1\textwidth}
        \includegraphics[width=\textwidth]{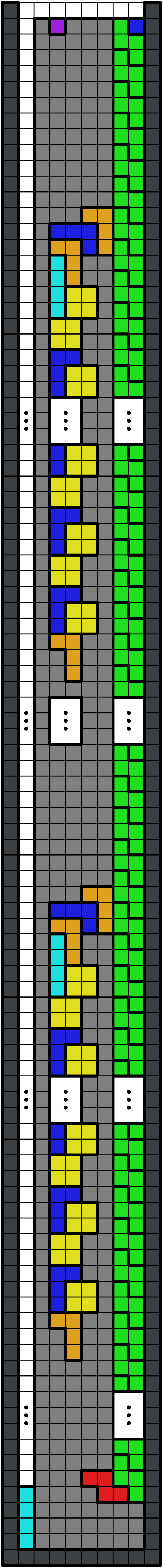}
        \caption{}
    \end{subfigure}
    \hspace{0.015\textwidth}
    \begin{subfigure}[t]{0.1\textwidth}
        \includegraphics[width=\textwidth]{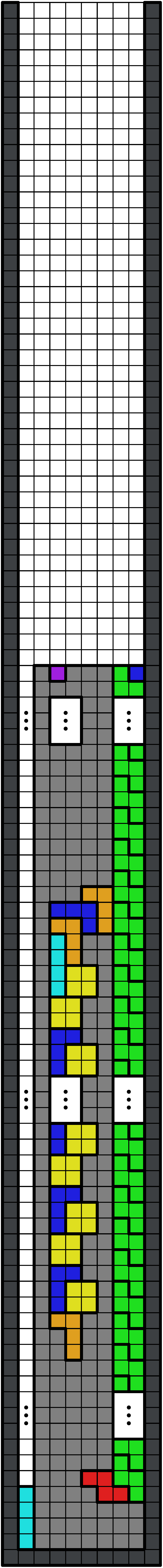}
        \caption{}
    \end{subfigure}
    \hspace{0.015\textwidth}
    \begin{subfigure}[t]{0.1\textwidth}
        \includegraphics[width=\textwidth]{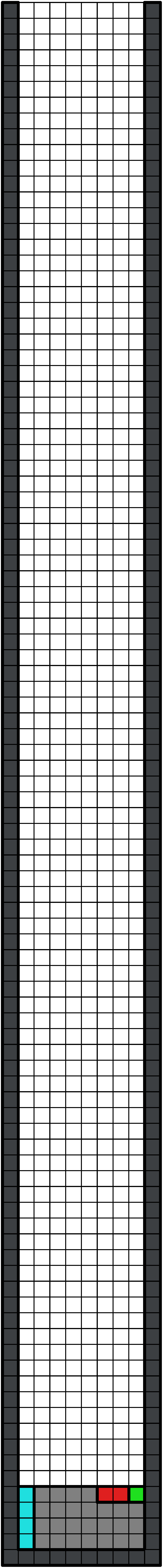}
        \caption{}
    \end{subfigure}
    \\
    \begin{subfigure}[h]{0.1\textwidth}
        \includegraphics[width=\textwidth]{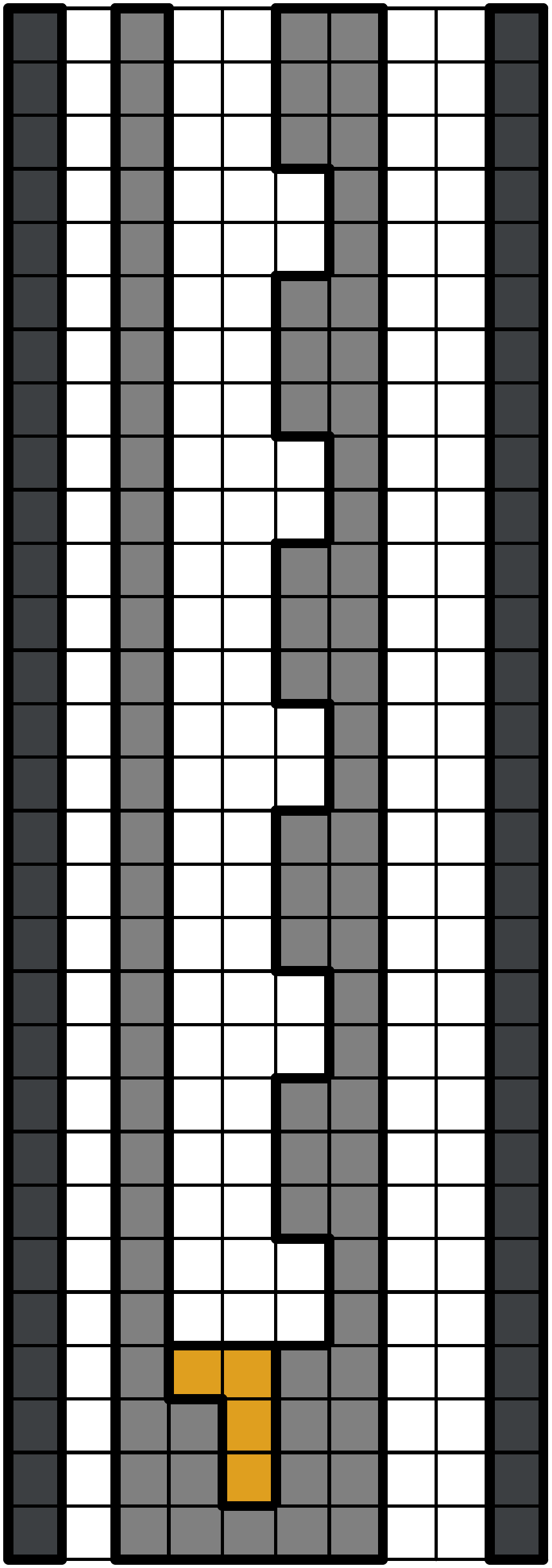}
        \caption{}
    \end{subfigure}
    \hspace{0.01\textwidth}
    \begin{subfigure}[h]{0.1\textwidth}
        \includegraphics[width=\textwidth]{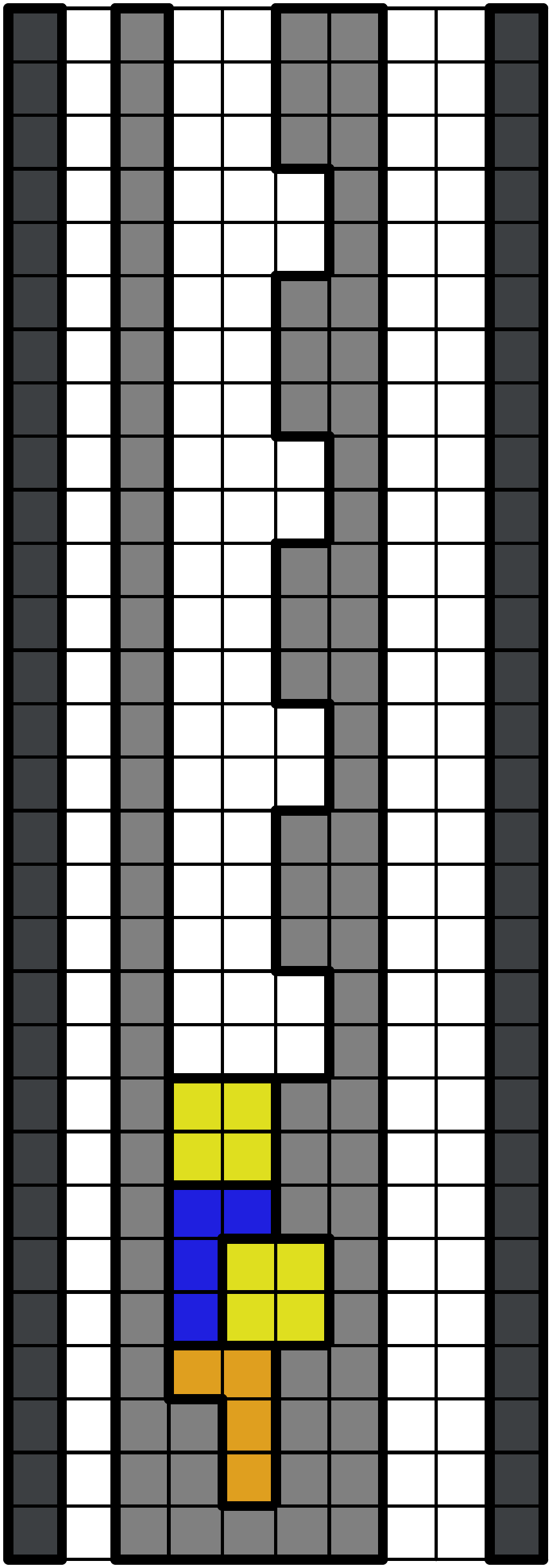}
        \caption{}
    \end{subfigure}
    \hspace{0.01\textwidth}
    \begin{subfigure}[h]{0.1\textwidth}
        \includegraphics[width=\textwidth]{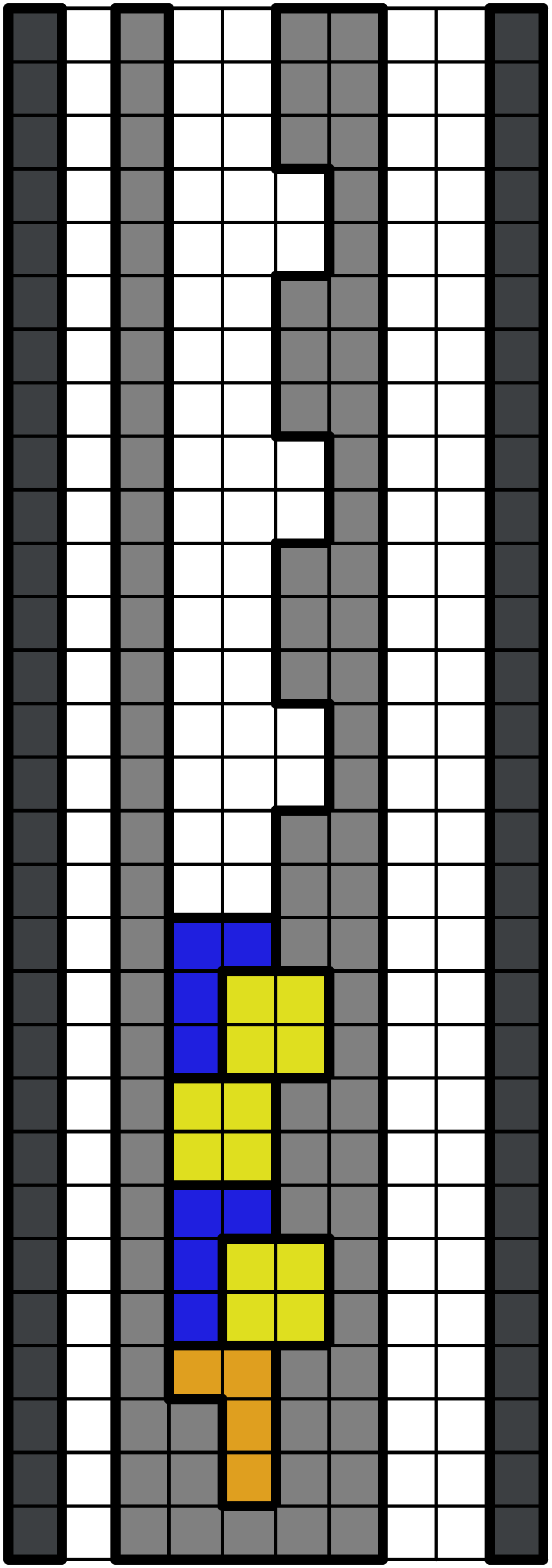}
        \caption{}
    \end{subfigure}
    \hspace{0.01\textwidth}
    \begin{subfigure}[h]{0.1\textwidth}
        \includegraphics[width=\textwidth]{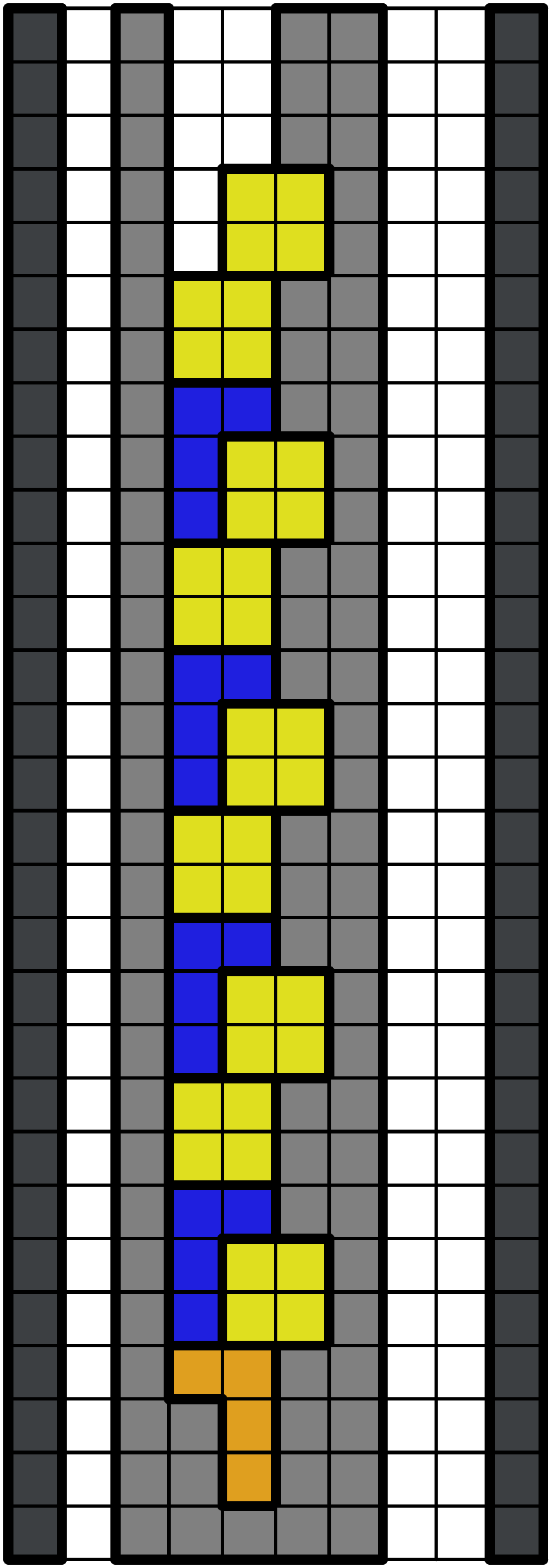}
        \caption{}
    \end{subfigure}
    \hspace{0.01\textwidth}
    \begin{subfigure}[h]{0.1\textwidth}
        \includegraphics[width=\textwidth]{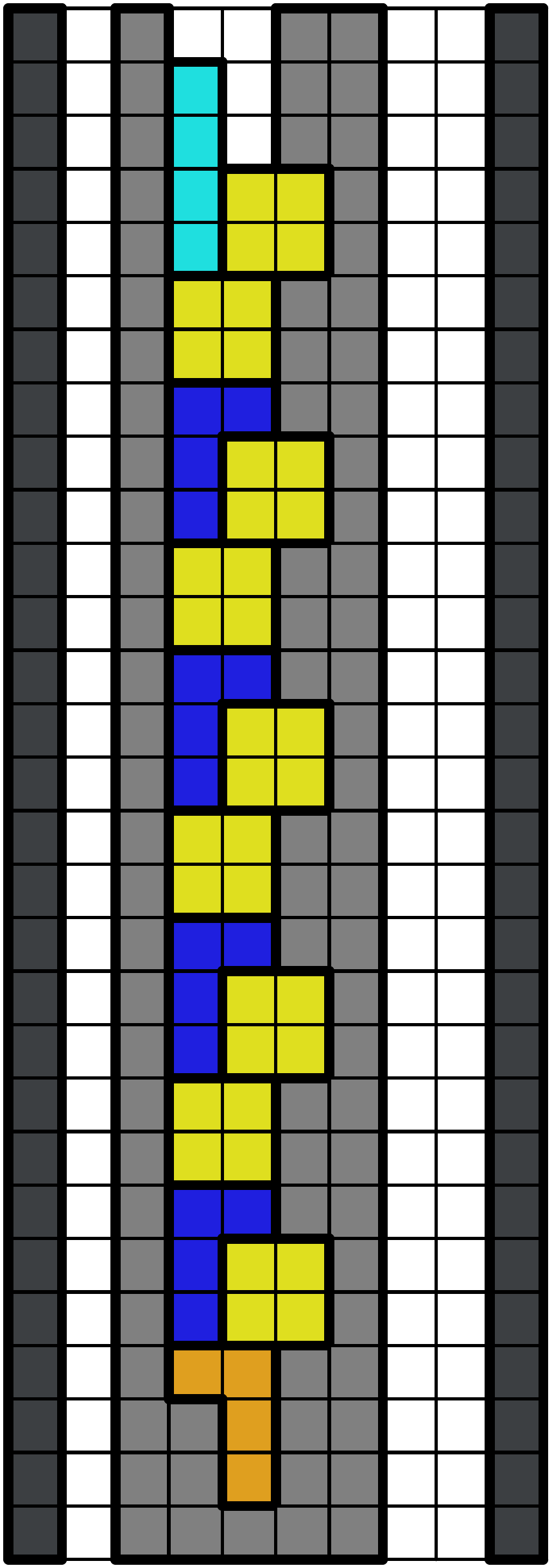}
        \caption{}
    \end{subfigure}
    \hspace{0.05\textwidth}
    \begin{subfigure}[h]{0.1\textwidth}
        \includegraphics[width=\textwidth]{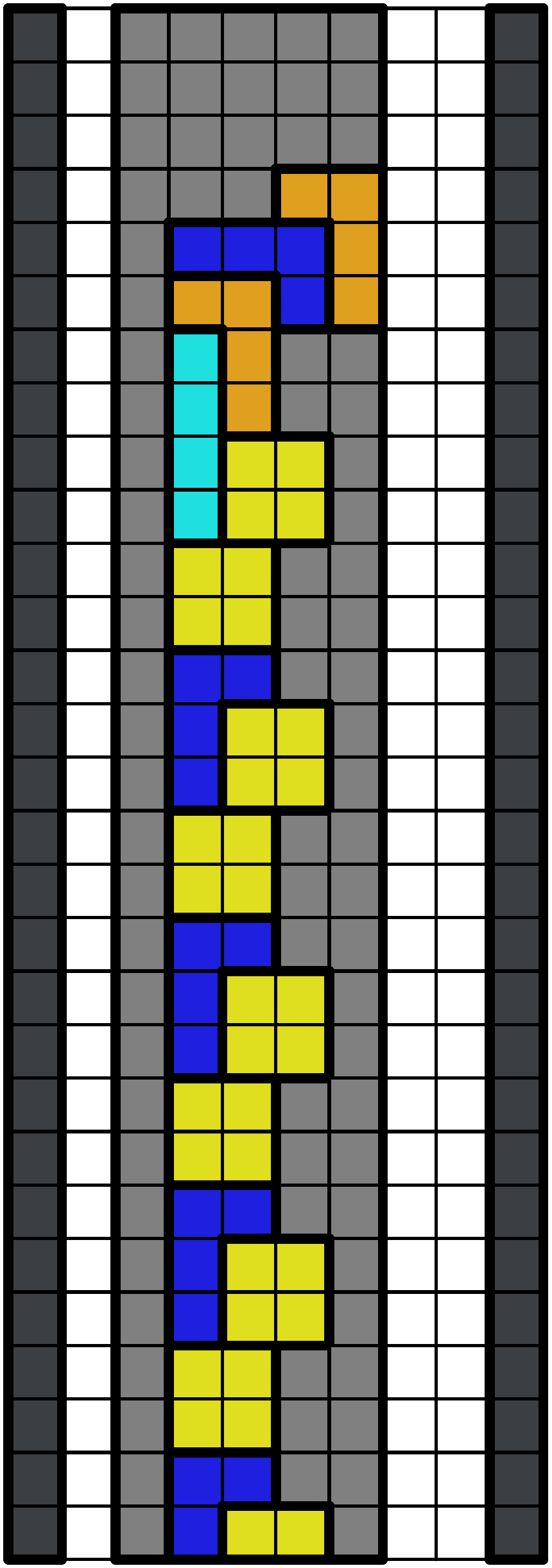}
        \caption{}
    \end{subfigure}
% \\
    % \begin{subfigure}[b]{0.08\textwidth}
    %     \includegraphics[width=\textwidth]{tetris/figures/corridor/corridor_L2.pdf}
    %     \caption{}
    % \end{subfigure}
    % \hspace{0.01\textwidth}
    % \begin{subfigure}[b]{0.08\textwidth}
    %     \includegraphics[width=\textwidth]{tetris/figures/corridor/corridor_L3.pdf}
    %     \caption{}
    % \end{subfigure}
    % \hspace{0.01\textwidth}
    % \begin{subfigure}[b]{0.08\textwidth}
    %     \includegraphics[width=\textwidth]{tetris/figures/corridor/corridor_L4.pdf}
    %     \caption{}
    % \end{subfigure}
    % \hspace{0.01\textwidth}
    % \begin{subfigure}[b]{0.08\textwidth}
    %     \includegraphics[width=\textwidth]{tetris/figures/corridor/corridor_O2.pdf}
    %     \caption{}
    % \end{subfigure}
    % \hspace{0.01\textwidth}
    % \begin{subfigure}[b]{0.08\textwidth}
    %     \includegraphics[width=\textwidth]{tetris/figures/corridor/corridor_J2.pdf}
    %     \caption{}
    % \end{subfigure}
    % \hspace{0.01\textwidth}
    % \begin{subfigure}[b]{0.08\textwidth}
    %     \includegraphics[width=\textwidth]{tetris/figures/corridor/corridor_J3.pdf}
    %     \caption{}
    % \end{subfigure}
    % \hspace{0.01\textwidth}
    % \begin{subfigure}[b]{0.08\textwidth}
    %     \includegraphics[width=\textwidth]{tetris/figures/corridor/corridor_J4.pdf}
    %     \caption{}
    % \end{subfigure}
    % \hspace{0.01\textwidth}
    % \begin{subfigure}[b]{0.08\textwidth}
    %     \includegraphics[width=\textwidth]{tetris/figures/corridor/corridor_I2.pdf}
    %     \caption{}
    % \end{subfigure}
    % \hspace{0.01\textwidth}
    % \begin{subfigure}[b]{0.08\textwidth}
    %     \includegraphics[width=\textwidth]{tetris/figures/corridor/corridor_I3.pdf}
    %     \caption{}
    % \end{subfigure}
\vspace{-\medskipamount}
\caption{(a) shows the initial board. (b--h) demonstrate filling and clearing the board in the final clearing sequence. (i--m) show a valid sequence of moves for $a_i = 5$. (n) shows our bucket terminator.
%(o--w) show invalid possibilities for various pieces in the bucket.
}
\vspace{-\bigskipamount}
\label{fig:bucketFilling}
\end{figure}

The piece sequence is as follows.
First, for each $a_i$, we send the following \emph{$a_i$ sequence}
(see Figures~\ref{fig:bucketFilling}(i--m)):
\[\langle \LL, \langle \OO, \JJ, \OO \rangle^{a_i}, \OO, \II \rangle.\]
After all these pieces, we send the following \emph{clearing sequence}
(see Figures~\ref{fig:bucketFilling}(n) and (b--h)):
\[\langle \langle \LL, \JJ, \LL \rangle^{s},
%\ZZ, \langle \SS \rangle^{4sT+12s+7}, \JJ, \TT, \langle \II \rangle^{2sT+6s+4}
\ZZ, \langle \SS \rangle^{6sT+24s+6}, \JJ, \TT, \langle \II \rangle^{3sT+12s+4}
\rangle. \]

Figures~\ref{fig:bucketFilling}(b--n) illustrate that a solution to {\partit} clears the Tetris board. To show the other direction, we progressively constrain any {\tet} solution to a form that directly encodes a {\partit} solution. Because the area of the pieces sent is exactly equal to $4(12sT+48s+13)$, no cell can be left empty. We enumerate all possible cases to show that this goal is impossible to meet (some cell must be left empty) if there is no {\partit} solution. Figures~\ref{fig:bucketFilling}(o--w) show some of the cases. 
%\
% \xxx[Jayson]{It took me a while to verify the area sum because I didn't account for set size correctly, perhaps we should break it down in the full paper.}
\end{proof}

% \xxx[Jayson]{Merge this proof sketch with the 'reduction' section which is closer to the actual proof}

\begin{comment}
First, for each $a_i$, the following pieces arrive, in this order, as in \cite{Tetris_IJCGA}:

\begin{itemize}
\item The \emph{initiator}, which is a single $\LL$ piece.
\item The \emph{filler}, which consists of the pieces $\langle \OO, \JJ, \OO \rangle^{a_i}$ (which we take to mean the sequence $\langle \OO, \JJ, \OO \rangle$ repeated $a_i$ times).
\item The \emph{terminator}, which consists of the pieces $\langle \OO, \II\rangle$.
\end{itemize}

After the pieces corresponding to each $a_i$, we then have the following pieces, which together we call the \emph{$a_i$-sequence}:

\begin{itemize}
\item The \emph{bucket closer}, which consists of the pieces $\langle \LL, \OO, \JJ, \OO, \OO, \JJ, \JJ, \LL\rangle^s$.
\item A single $\ZZ$ (note this is the first $\ZZ$ to arrive).
\item The \emph{corridor closer}, which consists of the pieces $\langle\langle \SS \rangle^{6sT+24s+6}, \JJ\rangle$.
\item A single $\TT$ (note this is the first $\TT$ to arrive).
\item The \emph{clearer}, which consists of the pieces $\langle \II \rangle^{3sT+12s+4}$.
\end{itemize}
\end{comment}

%{\partit} was proven in 1975 to be NP-hard \cite{partition}:

\import{}{reduction.tex}

\import{}{completeness.tex}

\import{}{soundness.tex}

%% file: tetris/reduction.tex
\subsection{Reduction}

In this section, we detail our polynomial reduction from an instance
$\mc P = \langle \{a_1, a_2, \dots, a_{3s}\}, T \rangle$ of {\partit} to
an instance $\mc G = \mc G(\mc P)$ of {\tet}.
In later sections, we prove that $\mc P$ and $\mc G$ have the same answer, i.e.,
there exists a valid 3-partition if and only if there is a sequence of moves
that survives or that clears the Tetris board.

\subsubsection{Initial board}

The initial board, illustrated in %Figure~\ref{fig:init} 
Figure~\ref{fig:bucketFilling}(a)
(where filled cells are grey and the rest of the cells are unfilled), has $8$ columns and 
%$8sT + 24s + 16$
$12sT + 48s + 17$
rows. The columns are numbered 1 to 8 from left to right.
(To prove hardness for $c > 8$ columns, we simply fill all columns beyond the 8th.)
The unfilled cells consist of five main parts:
%$12sT + 48s + 26$

% \begin{figure}
%     \centering
%     \includegraphics[width=0.15\textwidth]{tetris/figures/Tdrawings_columns_board.pdf}
%     \caption{The initial board, without the top 9 empty rows}
%     \label{fig:init}
% \end{figure}

\begin{itemize}
%\item 9 empty rows at the top of the board (not shown in Figure~\ref{fig:init}).
\item The \textit{corridor}, which consists of a $2 \times 
%(8sT + 24s + 14)
(12sT + 48s + 12)$ rectangle, as well as a 3-square cut out on the bottom left.
% in addition to an empty space in the shape of a $\TT$ piece immediately underneath.
\item The \textit{buckets} (of which there are $s$), which branch off the corridor to its left. These are similar in shape to the buckets used in \cite{Tetris_IJCGA}: except for the first few and last few rows, their shape is periodic with a period of 5 rows. Each bucket has a total height of $5T+20$, and contains $T+3$ \textit{notches} (the pairs of adjacent empty cells in column 5, not including the 3 empty cells at the top).
\item The \textit{T-lock}, which is in the shape of a $\TT$ piece. The buckets are each separated by two rows.
\item The \textit{alley}, which is a $1 \times 
%(8sT + 24s + 16)
(12sT + 48s + 16)$ rectangle which is ``unlocked'' by the T-lock. %$5T+26$, and contains $T+3$
%we need a padding of +3TS to make things nicely divisible
\end{itemize}

We also define the \textit{horizon}, which is the horizontal line separating the empty rows at the top of the board from the topmost filled cell. When a row is cleared, the horizon moves down by one row per row that is cleared. We also define the \textit{bucket line} as the horizontal line immediately below the bottommost cell of the bottommost bucket. 
%Note that the bucket line is (initially) $5sT + 21s + 3$ cells below the horizon.

\subsubsection{Piece sequence} \label{seq}
The pieces arrive in the following order.

First, for each $a_i$, the following pieces arrive in the following order,
called the \emph{$a_i$-sequence}.
This part is identical to the reduction in \cite{Tetris_IJCGA}.

\begin{itemize}
\item The \textit{initiator}, which is a single $\LL$ piece.
\item The \textit{filler}, which consists of the pieces $\langle \OO, \JJ, \OO \rangle^{a_i}$ (which we take to mean the sequence $\langle \OO, \JJ, \OO \rangle$ repeated $a_i$ times).
\item The \textit{terminator}, which consists of the pieces $\langle \OO, \II\rangle$.
\end{itemize}

After all the $a_i$-sequences, we then have the following pieces
in the following order, called the \emph{closing sequence}:

\begin{itemize}
\item The \emph{bucket closers}, which consists of the pieces $\langle \LL, \JJ, \LL\rangle^s$.
\item A single $\ZZ$ (note this is the first $\ZZ$ to arrive).
\item The \emph{corridor closer}, which consists of the pieces $\langle\langle \SS \rangle^
%{4sT+12s+7}
{6sT+24s+6}
, \JJ\rangle$.
\item A single $\TT$ (note this is the first $\TT$ to arrive).
\item The \emph{clearer}, which consists of the pieces $\langle \II \rangle^
%{2sT+6s+4}
{3sT+12s+4}$.
\end{itemize}

%Math
%B = blowup, s = slots, 3s = max i, T = target, sT = total sum of numbers
%Slot height = 5BT + 3*3 [3 a_i] + 13 [clear sequence]
%Total height = s*(slot_height + 2 [space between slots]) + 2 [z lock] + 1 [Top row] + 3 [fixing pairty] +11 [just because?]
%= 40Ts + 24s + 17
%\subsubsection{Polynomial size of reduction}

The total size of the board is 
%$8(8sT + 24s + 16)$
$8(12sT + 48s + 17)$
and the total number of pieces is
%
%\[\sum_{i=1}^{3s} (3 + 3a_i) + 3s + 1 + (4sT + 12s+7) + 1 + 1 +(2sT + 6s + 4) = 10sT + 53s + 12,\]
\[\sum_{i=1}^{3s} (3 + 3a_i) + 3s + 1 + (6sT + 24s+ 6) + 1 + 1 +(3sT + 12s + 4) = 12sT + 48s + 13,\]
which are both polynomial in the size of the {\partit} instance.

%% file: tetris/completeness.tex
\subsection{{\partit} solvable \texorpdfstring{$\Rightarrow$}{=>} {\tet} solvable} \label{completeness}

In this section, we show one side of the bijection: for a ``yes'' instance of $\partit$, we can clear the game board.

\begin{thm}
For a ``yes'' instance of $\partit$, there is a trajectory sequence $\Sigma$ that clears the entire gameboard of $\mc G(\mc P)$ without triggering a loss.
\end{thm}
\begin{proof}
Since $\mc P$ is a ``yes'' instance, there is a partitioning of $\{1,2,\dots,3s\}$ into sets $A_1,A_2,\dots,A_s$ so that $\sum_{i\in A_j} a_i = T$ We have ensured that $|A_j| = 3$ for all $j$. All pieces associated with set $A_j = \{x, y, z\}$ should be placed into the $j$th bucket of the gameboard.

We place the $a_x$-sequence into bucket $j$ as in Figures~\ref{fig:bucketFilling}(i--l). After all pieces associated with the number $a_x$ have been placed into bucket $j$, the bucket has $a_x + 1$ fewer notches, but otherwise still has the shape of a bucket. Similarly, $a_y, a_z$ are placed in bucket $j$, for a total of $(a_x + 1) + (a_y + 1) + (a_z + 1) = T + 3$ notches being filled, so each bucket has 0 notches left and may then be filled by the bucket closers, as in 
%\ref{fig:bucketclose}. 
Figure~\ref{fig:bucketFilling}(n).
After all the $a_i$-sequences arrive, then, we fill all the buckets with the bucket closers.

% \begin{figure}
% \centering
% \begin{subfigure}[b]{0.12\textwidth}
%     \includegraphics[width=\textwidth]{tetris/figures/FinishingBucket/bucketend_s1.pdf}
%     \caption{}
% \end{subfigure}
% \hspace{0.015\textwidth}
% \begin{subfigure}[b]{0.12\textwidth}
%     \includegraphics[width=\textwidth]{tetris/figures/FinishingBucket/bucketend_s2.pdf}
%     \caption{}
% \end{subfigure}
% \hspace{0.015\textwidth}
% \begin{subfigure}[b]{0.12\textwidth}
%     \includegraphics[width=\textwidth]{tetris/figures/FinishingBucket/bucketend_s3.pdf}
%     \caption{}
% \end{subfigure}
% \hspace{0.015\textwidth}
% \begin{subfigure}[b]{0.12\textwidth}
%     \includegraphics[width=\textwidth]{tetris/figures/FinishingBucket/bucketend_s4.pdf}
%     \caption{}
% \end{subfigure}
% \hspace{0.015\textwidth}
% \caption{Filling a bucket with the bucket closer}
% \label{fig:bucketclose}
% \end{figure}

Next, now that the buckets are all filled, the remaining moves are straightforward. We drop a $\ZZ$ into the corridor to fill the bottom 4 cells of the corridor. Now we use the $6sT+24s+6$ $\SS$s in the corridor closer to fill the corridor, as depicted in Figure~\ref{fig:corridorFill}. We then drop the $\TT$ into the T-lock, which fills the top row immediately below the horizon and clears it, opening the alley. Now we drop a sequence of $3sT+12s+4$ $\II$s which clear all the rows of the board since they clear $4(3sT+12s+4) = 12sT+48s+16$ rows, as shown in 
%Figure~\ref{fig:clear}. 
Figures~\ref{fig:bucketFilling}(f--h).
This clears the whole board, as desired.
\end{proof}

%% file: tetris/soundness.tex
\subsection{{\tet} solvable \texorpdfstring{$\Rightarrow$}{=>} {\partit} solvable} 
\label{soundness}

Here we show that if $\mc G(\mc P)$ has a sequence of moves that survive, then the {\partit} instance $\mc P$ must also have a solution (i.e., a valid partition).
Suppose there is such a surviving sequence of moves.
By a sequence of claims, we progressively constrain this survival strategy into a form that directly encodes a {\partit} solution.

\begin{claim} 
The top row must be the first row to be cleared.
\end{claim}
\begin{proof}
Every row except the top has an empty square in the alley. The alley is completely surrounded by pieces, and thus no part of the alley can be filled until a row is filled.
\end{proof}

\begin{claim} 
\label{claim:only-t-in-lock}
Only a $\TT$ can go in the T-lock. 
\end{claim}
\begin{proof}
By the previous claim, no rows can be cleared before the top row, and thus the T-lock will remain at the top of the board until it is filled. There are only four empty cells in the connected component of the T-lock within the board, so any piece placed other than a $\TT$ will have at least one block above the board, causing a loss by partial lock out.
\end{proof}

\begin{claim} 
\label{claim:none-before-t}
No rows can be cleared before the $\TT$ is given.
\end{claim}
\begin{proof}
Follows from the prior two claims.
\end{proof}

\begin{claim}\label{claim:area-condition}
All squares not in the alley and T-lock must be completely filled before the $\TT$ arrives.
\end{claim}
\begin{proof}
Cells in the alley and cells above the horizon cannot be filled before the $\TT$ arrives by Claim~\ref{claim:none-before-t}. Cells in the T-lock cannot be filled before the $\TT$ arrives by Claim~\ref{claim:only-t-in-lock}. The total number of empty cells outside those two areas is $48sT+192s+52 = 4(12sT + 48s + 13)$, which is exactly four times the number of pieces that arrive before the $\TT$. Each piece fills four cells, so every cell not in the alley and T-lock must be filled; otherwise, some piece will extend above the board, causing a loss by partial lock out.
\end{proof}

By Claim~\ref{claim:area-condition}, the surviving trajectory sequence cannot leave any unfillable holes behind when placing pieces.
Henceforth, we focus on the prefix of the surviving trajectory sequence that places the $a_i$-sequences without such holes; we only need the surviving trajectory sequence for the closing sequence in order to guarantee Claim~\ref{claim:area-condition}.

\begin{claim}\label{claim:no-corridor}
During the $a_i$-sequences,
%Before the $\ZZ$ arrives.
no piece other than an $\LL$ can be placed first in the corridor.
\end{claim}
\begin{proof}
The casework in Figure~\ref{fig:no_piece_before_z} shows that any $\OO$, $\SS$, $\II$, or $\JJ$ placed in the bottom of the corridor will leave empty squares, which is not allowed by Claim~\ref{claim:area-condition}.
%No $\TT$ arrives before the $\ZZ$, leaving just $\LL$ and $\ZZ$.
No $\TT$, $\SS$, or $\ZZ$ arrives during the $a_i$-sequences, leaving just $\LL$.

\begin{figure}
\centering
\subcaptionbox{}{\includegraphics[scale=0.3]{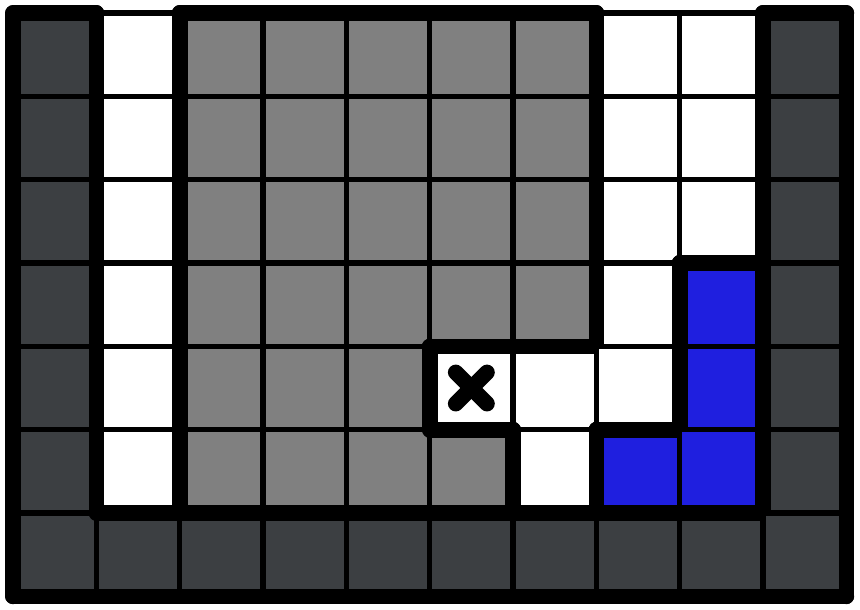}}
\subcaptionbox{}{\includegraphics[scale=0.3]{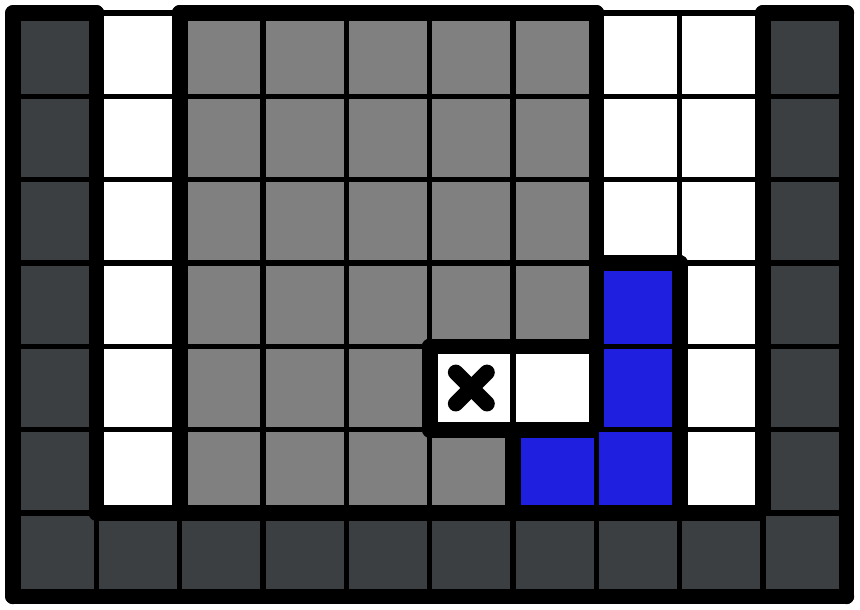}}
\subcaptionbox{}{\includegraphics[scale=0.3]{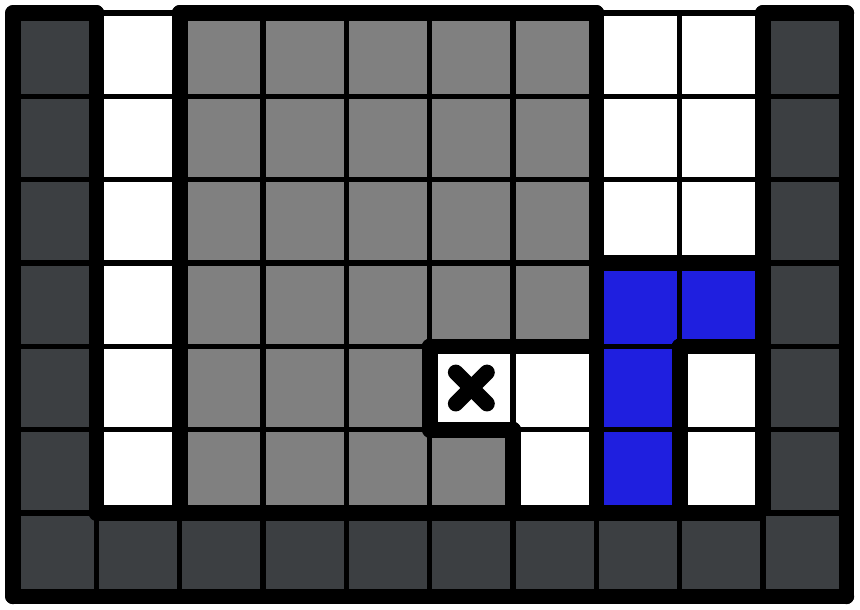}}
\subcaptionbox{}{\includegraphics[scale=0.3]{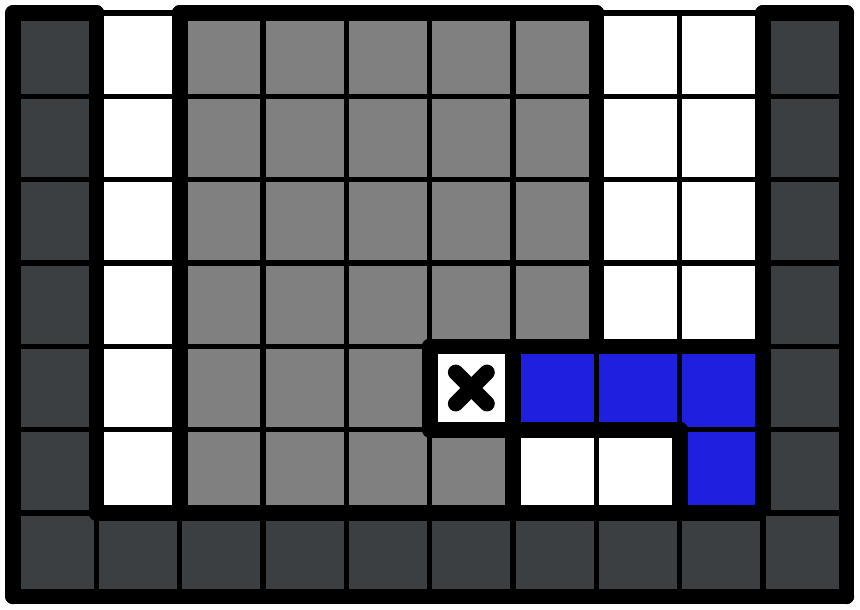}}
\subcaptionbox{}{\includegraphics[scale=0.3]{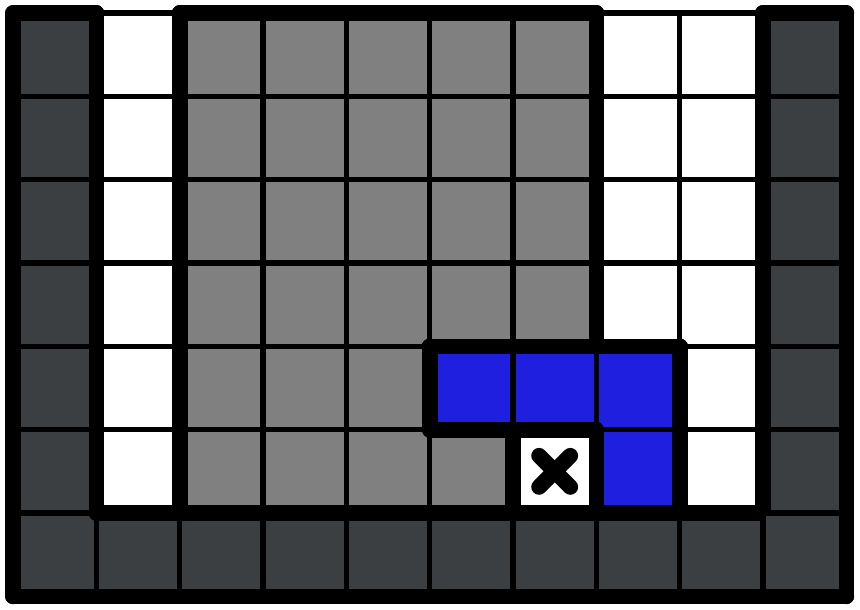}}
\subcaptionbox{}{\includegraphics[scale=0.3]{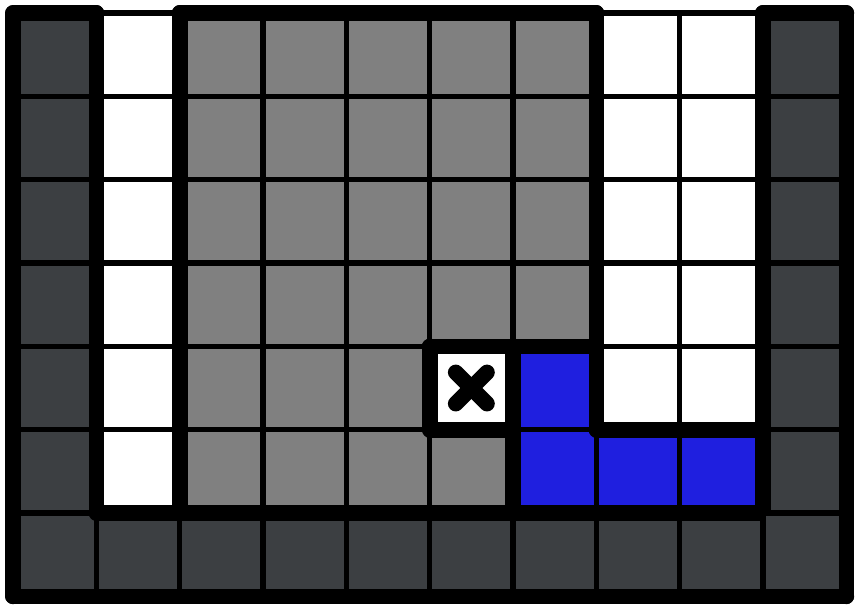}}

\subcaptionbox{}{\includegraphics[scale=0.3]{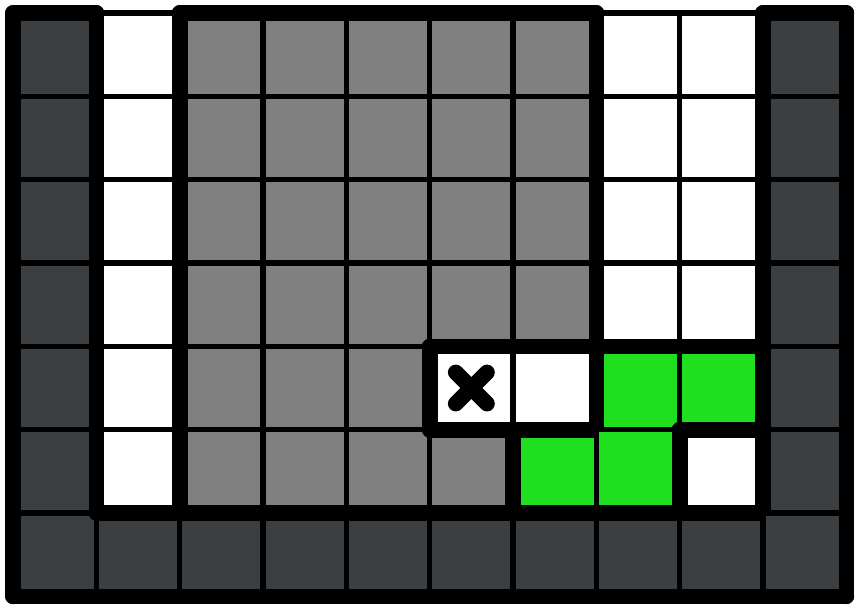}}
\subcaptionbox{}{\includegraphics[scale=0.3]{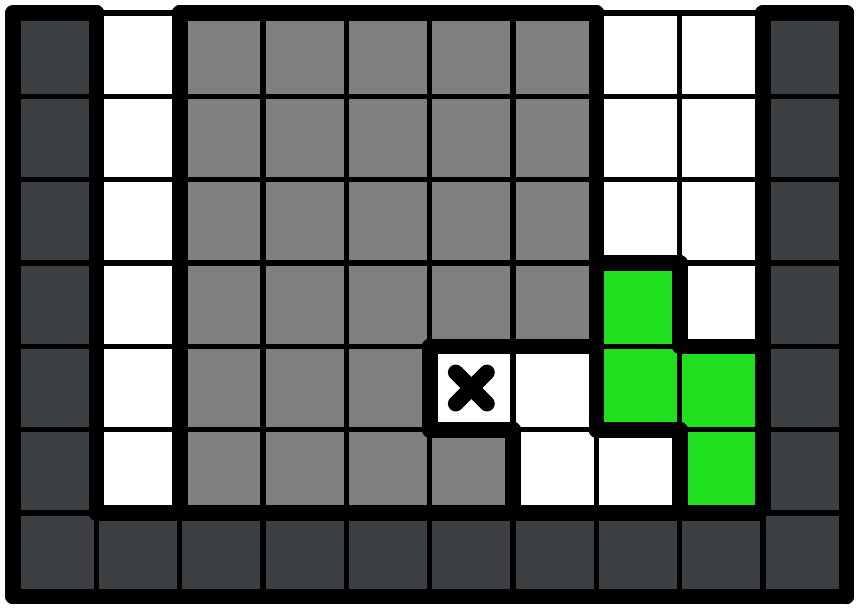}}
\qquad
\subcaptionbox{}{\includegraphics[scale=0.3]{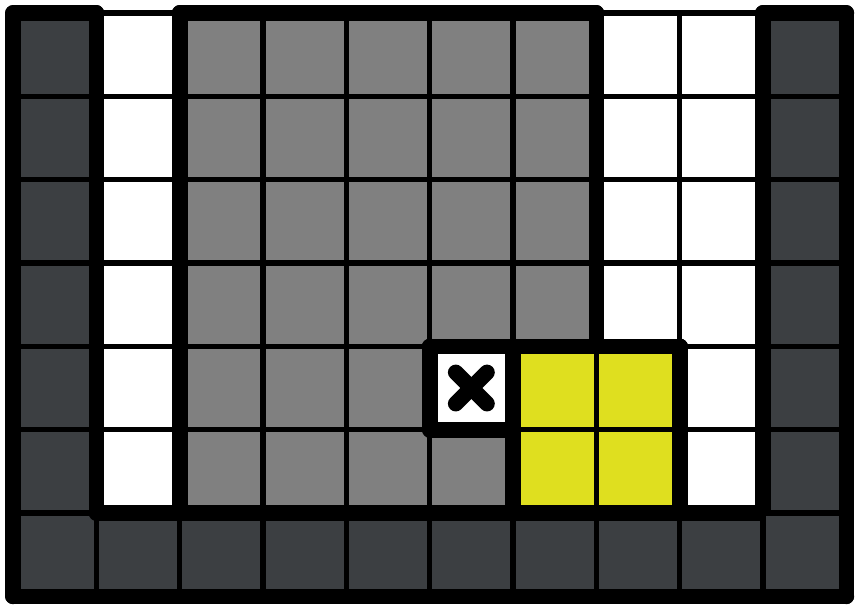}}
\subcaptionbox{}{\includegraphics[scale=0.3]{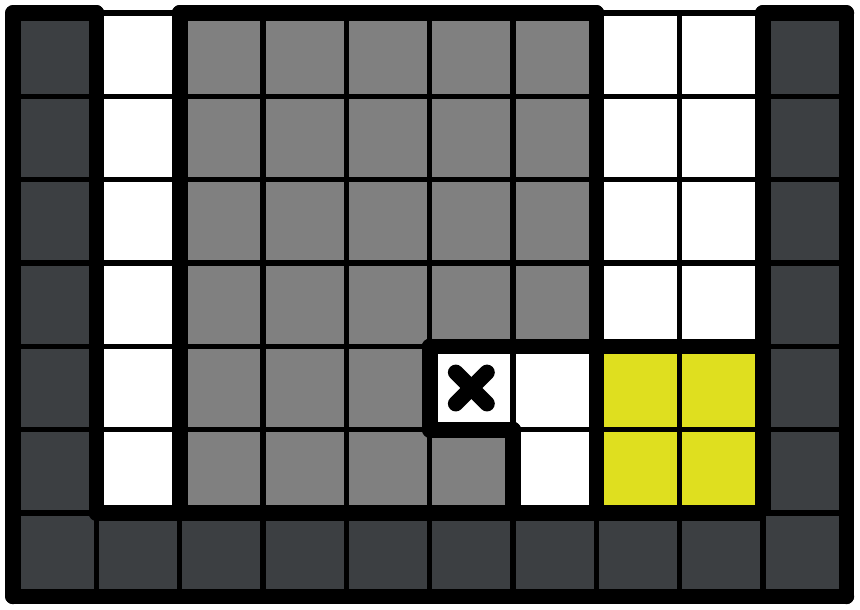}}

\subcaptionbox{}{\includegraphics[scale=0.3]{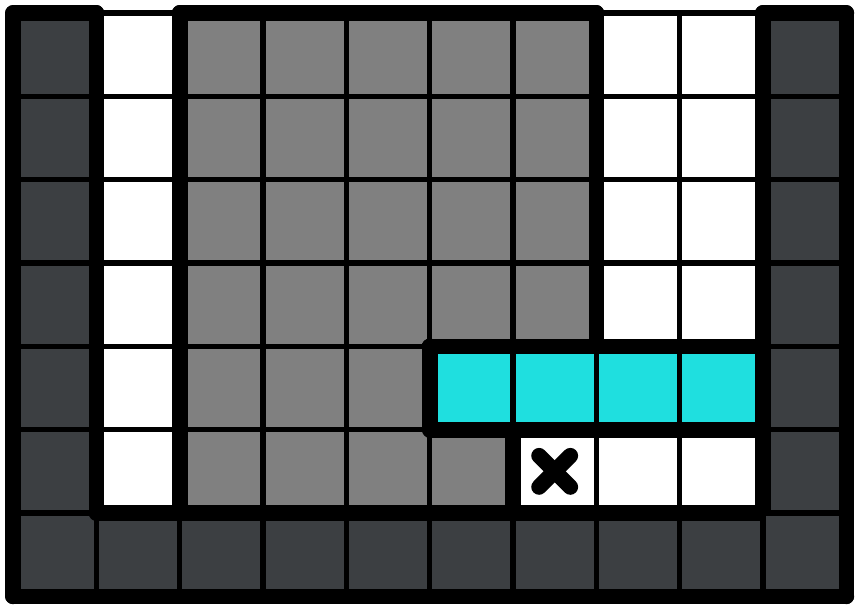}}
\subcaptionbox{}{\includegraphics[scale=0.3]{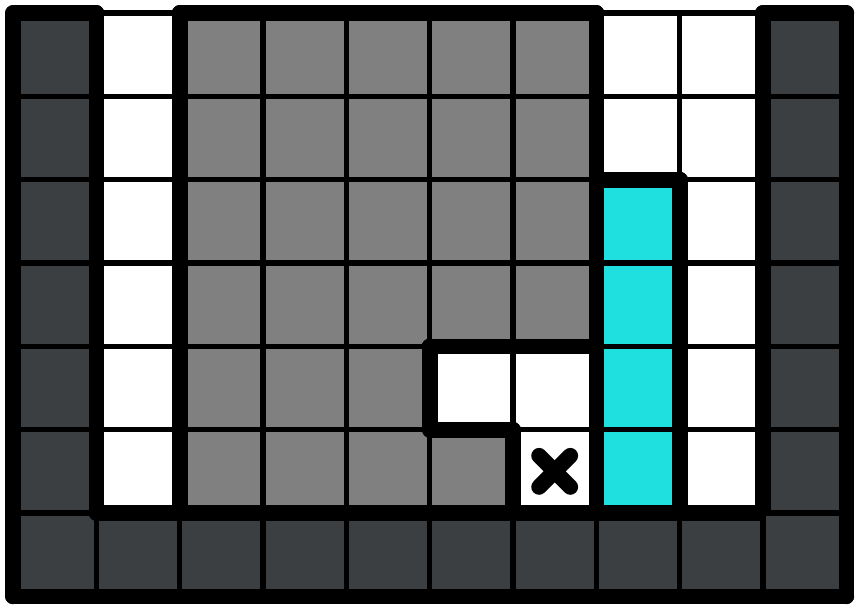}}
\subcaptionbox{}{\includegraphics[scale=0.3]{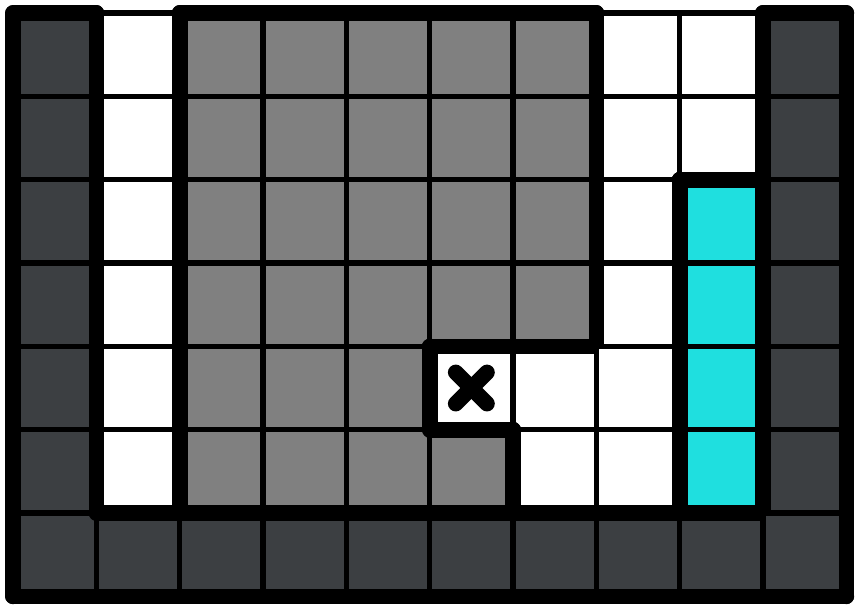}}
\caption{Possibilities for placing various pieces other than $\protect\ZZ$ and $\protect\LL$ into the corridor, all of which make the puzzle unsolvable.}
\label{fig:no_piece_before_z}
\end{figure}

Any piece placed higher up in the corridor creates a choke point of width 1 through which no pieces but $\II$ can pass. Since an $\II$ placed at the bottom of the corridor leaves empty squares, this makes the bottom of the corridor unfillable.
\end{proof}

\begin{figure}
\centering
\begin{minipage}[b]{0.45\linewidth}
\centering
\begin{subfigure}[b]{0.22\textwidth}
    \includegraphics[width=\textwidth]{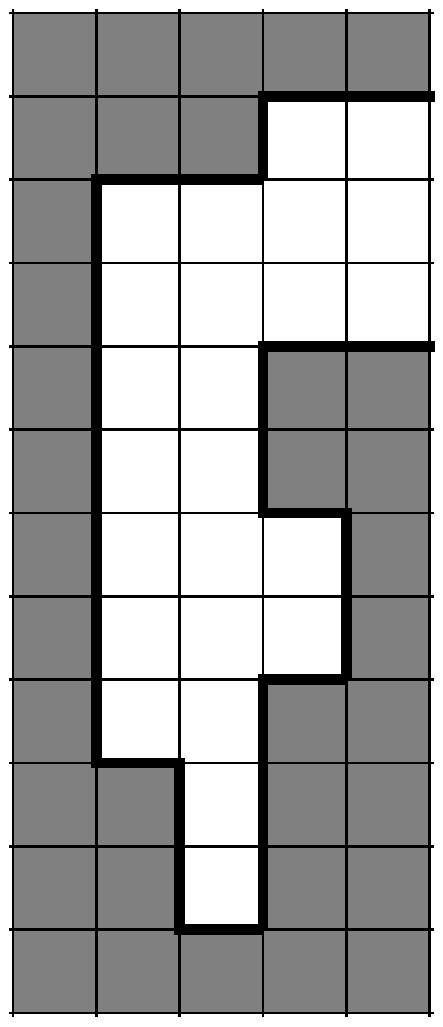}
    \caption{}
\end{subfigure}
\hspace{0.015\textwidth}
\begin{subfigure}[b]{0.22\textwidth}
    \includegraphics[width=\textwidth]{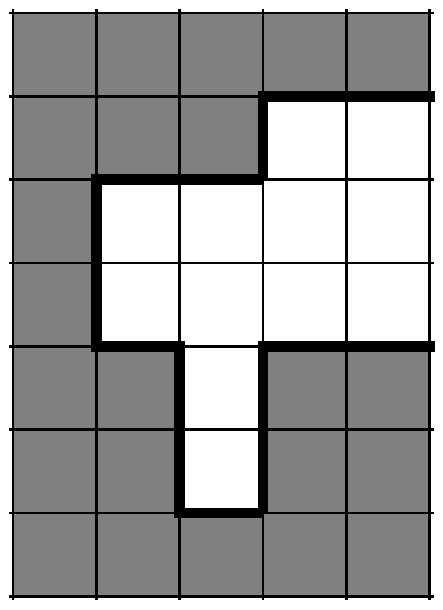}
    \caption{}
\end{subfigure}
\hspace{0.015\textwidth}
\caption{Unprepped buckets of heights 1 and 0, respectively}
\label{fig:unprep}
\end{minipage}\hfil\hfil
\begin{minipage}[b]{0.45\linewidth}
\centering
\begin{subfigure}[b]{0.22\textwidth}
    \includegraphics[width=\textwidth]{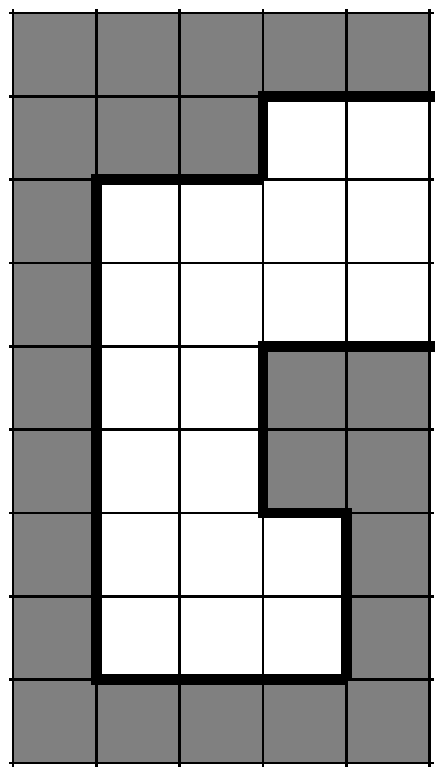}
    \caption{}
\end{subfigure}
\hspace{0.015\textwidth}
\begin{subfigure}[b]{0.22\textwidth}
    \includegraphics[width=\textwidth]{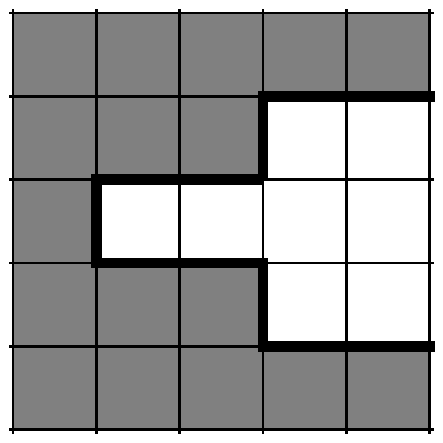}
    \caption{}
\end{subfigure}
\caption{Prepped buckets of heights 1 and 0, respectively}
\label{fig:prep}
\end{minipage}
\end{figure}

Next we show that the buckets must be filled in the manner given by Section~\ref{completeness}. We define prepped and unprepped buckets as in \cite{Tetris_IJCGA}. An \textit{unprepped} bucket is one that takes the form of a bucket in the initial board, but possibly with fewer notches, as shown in Figure~\ref{fig:unprep}(a). The \textit{height} of an unprepped bucket is the number of notches in the bucket; the buckets all initially have height $T+3$. A special case is an unprepped bucket of height 0; this is also shown in Figure~\ref{fig:unprep}(b).

We also define a \textit{prepped} bucket as one in which all cells below some notch are filled, as in Figure~\ref{fig:prep}(a). The \textit{height} of a prepped bucket is again its number of notches. Again, there is the special case of a prepped bucket of height 0, also shown in Figure~\ref{fig:prep}(b).

\begin{figure}
\centering
\begin{subfigure}[b]{0.1\textwidth}
    \includegraphics[width=\textwidth]{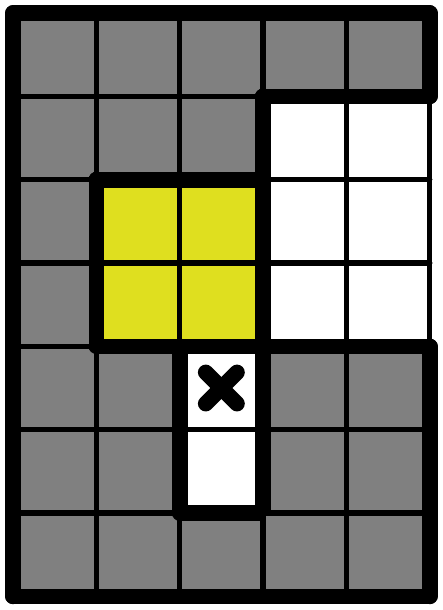}
    \caption{}
\end{subfigure}
\hspace{0.015\textwidth}
\begin{subfigure}[b]{0.1\textwidth}
    \includegraphics[width=\textwidth]{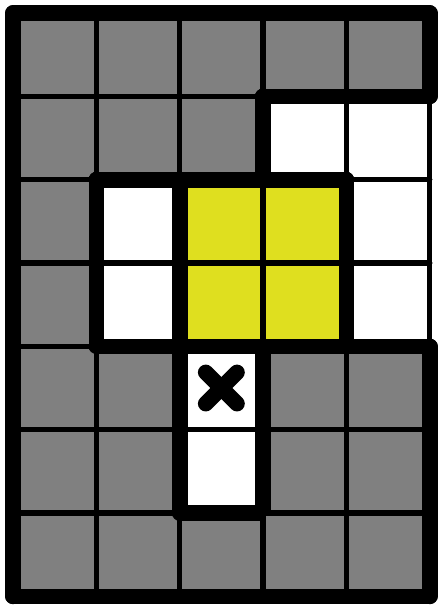}
    \caption{}
\end{subfigure}
\hspace{0.015\textwidth}
\begin{subfigure}[b]{0.1\textwidth}
    \includegraphics[width=\textwidth]{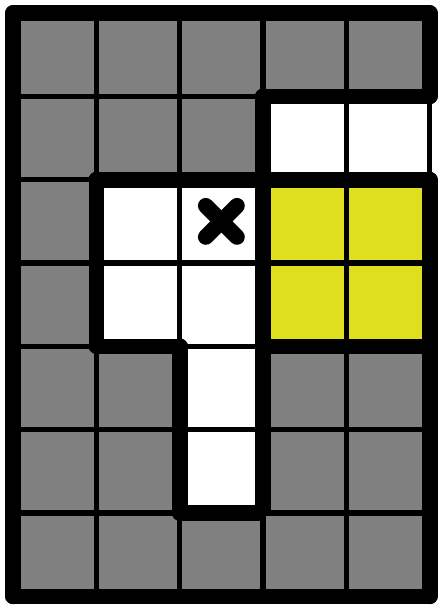}
    \caption{}
\end{subfigure}
\hspace{0.015\textwidth}
\begin{subfigure}[b]{0.1\textwidth}
    \includegraphics[width=\textwidth]{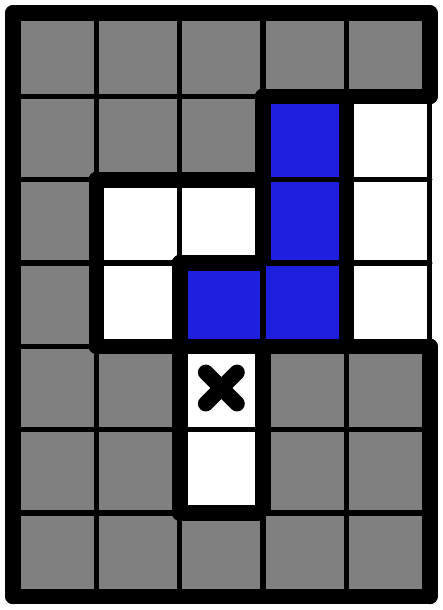}
    \caption{}
\end{subfigure}
\hspace{0.015\textwidth}
\begin{subfigure}[b]{0.1\textwidth}
    \includegraphics[width=\textwidth]{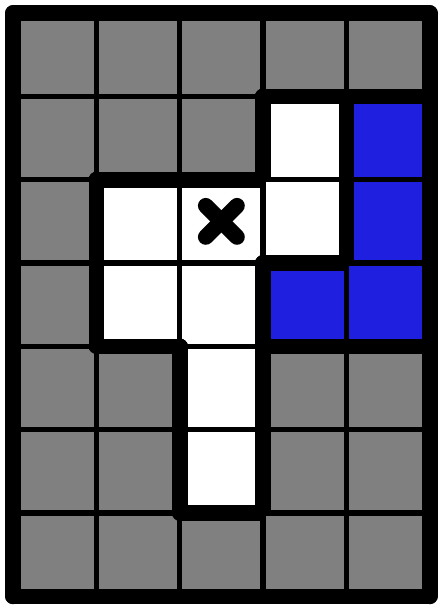}
    \caption{}
\end{subfigure}
\hspace{0.015\textwidth}
\begin{subfigure}[b]{0.1\textwidth}
    \includegraphics[width=\textwidth]{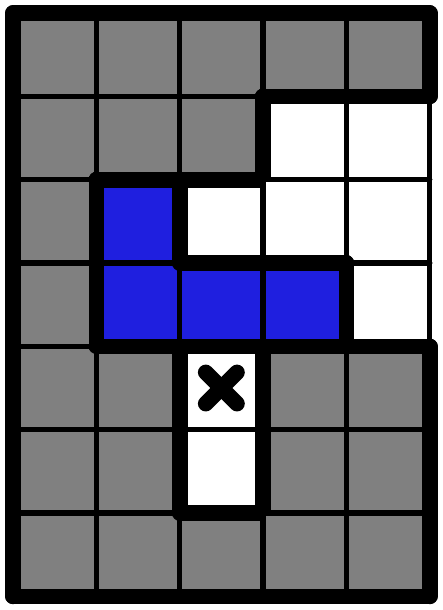}
    \caption{}
\end{subfigure}
\hspace{0.015\textwidth}
\begin{subfigure}[b]{0.1\textwidth}
    \includegraphics[width=\textwidth]{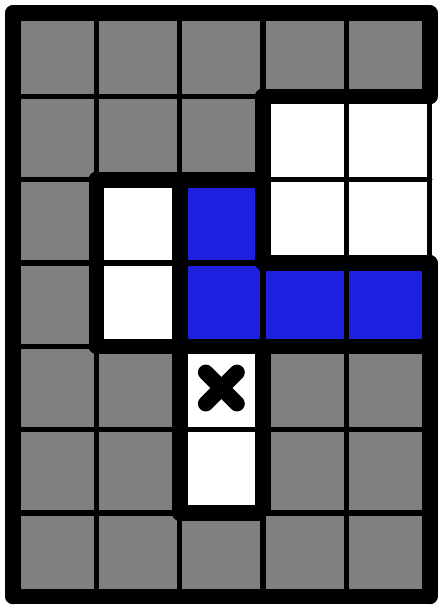}
    \caption{}
\end{subfigure}
\\
\vspace{0.4cm}
\begin{subfigure}[b]{0.1\textwidth}
    \includegraphics[width=\textwidth]{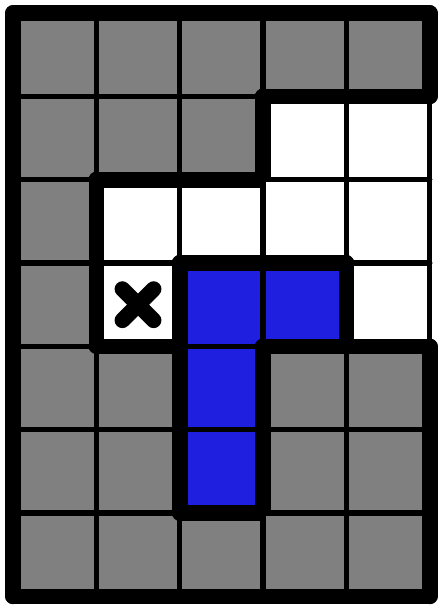}
    \caption{}
\end{subfigure}
\hspace{0.015\textwidth}
\begin{subfigure}[b]{0.1\textwidth}
    \includegraphics[width=\textwidth]{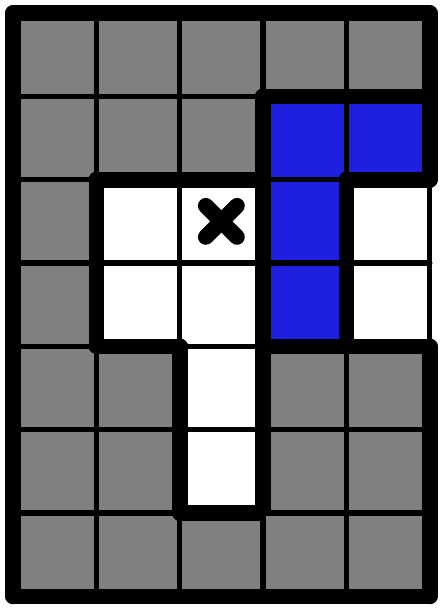}
    \caption{}
\end{subfigure}
\hspace{0.015\textwidth}
\begin{subfigure}[b]{0.1\textwidth}
    \includegraphics[width=\textwidth]{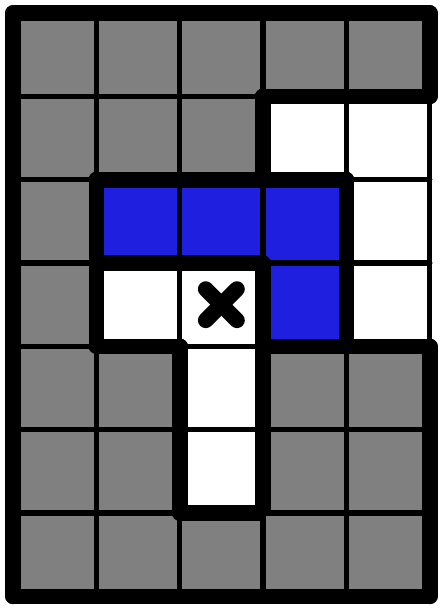}
    \caption{}
\end{subfigure}
\hspace{0.015\textwidth}
\begin{subfigure}[b]{0.1\textwidth}
    \includegraphics[width=\textwidth]{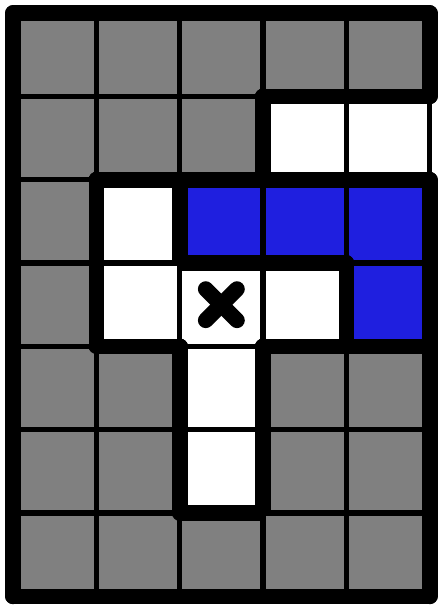}
    \caption{}
\end{subfigure}
\hspace{0.015\textwidth}
\begin{subfigure}[b]{0.1\textwidth}
    \includegraphics[width=\textwidth]{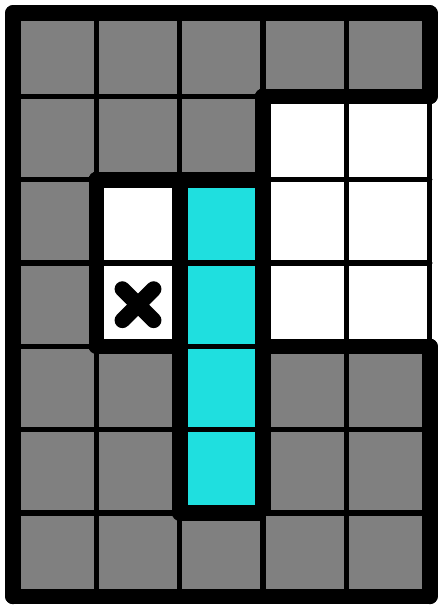}
    \caption{}
\end{subfigure}
\hspace{0.015\textwidth}
\begin{subfigure}[b]{0.1\textwidth}
    \includegraphics[width=\textwidth]{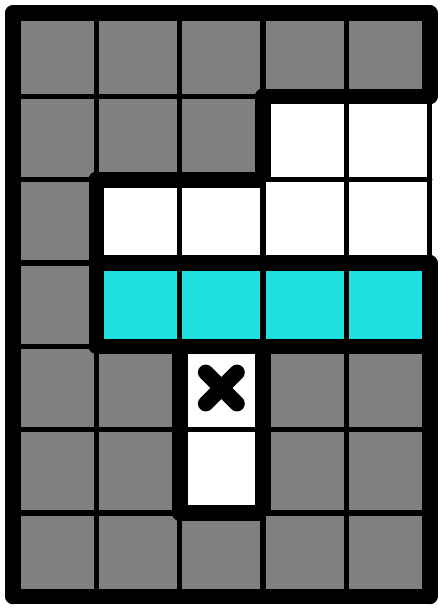}
    \caption{}
\end{subfigure}
\caption{Possibilities for placing an $\protect\OO$, $\protect\JJ$, or $\protect\II$ into an unprepped bucket of height 0. All leave the puzzle unsolvable.}
\label{fig:OJIunpreph0}
\end{figure}

\begin{figure}
\centering
\begin{subfigure}[b]{0.1\textwidth}
    \includegraphics[width=\textwidth]{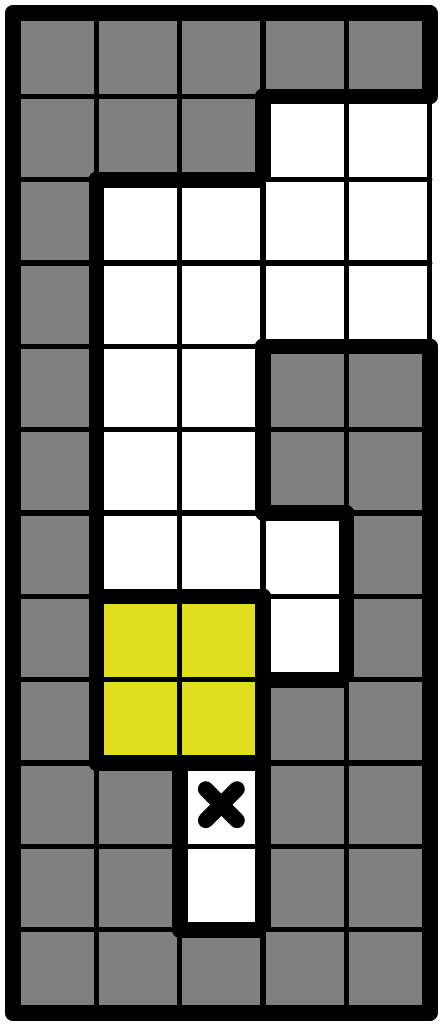}
    \caption{}
\end{subfigure}
\hspace{0.015\textwidth}
\begin{subfigure}[b]{0.1\textwidth}
    \includegraphics[width=\textwidth]{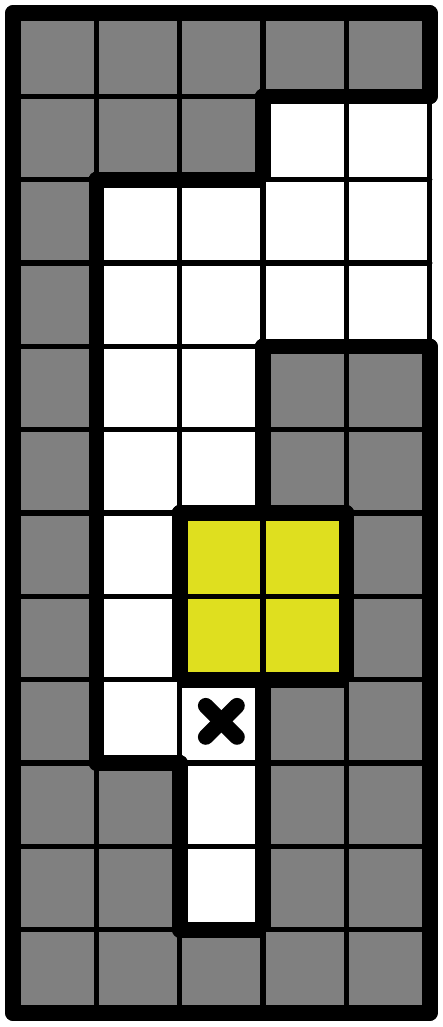}
    \caption{}
\end{subfigure}
\hspace{0.015\textwidth}
\begin{subfigure}[b]{0.1\textwidth}
    \includegraphics[width=\textwidth]{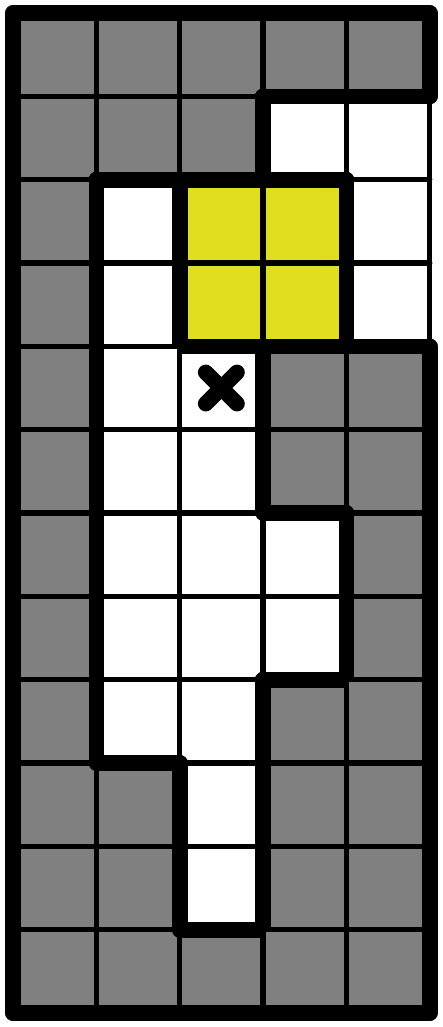}
    \caption{}
\end{subfigure}
\hspace{0.015\textwidth}
\begin{subfigure}[b]{0.1\textwidth}
    \includegraphics[width=\textwidth]{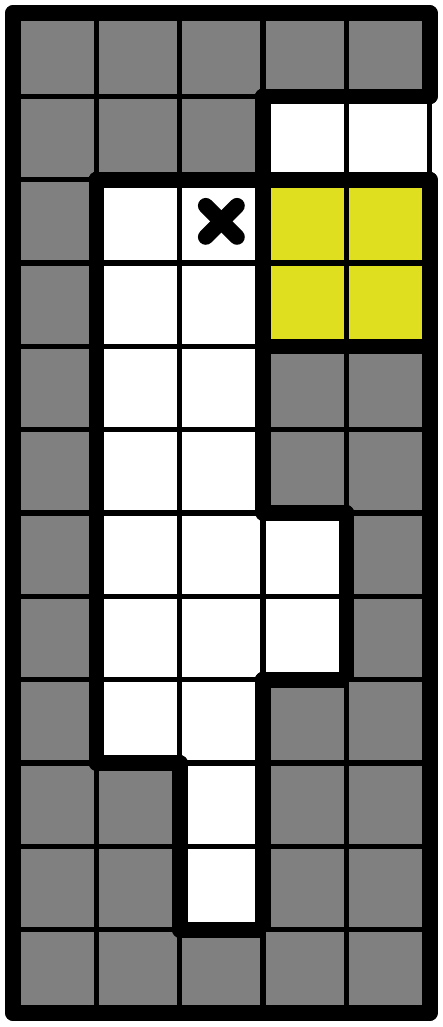}
    \caption{}
\end{subfigure}
\hspace{0.015\textwidth}
\begin{subfigure}[b]{0.1\textwidth}
    \includegraphics[width=\textwidth]{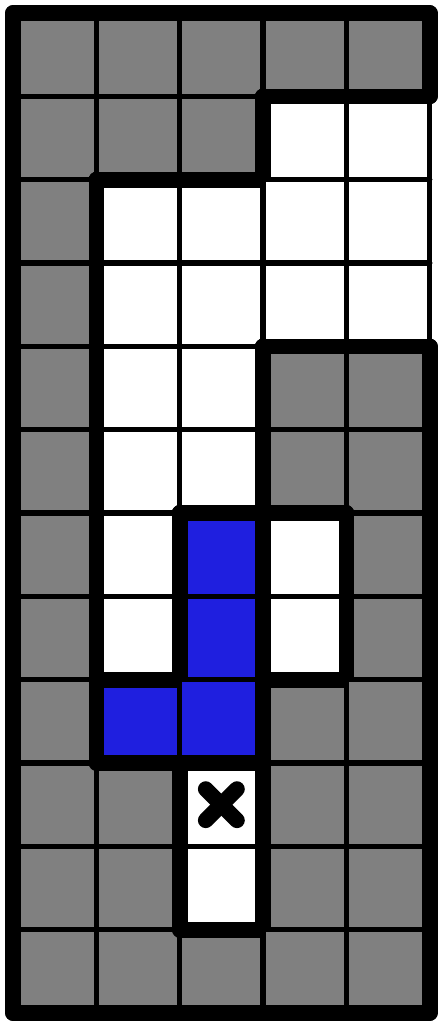}
    \caption{}
\end{subfigure}
\hspace{0.015\textwidth}
\begin{subfigure}[b]{0.1\textwidth}
    \includegraphics[width=\textwidth]{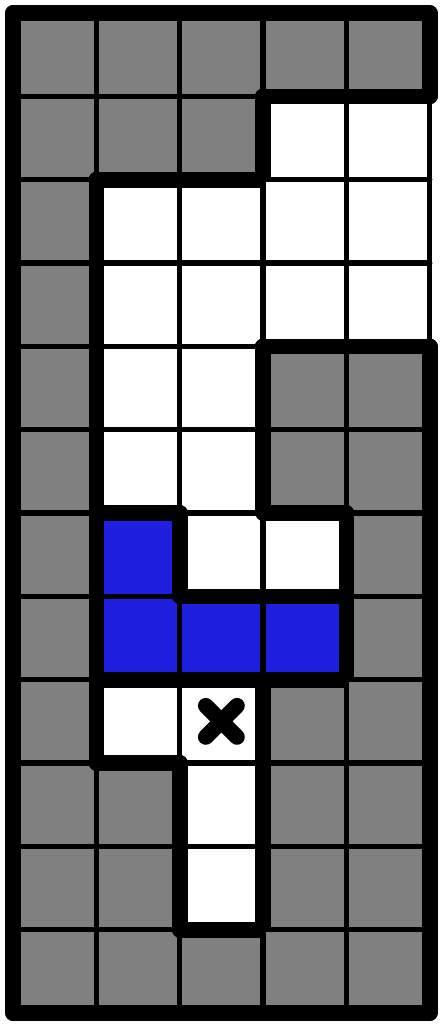}
    \caption{}
\end{subfigure}
\hspace{0.015\textwidth}
\begin{subfigure}[b]{0.1\textwidth}
    \includegraphics[width=\textwidth]{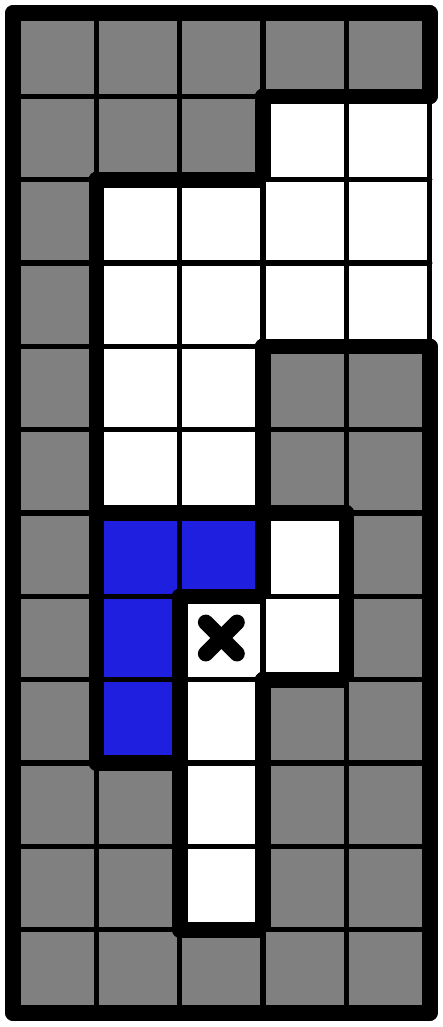}
    \caption{}
\end{subfigure}
\\
\vspace{0.4cm}
\begin{subfigure}[b]{0.1\textwidth}
    \includegraphics[width=\textwidth]{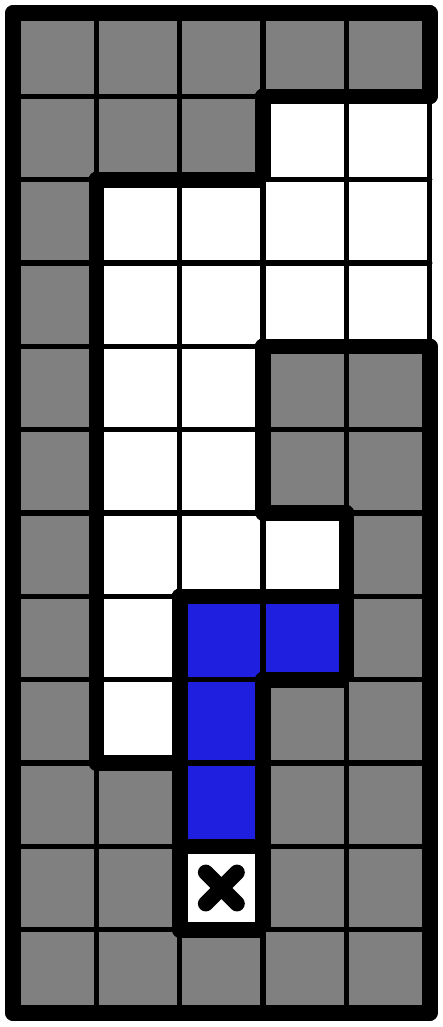}
    \caption{}
\end{subfigure}
\hspace{0.015\textwidth}
\begin{subfigure}[b]{0.1\textwidth}
    \includegraphics[width=\textwidth]{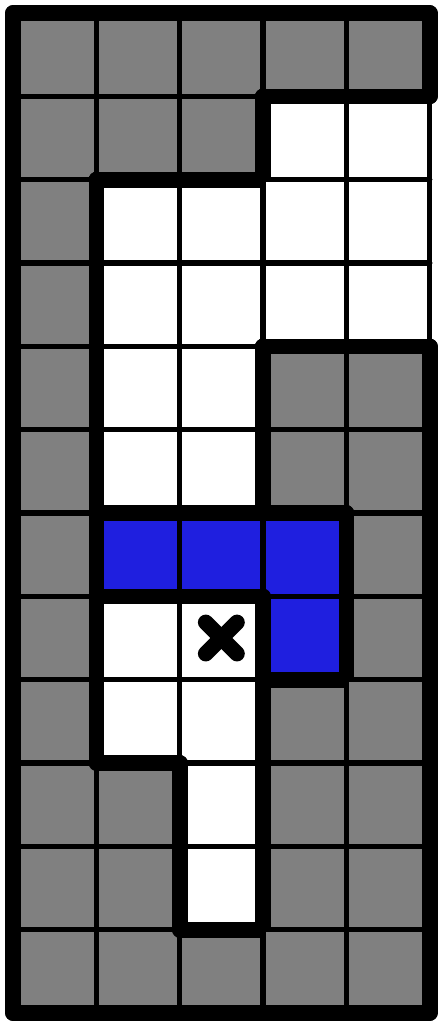}
    \caption{}
\end{subfigure}
\hspace{0.015\textwidth}
\begin{subfigure}[b]{0.1\textwidth}
    \includegraphics[width=\textwidth]{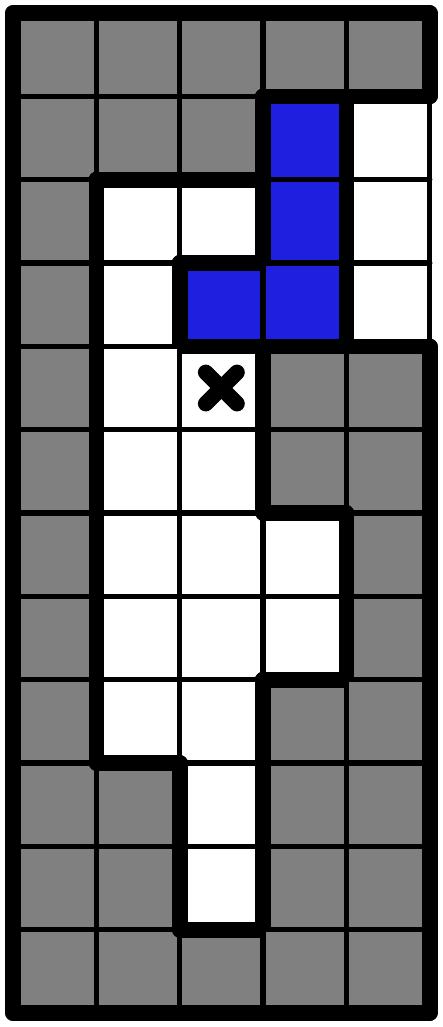}
    \caption{}
\end{subfigure}
\hspace{0.015\textwidth}
\begin{subfigure}[b]{0.1\textwidth}
    \includegraphics[width=\textwidth]{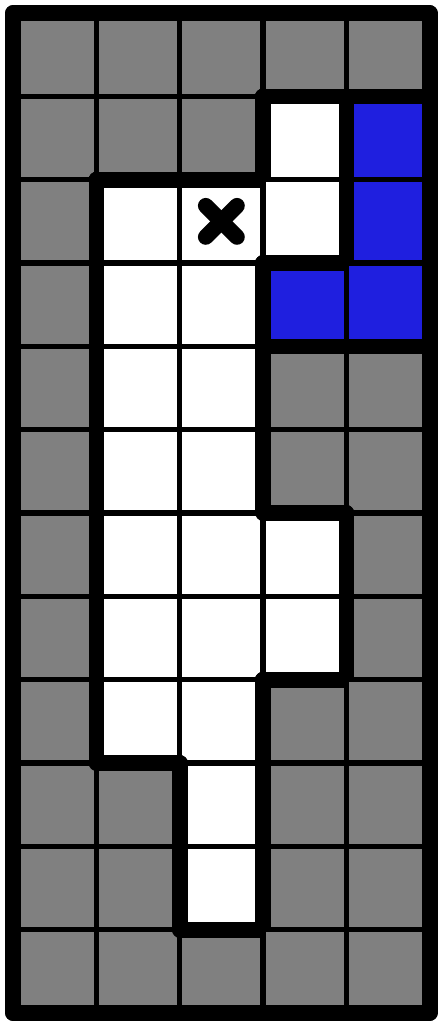}
    \caption{}
\end{subfigure}
\hspace{0.015\textwidth}
\begin{subfigure}[b]{0.1\textwidth}
    \includegraphics[width=\textwidth]{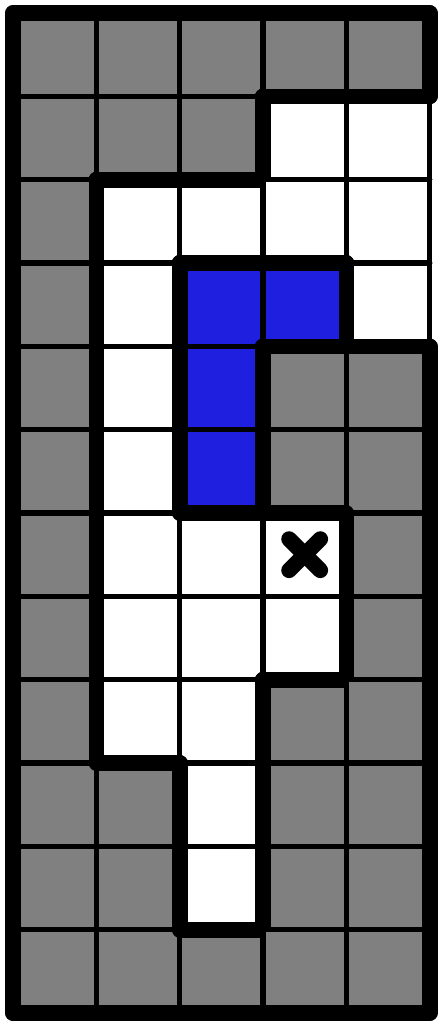}
    \caption{}
\end{subfigure}
\hspace{0.015\textwidth}
\begin{subfigure}[b]{0.1\textwidth}
    \includegraphics[width=\textwidth]{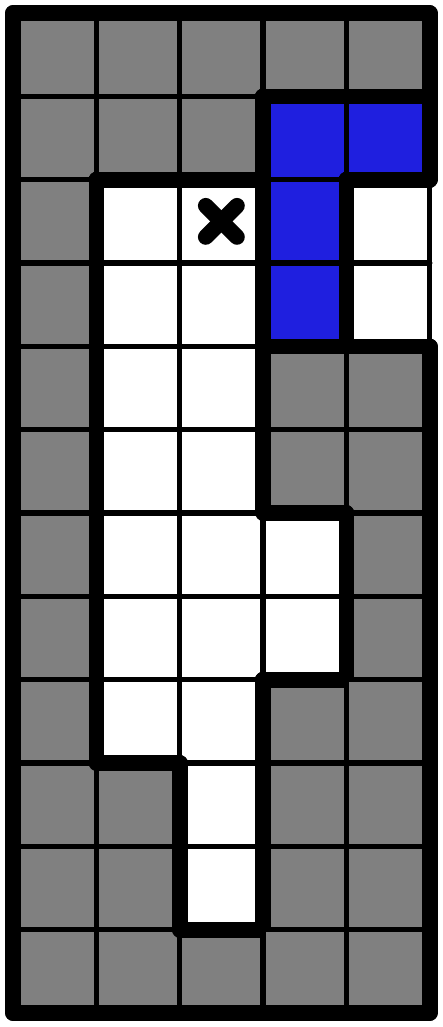}
    \caption{}
\end{subfigure}
\hspace{0.015\textwidth}
\begin{subfigure}[b]{0.1\textwidth}
    \includegraphics[width=\textwidth]{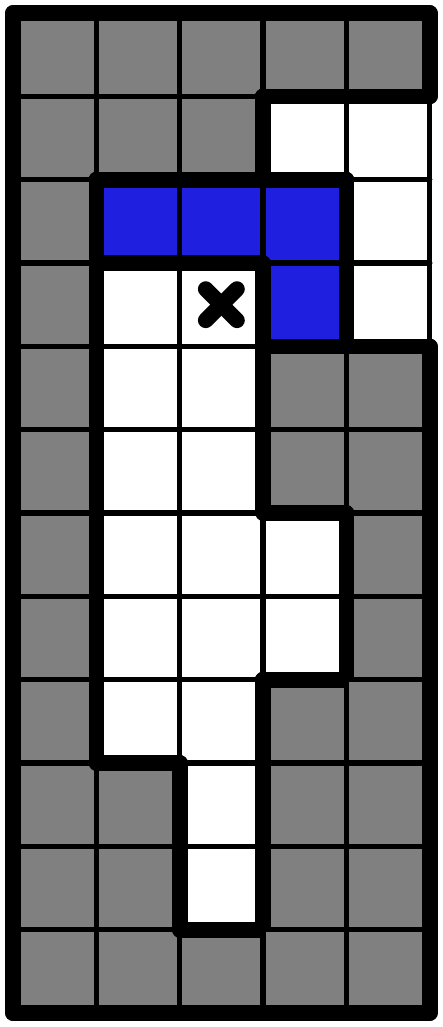}
    \caption{}
\end{subfigure}
\\
\vspace{0.4cm}
\begin{subfigure}[b]{0.1\textwidth}
    \includegraphics[width=\textwidth]{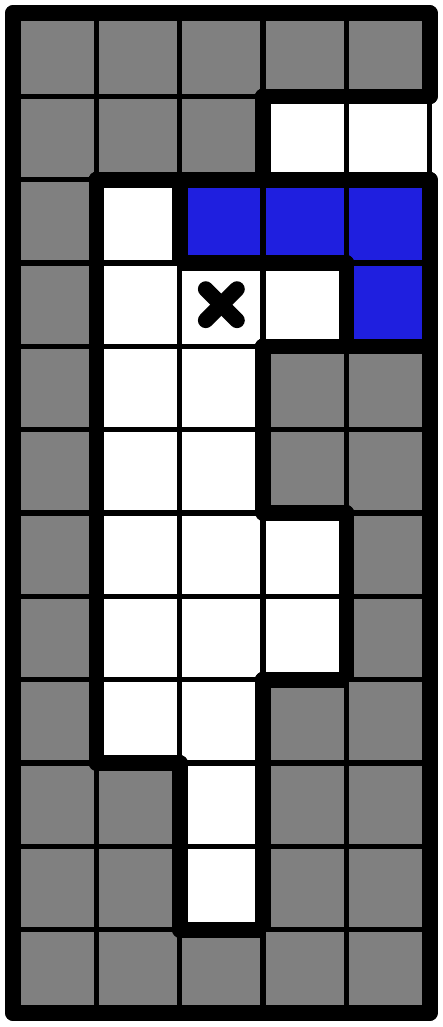}
    \caption{}
\end{subfigure}
\hspace{0.015\textwidth}
\begin{subfigure}[b]{0.1\textwidth}
    \includegraphics[width=\textwidth]{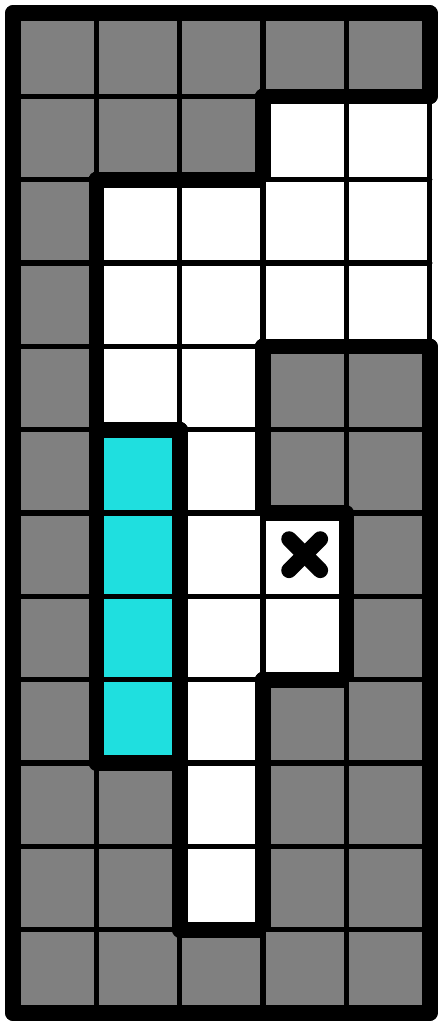}
    \caption{}
\end{subfigure}
\hspace{0.015\textwidth}
\begin{subfigure}[b]{0.1\textwidth}
    \includegraphics[width=\textwidth]{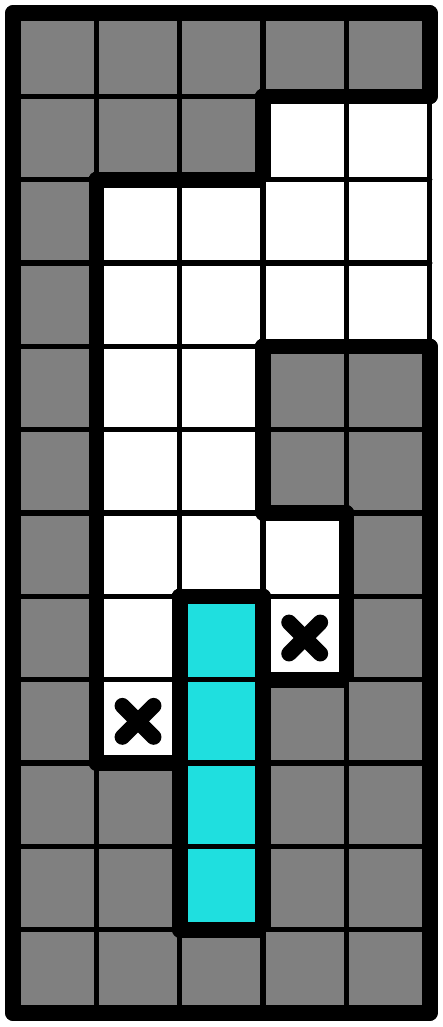}
    \caption{}
\end{subfigure}
\hspace{0.015\textwidth}
\begin{subfigure}[b]{0.1\textwidth}
    \includegraphics[width=\textwidth]{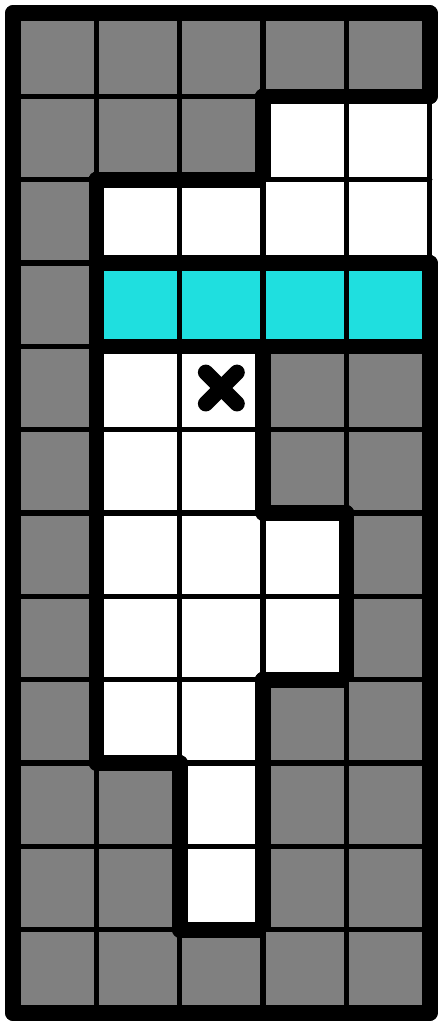}
    \caption{}
\end{subfigure}
\caption{Possibilities for placing an $\protect\OO$, $\protect\JJ$, or $\protect\II$ in an unprepped bucket of positive height. All leave the puzzle unsolvable.}
\label{fig:OJIunpreph1}
\end{figure}

\begin{claim} \label{struc2}
None of $\OO, \JJ, \II$ may be placed in an unprepped bucket.
\end{claim}
\begin{proof}
Figures \ref{fig:OJIunpreph0} and \ref{fig:OJIunpreph1}
show all possible placements, and the crosses show cells that cannot be filled. In Figure~\ref{fig:OJIunpreph1}(q), there are two cells with crosses; these two cells cannot both be filled (noting that there is no $\SS$ in the piece sequence). Thus we have a contradiction by Claim \ref{claim:area-condition}, since all the crossed cells must be filled eventually.
\end{proof}

\begin{figure}
\centering
\begin{subfigure}[b]{0.1\textwidth}
    \includegraphics[width=\textwidth]{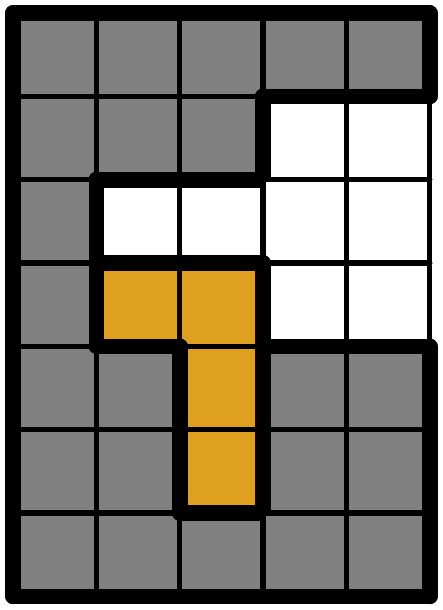}
    \caption{}
\end{subfigure}
\hspace{0.015\textwidth}
\begin{subfigure}[b]{0.1\textwidth}
    \includegraphics[width=\textwidth]{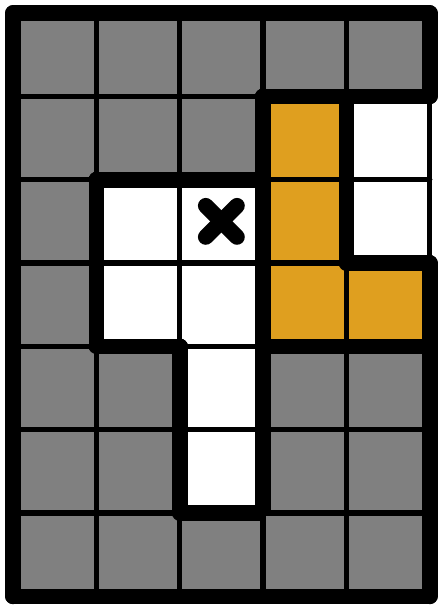}
    \caption{}
\end{subfigure}
\hspace{0.015\textwidth}
\begin{subfigure}[b]{0.1\textwidth}
    \includegraphics[width=\textwidth]{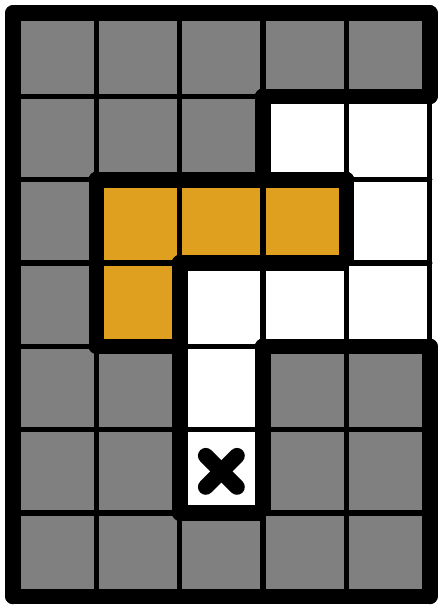}
    \caption{}
\end{subfigure}
\hspace{0.015\textwidth}
\begin{subfigure}[b]{0.1\textwidth}
    \includegraphics[width=\textwidth]{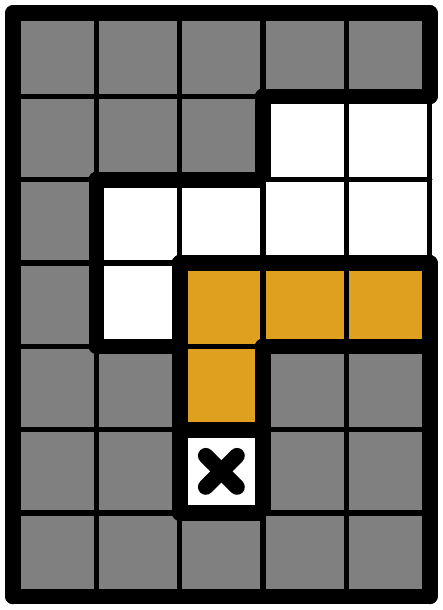}
    \caption{}
\end{subfigure}
\hspace{0.015\textwidth}
\begin{subfigure}[b]{0.1\textwidth}
    \includegraphics[width=\textwidth]{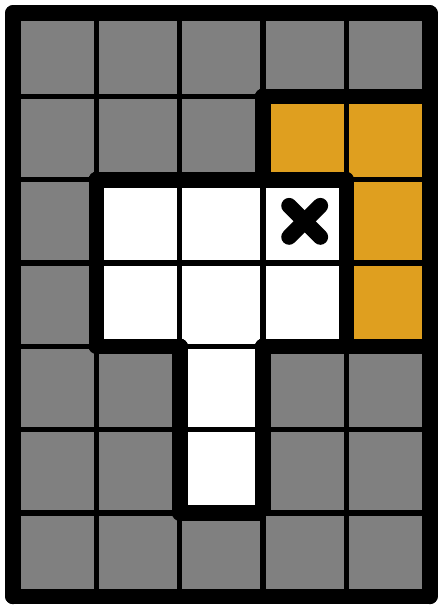}
    \caption{}
\end{subfigure}
\hspace{0.015\textwidth}
\begin{subfigure}[b]{0.1\textwidth}
    \includegraphics[width=\textwidth]{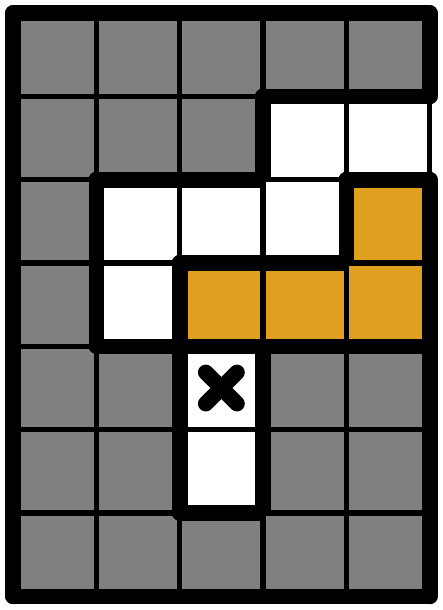}
    \caption{}
\end{subfigure}
\hspace{0.015\textwidth}
\begin{subfigure}[b]{0.1\textwidth}
    \includegraphics[width=\textwidth]{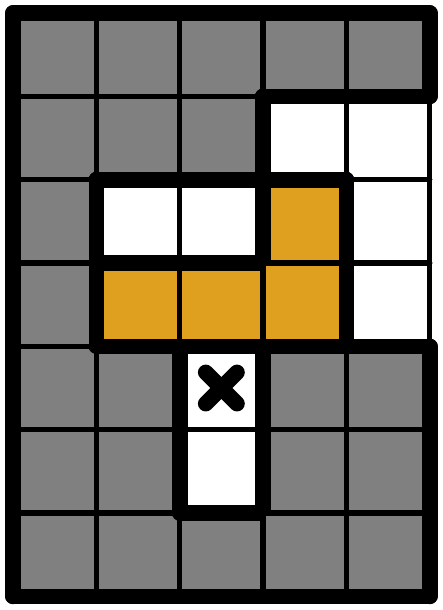}
    \caption{}
\end{subfigure}
\caption{Possibilities for placing an $\protect\LL$ into an unprepped bucket of height 0. All but the first leave the puzzle unsolvable.}
\label{fig:Lunpreph0}
\end{figure}

\begin{figure}
\centering
\begin{subfigure}[b]{0.1\textwidth} \label{goodL1}
    \includegraphics[width=\textwidth]{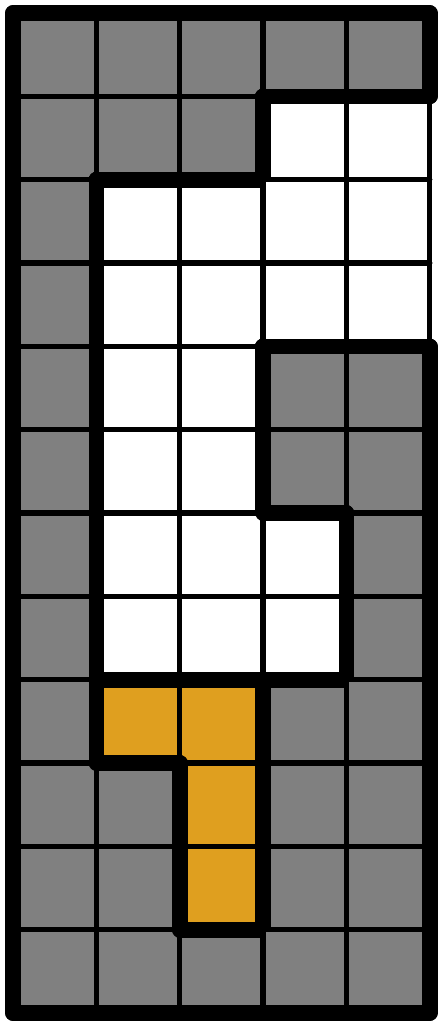}
    \caption{}
\end{subfigure}
\hspace{0.015\textwidth}
\begin{subfigure}[b]{0.1\textwidth}
    \includegraphics[width=\textwidth]{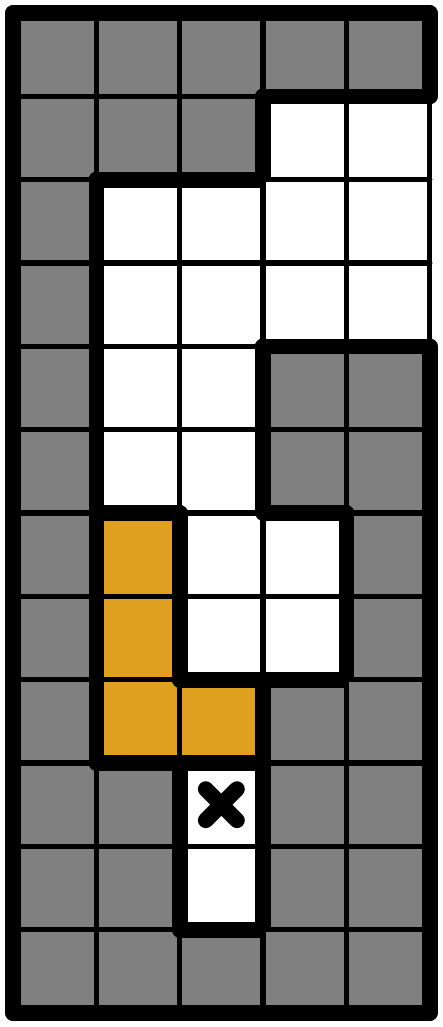}
    \caption{}
\end{subfigure}
\hspace{0.015\textwidth}
\begin{subfigure}[b]{0.1\textwidth}
    \includegraphics[width=\textwidth]{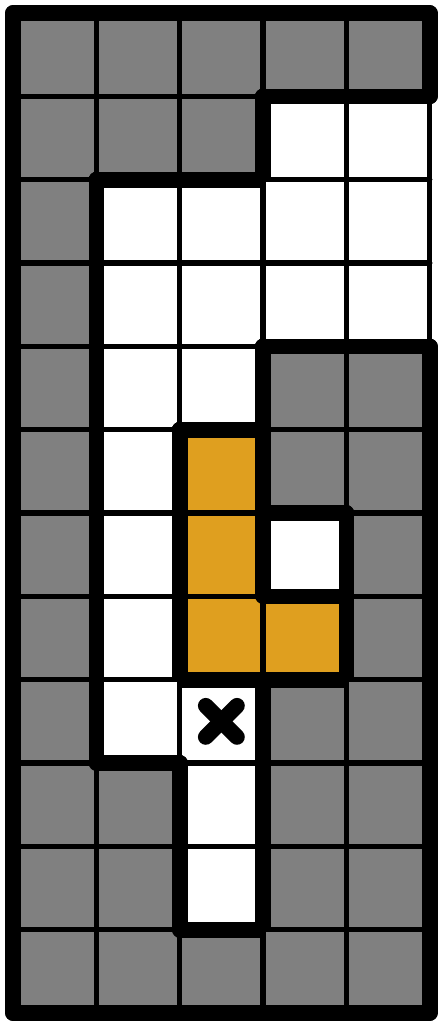}
    \caption{}
\end{subfigure}
\hspace{0.015\textwidth}
\begin{subfigure}[b]{0.1\textwidth}
    \includegraphics[width=\textwidth]{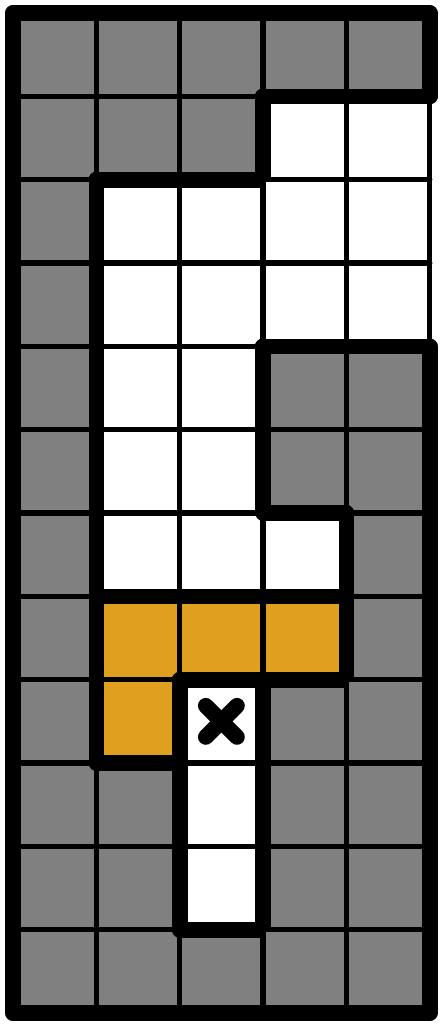}
    \caption{}
\end{subfigure}
\hspace{0.015\textwidth}
\begin{subfigure}[b]{0.1\textwidth}
    \includegraphics[width=\textwidth]{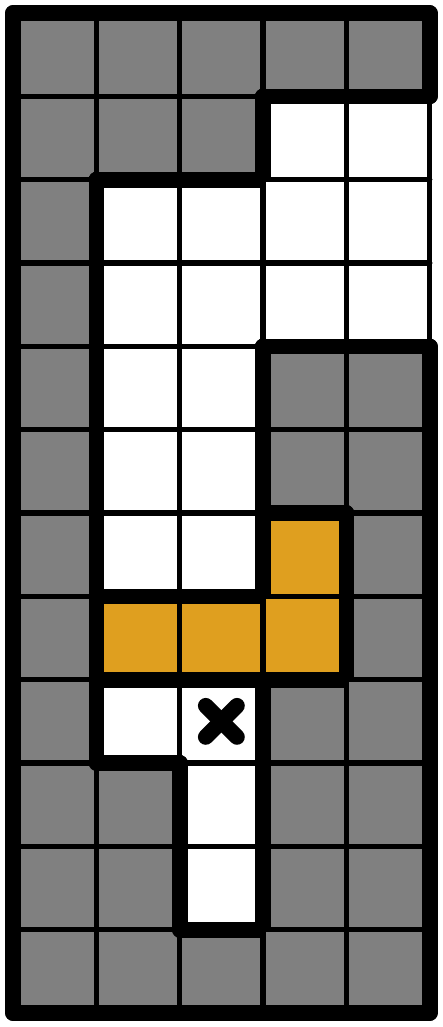}
    \caption{}
\end{subfigure}
\hspace{0.015\textwidth}
\begin{subfigure}[b]{0.1\textwidth}
    \includegraphics[width=\textwidth]{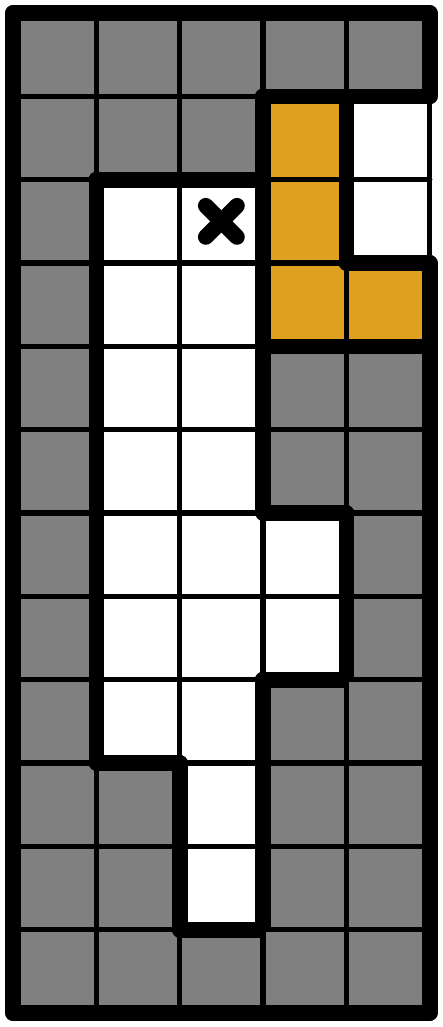}
    \caption{}
\end{subfigure}
\hspace{0.015\textwidth}
\begin{subfigure}[b]{0.1\textwidth}
    \includegraphics[width=\textwidth]{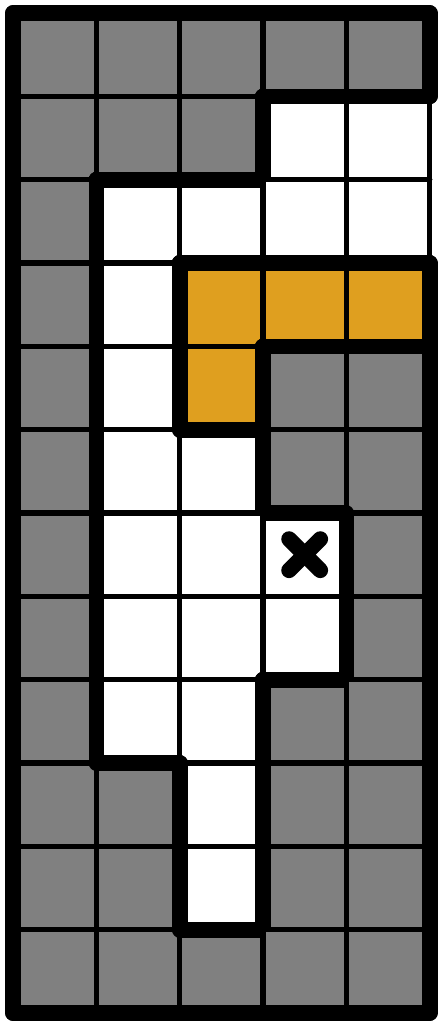}
    \caption{}
\end{subfigure}
\\
\vspace{0.4cm}
\begin{subfigure}[b]{0.1\textwidth}
    \includegraphics[width=\textwidth]{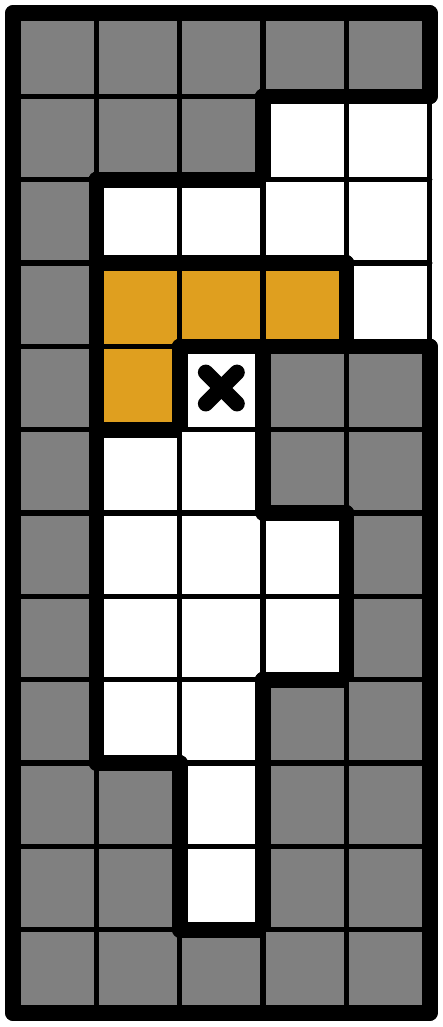}
    \caption{}
\end{subfigure}
\hspace{0.015\textwidth}
\begin{subfigure}[b]{0.1\textwidth}
    \includegraphics[width=\textwidth]{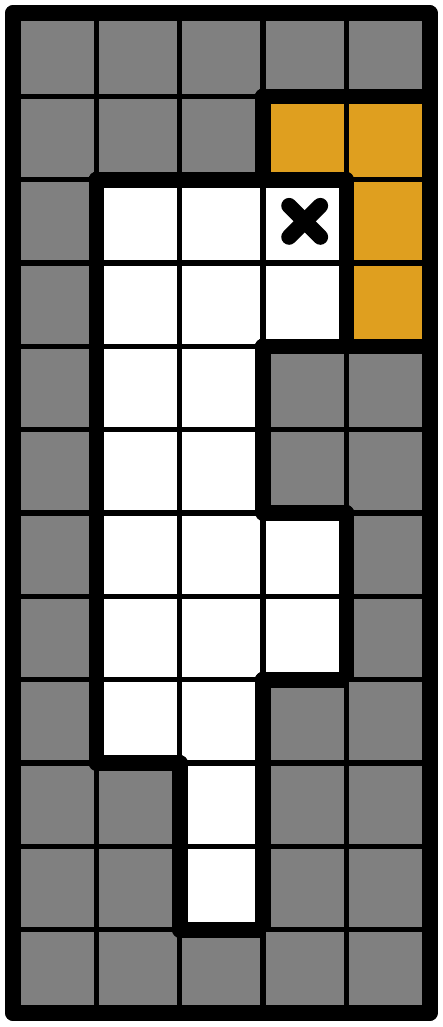}
    \caption{}
\end{subfigure}
\hspace{0.015\textwidth}
\begin{subfigure}[b]{0.1\textwidth}
    \includegraphics[width=\textwidth]{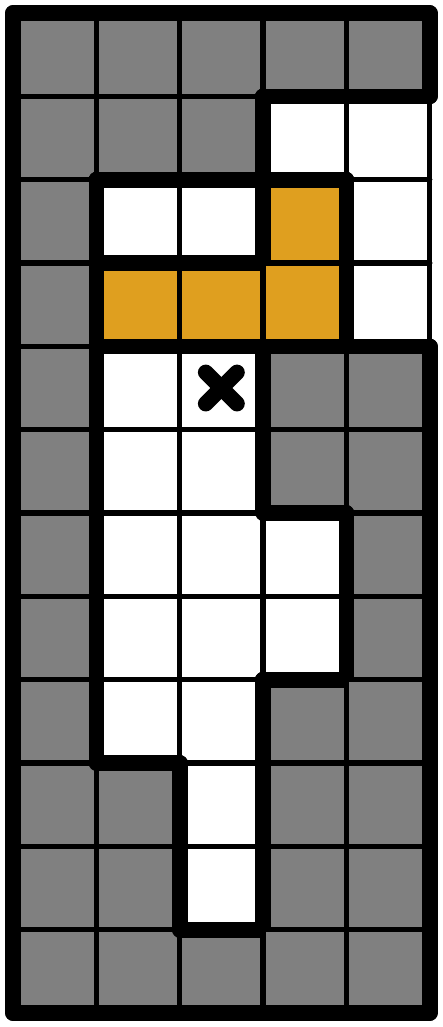}
    \caption{}
\end{subfigure}
\hspace{0.015\textwidth}
\begin{subfigure}[b]{0.1\textwidth}
    \includegraphics[width=\textwidth]{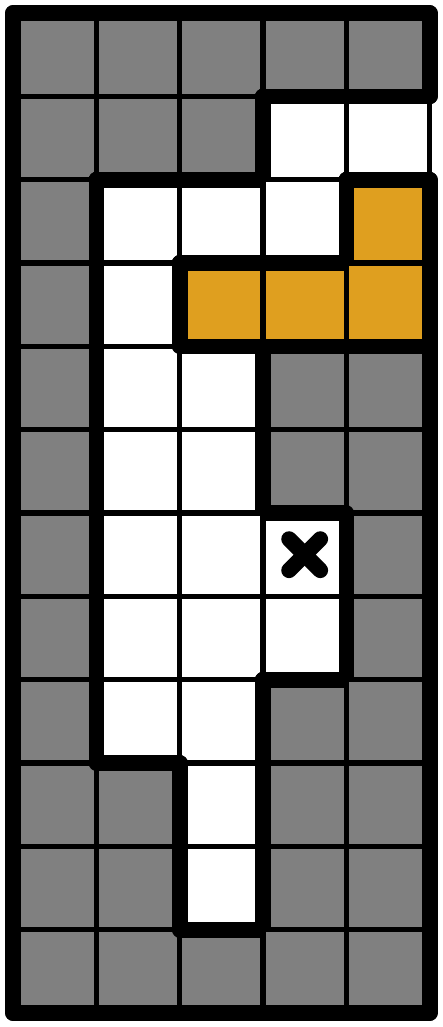}
    \caption{}
\end{subfigure}
\caption{Possibilities for placing an $\protect\LL$ into an unprepped bucket of positive height. All but the first leave the puzzle unsolvable.}
\label{fig:Lunpreph1}
\end{figure}

\begin{claim} \label{struc1}
If an $\LL$ is placed in an unprepped bucket, it must form a prepped bucket of the same height.
\end{claim}
\begin{proof}
We do casework on the possible placements of the $\LL$, showing that in each other case, there is a cell that can never be filled before any row is cleared. This would contradict Claim \ref{claim:area-condition}.

All possible placements are shown in Figures \ref{fig:Lunpreph0}
and \ref{fig:Lunpreph1}, where we have split into cases based on whether or not the bucket has height 0. (The cases where the bucket does not have height 0 are essentially the same even as the height varies). Most cases are marked with a cross, which indicates a cell which can never be filled, making that placement invalid. In the only valid cases, Figures \ref{fig:Lunpreph0}(d) and \ref{fig:Lunpreph1}(d), an unprepped bucket of the same height results.
\end{proof}

\xxx{It'd be nice to modify the figures showing exactly one valid placement of a piece so that that one valid placement stands out somehow---e.g. is highlighted.}

\begin{claim} \label{struc3}
An $\OO$ cannot be placed in a prepped bucket of height 0.
\end{claim}
\begin{proof}
There is only one possible placement of an $\OO$ in a prepped bucket of height 0, shown in Figure~\ref{fig:Oprep0}. The cross marks an unfillable cell; hence, this is not permissible.

\begin{figure}
    \centering
    \includegraphics[width=0.1\textwidth]{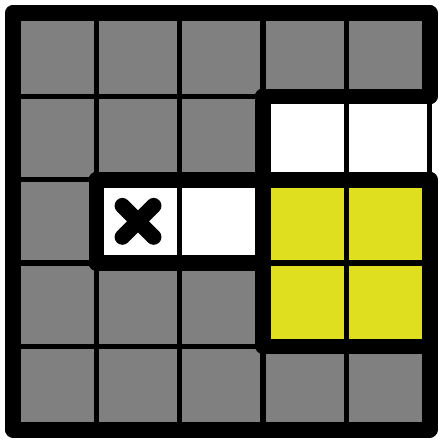}
    \caption{The only way to place an $\protect\OO$ in a prepped bucket of height 0. This leaves the puzzle unsolvable.}
    \label{fig:Oprep0}
\end{figure}
\end{proof}

\begin{figure}
\centering
\begin{subfigure}[b]{0.1\textwidth}
    \includegraphics[width=\textwidth]{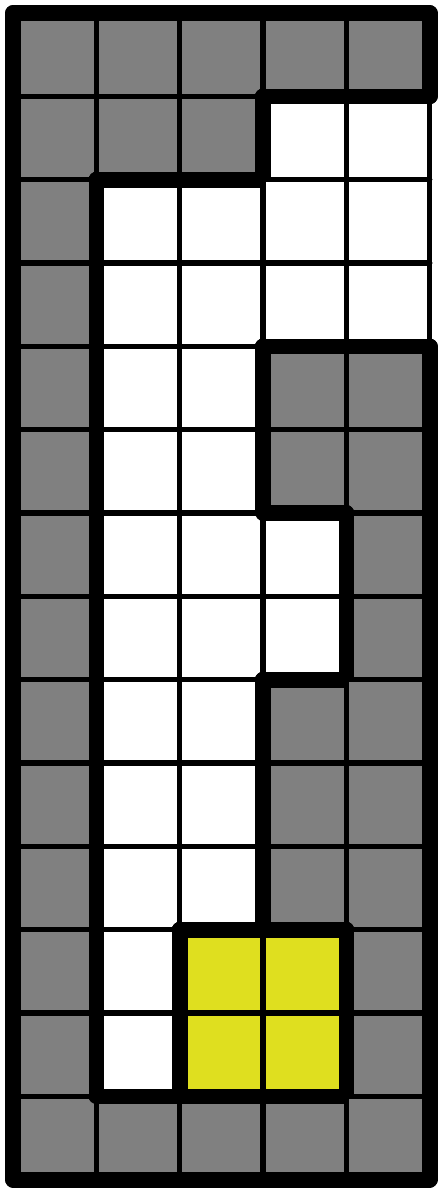}
    \caption{}
\end{subfigure}
\hspace{0.015\textwidth}
\begin{subfigure}[b]{0.1\textwidth}
    \includegraphics[width=\textwidth]{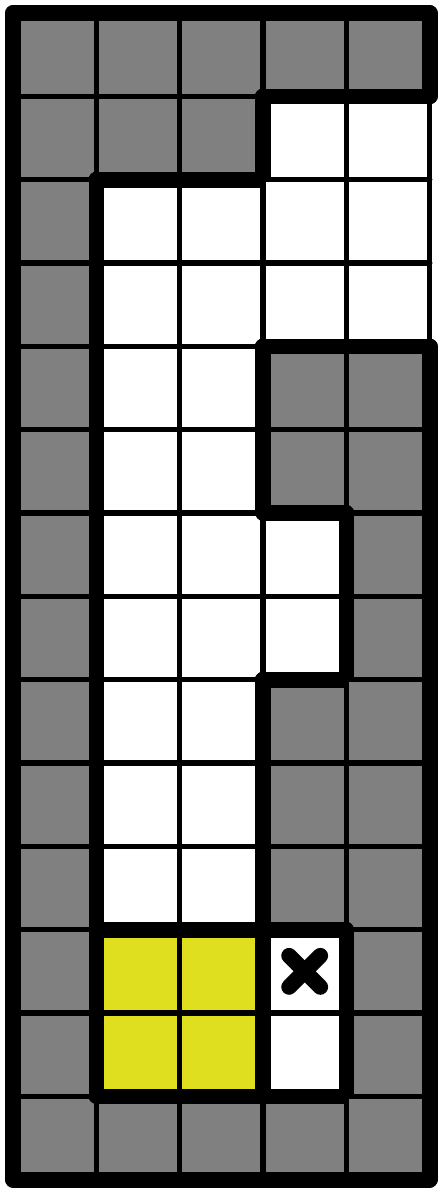}
    \caption{}
\end{subfigure}
\hspace{0.015\textwidth}
\begin{subfigure}[b]{0.1\textwidth}
    \includegraphics[width=\textwidth]{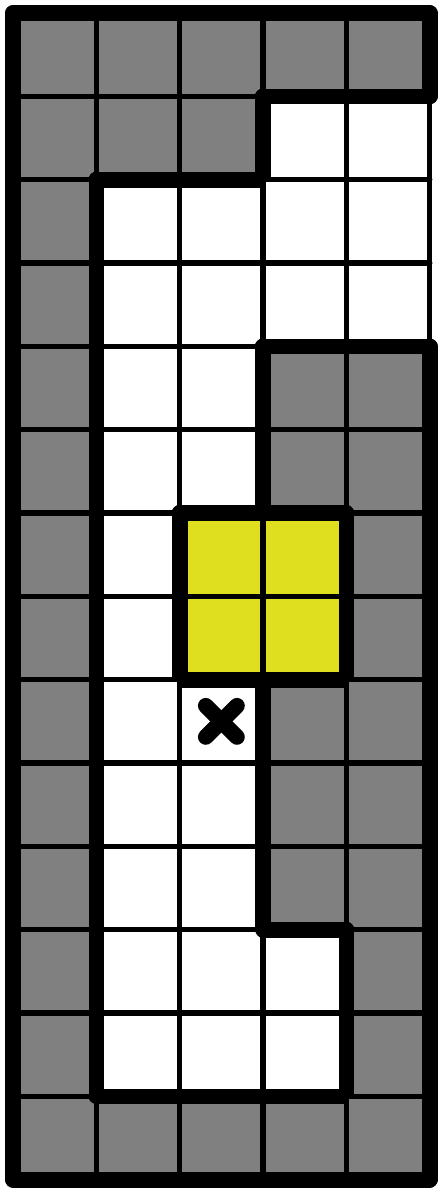}
    \caption{}
\end{subfigure}
\hspace{0.015\textwidth}
\begin{subfigure}[b]{0.1\textwidth}
    \includegraphics[width=\textwidth]{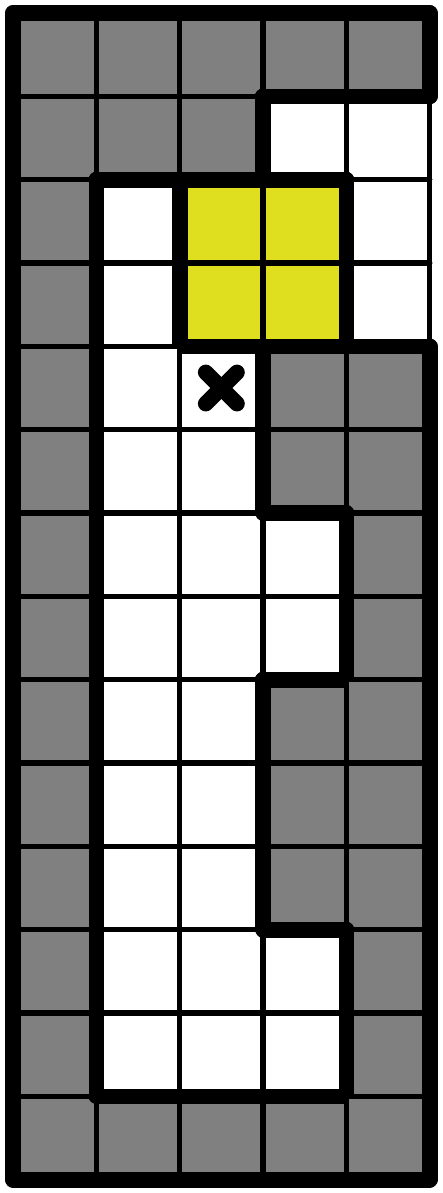}
    \caption{}
\end{subfigure}
\hspace{0.015\textwidth}
\begin{subfigure}[b]{0.1\textwidth}
    \includegraphics[width=\textwidth]{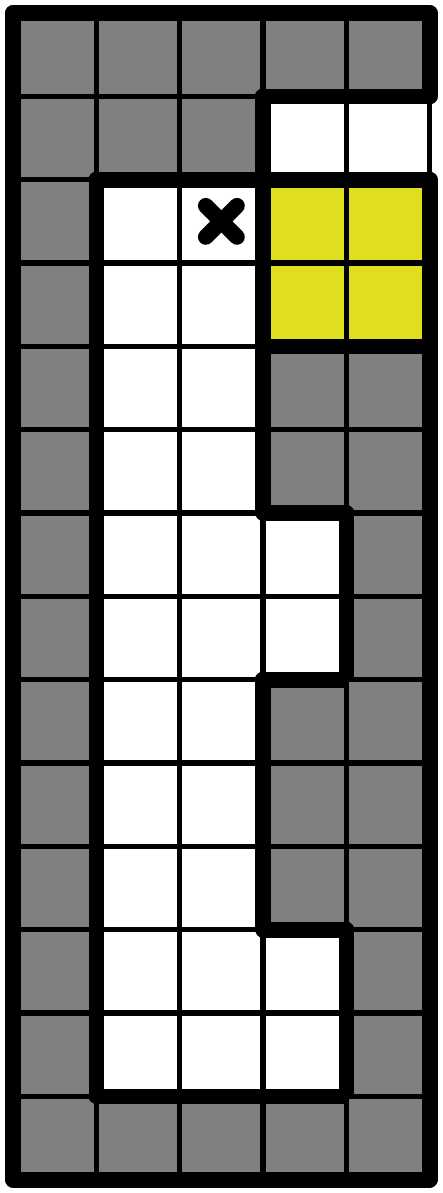}
    \caption{}
\end{subfigure}
\caption{Possibilities for placing the first $\protect\OO$ in $\langle \protect\OO, \protect\JJ, \protect\OO\rangle$. All but the first leave the puzzle unsolvable.}
\label{fig:O1inseq}
\end{figure}

\begin{figure}
\centering
\begin{subfigure}[b]{0.1\textwidth}
    \includegraphics[width=\textwidth]{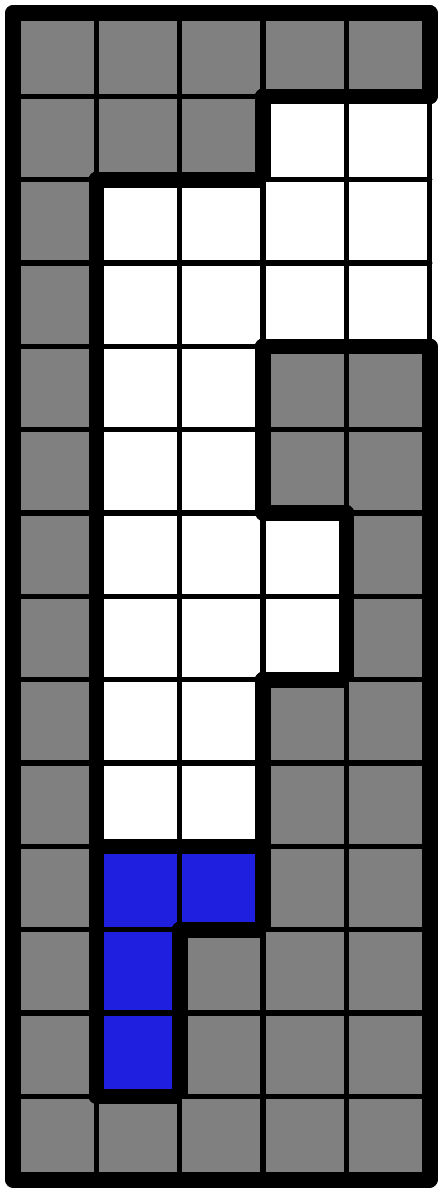}
    \caption{}
\end{subfigure}
\hspace{0.015\textwidth}
\begin{subfigure}[b]{0.1\textwidth}
    \includegraphics[width=\textwidth]{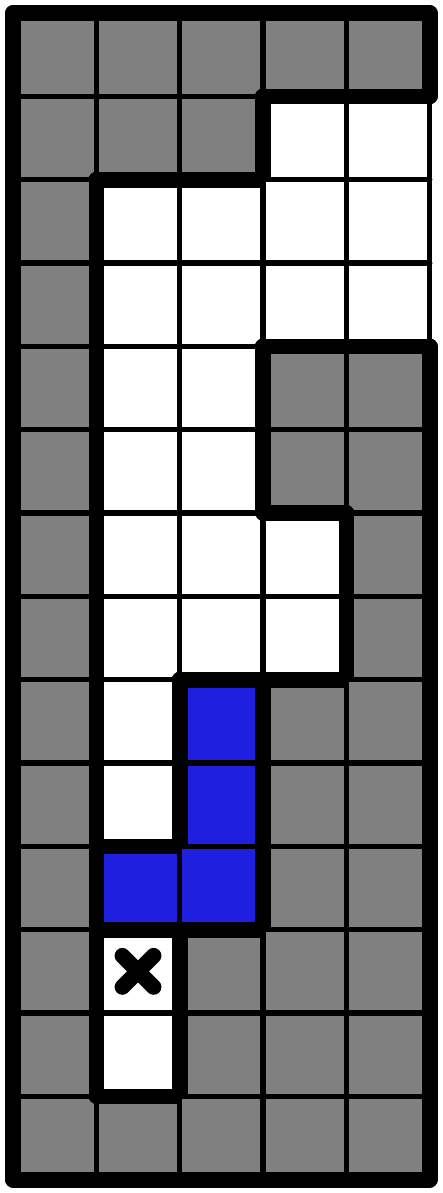}
    \caption{}
\end{subfigure}
\hspace{0.015\textwidth}
\begin{subfigure}[b]{0.1\textwidth}
    \includegraphics[width=\textwidth]{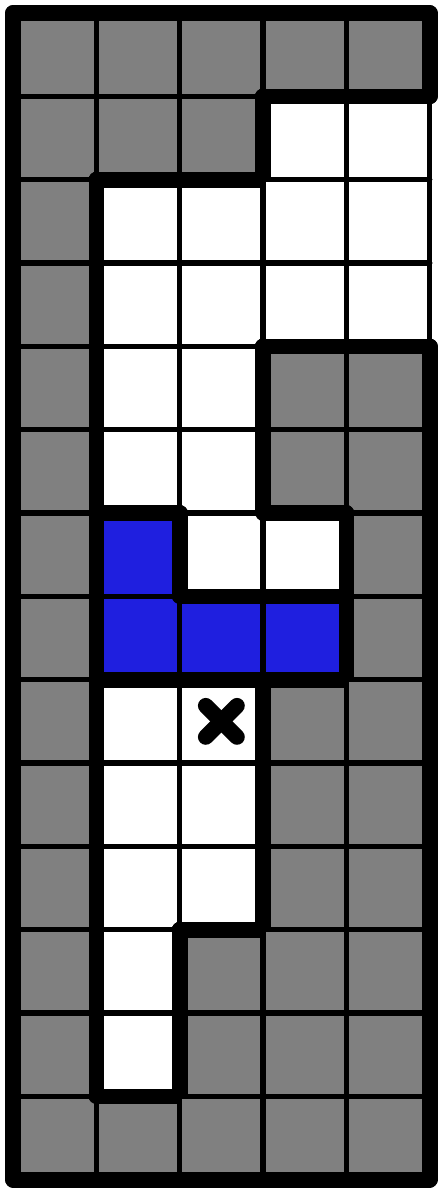}
    \caption{}
\end{subfigure}
\hspace{0.015\textwidth}
\begin{subfigure}[b]{0.1\textwidth}
    \includegraphics[width=\textwidth]{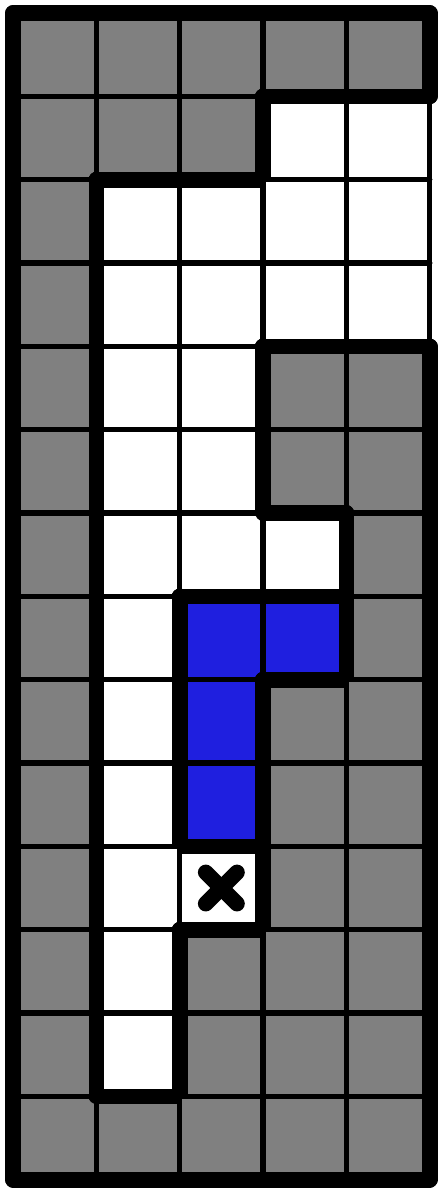}
    \caption{}
\end{subfigure}
\hspace{0.015\textwidth}
\begin{subfigure}[b]{0.1\textwidth}
    \includegraphics[width=\textwidth]{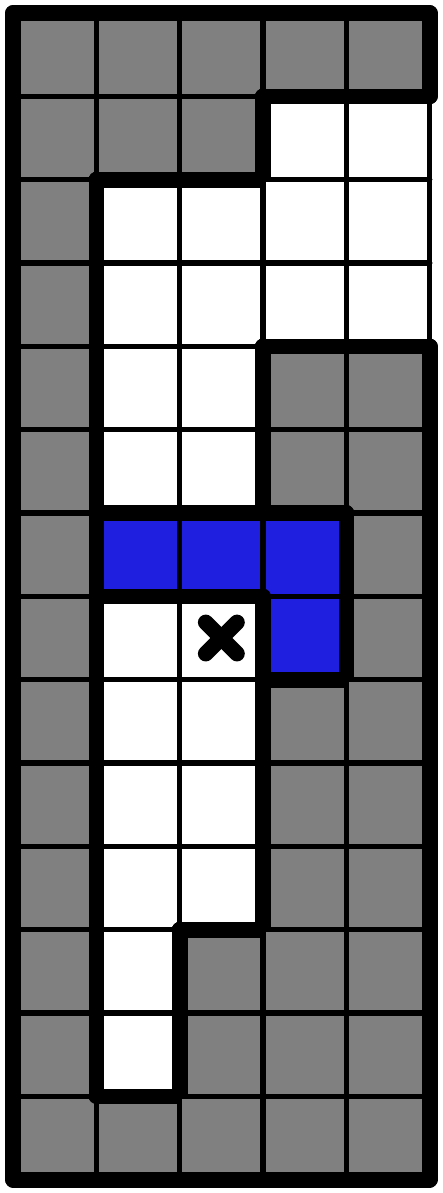}
    \caption{}
\end{subfigure}
\hspace{0.015\textwidth}
\begin{subfigure}[b]{0.1\textwidth}
    \includegraphics[width=\textwidth]{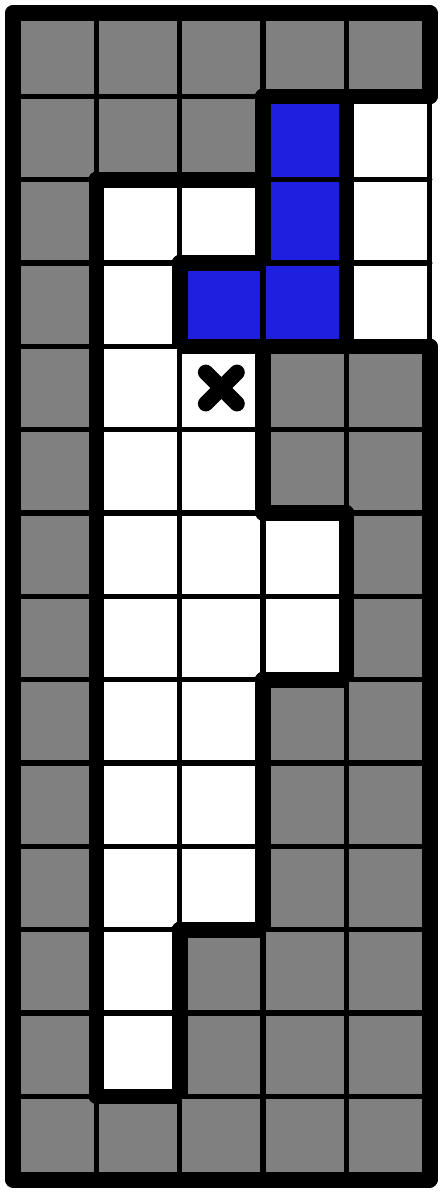}
    \caption{}
\end{subfigure}
\hspace{0.015\textwidth}
\begin{subfigure}[b]{0.1\textwidth}
    \includegraphics[width=\textwidth]{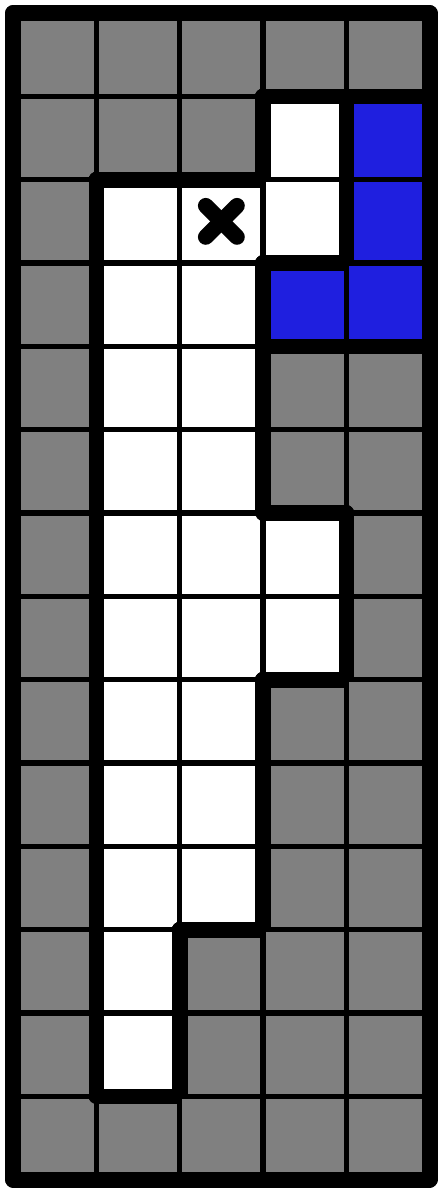}
    \caption{}
\end{subfigure}
\\
\vspace{0.4cm}
\begin{subfigure}[b]{0.1\textwidth}
    \includegraphics[width=\textwidth]{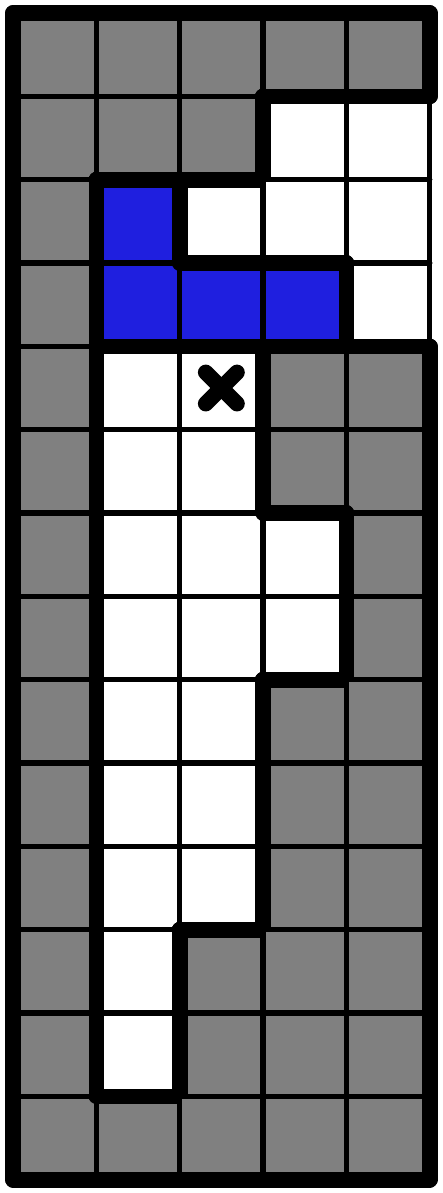}
    \caption{}
\end{subfigure}
\hspace{0.015\textwidth}
\begin{subfigure}[b]{0.1\textwidth}
    \includegraphics[width=\textwidth]{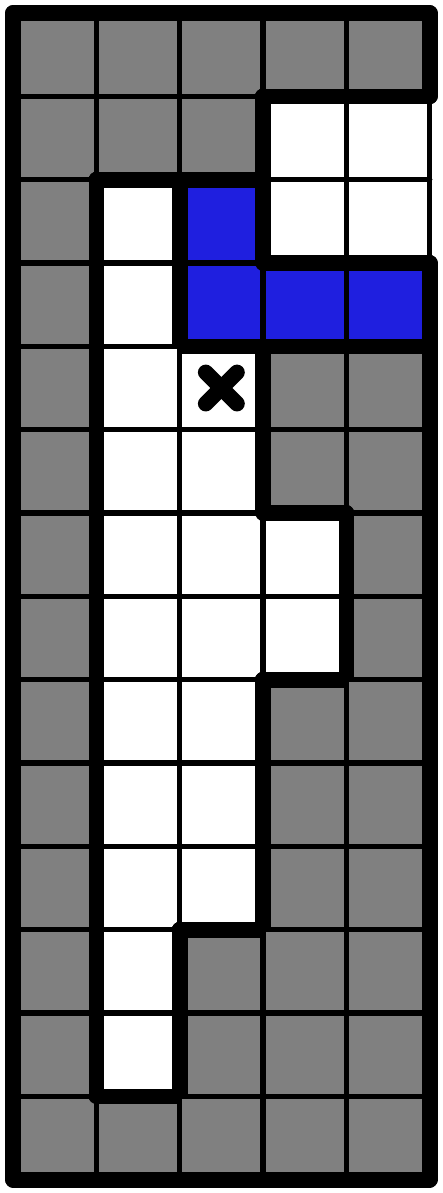}
    \caption{}
\end{subfigure}
\hspace{0.015\textwidth}
\begin{subfigure}[b]{0.1\textwidth}
    \includegraphics[width=\textwidth]{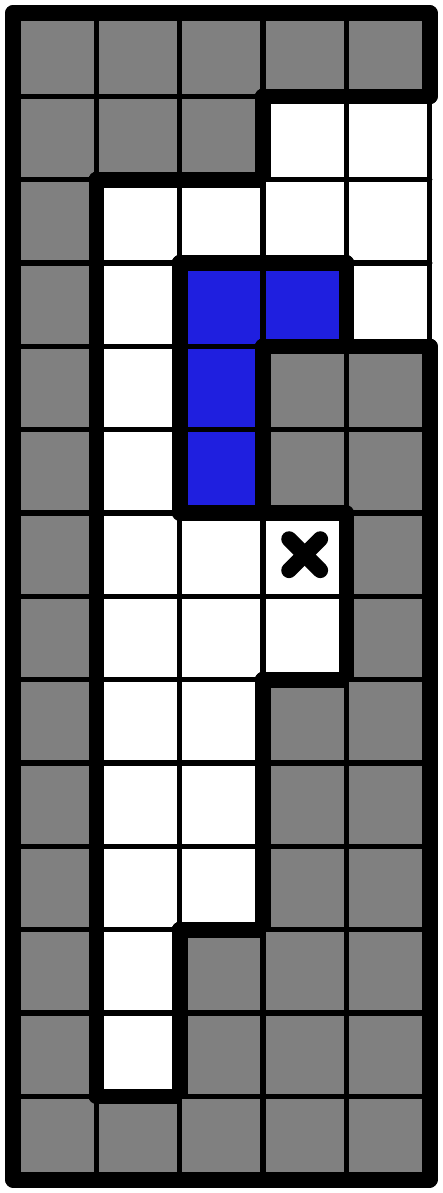}
    \caption{}
\end{subfigure}
\hspace{0.015\textwidth}
\begin{subfigure}[b]{0.1\textwidth}
    \includegraphics[width=\textwidth]{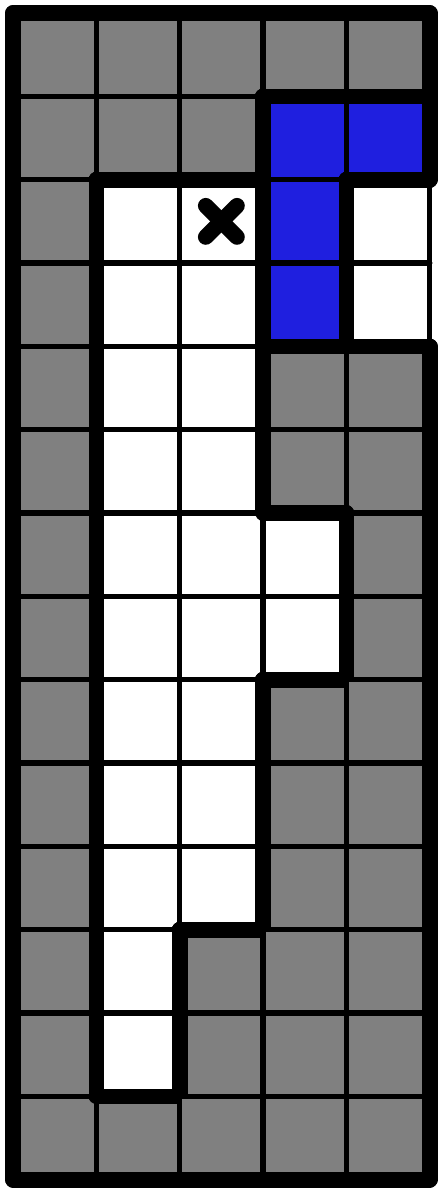}
    \caption{}
\end{subfigure}
\hspace{0.015\textwidth}
\begin{subfigure}[b]{0.1\textwidth}
    \includegraphics[width=\textwidth]{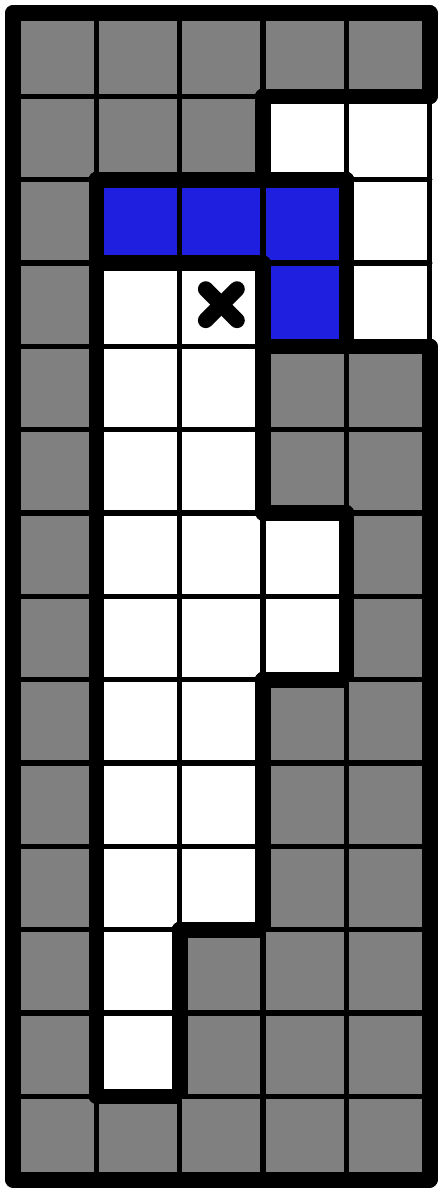}
    \caption{}
\end{subfigure}
\hspace{0.015\textwidth}
\begin{subfigure}[b]{0.1\textwidth}
    \includegraphics[width=\textwidth]{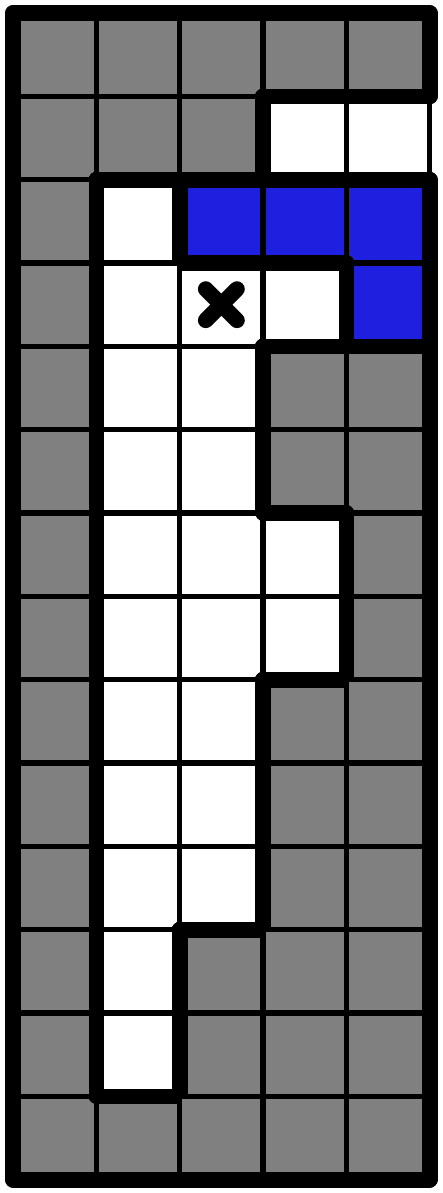}
    \caption{}
\end{subfigure}
\caption{Possibilities for placing the $\protect\JJ$ in $\langle \protect\OO, \protect\JJ, \protect\OO\rangle$. All but the first leave the puzzle unsolvable.}
\label{fig:Jinseq}
\end{figure}

\begin{figure}
\centering
\begin{subfigure}[b]{0.1\textwidth}
    \includegraphics[width=\textwidth]{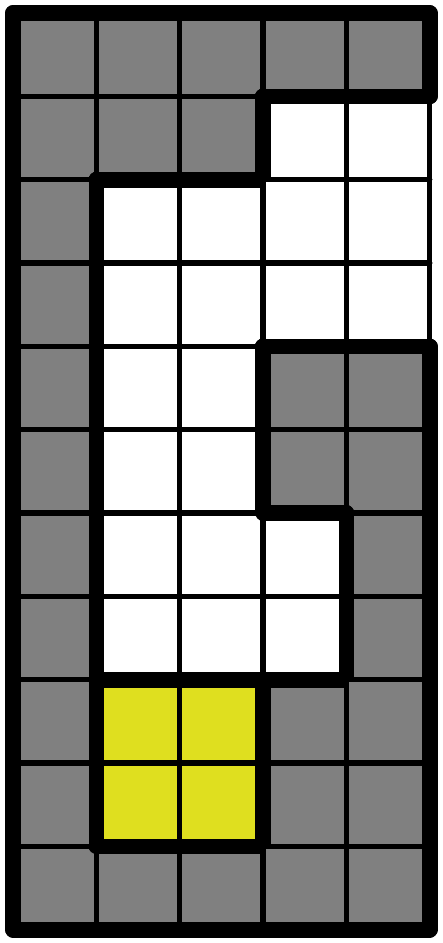}
    \caption{}
\end{subfigure}
\hspace{0.015\textwidth}
\begin{subfigure}[b]{0.1\textwidth}
    \includegraphics[width=\textwidth]{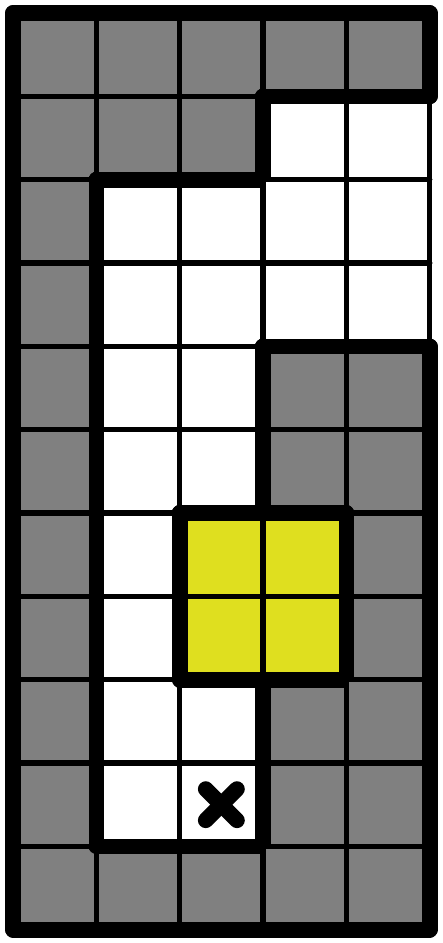}
    \caption{}
\end{subfigure}
\hspace{0.015\textwidth}
\begin{subfigure}[b]{0.1\textwidth}
    \includegraphics[width=\textwidth]{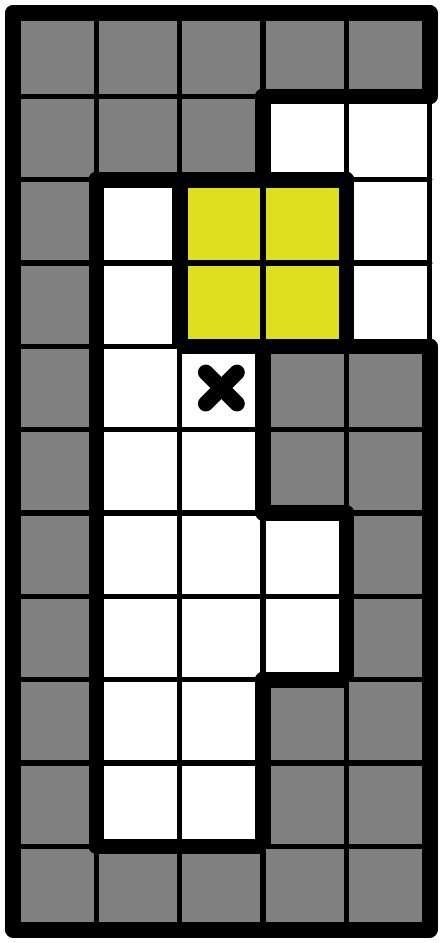}
    \caption{}
\end{subfigure}
\hspace{0.015\textwidth}
\begin{subfigure}[b]{0.1\textwidth}
    \includegraphics[width=\textwidth]{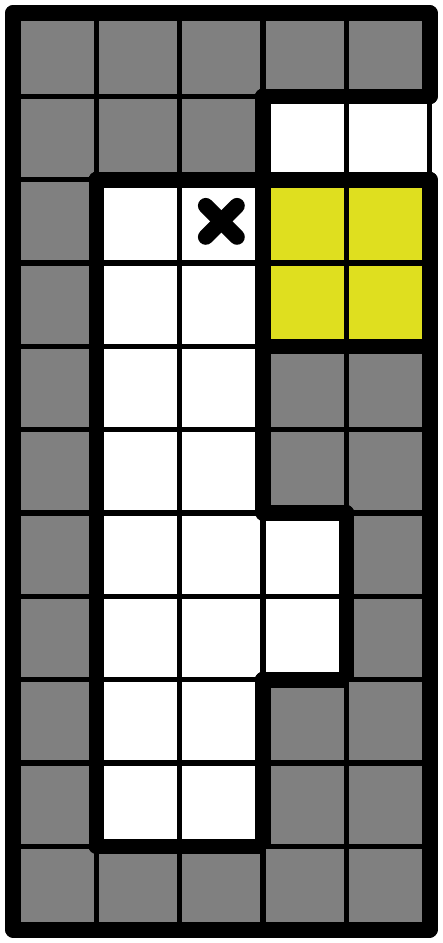}
    \caption{}
\end{subfigure}
\caption{Possibilities for placing the second $\protect\OO$ in $\langle \protect\OO, \protect\JJ, \protect\OO\rangle$. All but the first leave the puzzle unsolvable.}
\label{fig:O2inseq}
\end{figure}

\begin{claim} \label{struc4}
When the sequence $\langle \OO, \JJ, \OO \rangle$ is placed in a prepped bucket of height $h$, the bucket must end up as an prepped bucket of height $h-1$. (We know $h \ge 1$ by Claim \ref{struc3}.)
\end{claim}
\begin{proof}
Here we show that each of the parts of the sequence must be placed in a specific way. First, Figure~\ref{fig:O1inseq} shows all the ways an $\OO$ can be placed. Only Figure~\ref{fig:O1inseq}(b) shows a valid placement.

After the $\OO$ has been placed as in Figure~\ref{fig:O1inseq}(b), Figure~\ref{fig:Jinseq} shows all the ways the $\JJ$ can be placed afterward. Again, only Figure~\ref{fig:Jinseq}(c) shows a valid placement.

Finally, after the $\JJ$ has been placed as in \ref{fig:Jinseq}(c), Figure~\ref{fig:O2inseq} shows all the ways the second $\OO$ can be placed. In the only valid case, Figure~\ref{fig:O2inseq}(a), what remains is a bucket of height $h-1$, as desired.
\end{proof}

\begin{figure}
\centering
\begin{subfigure}[b]{0.1\textwidth}
    \includegraphics[width=\textwidth]{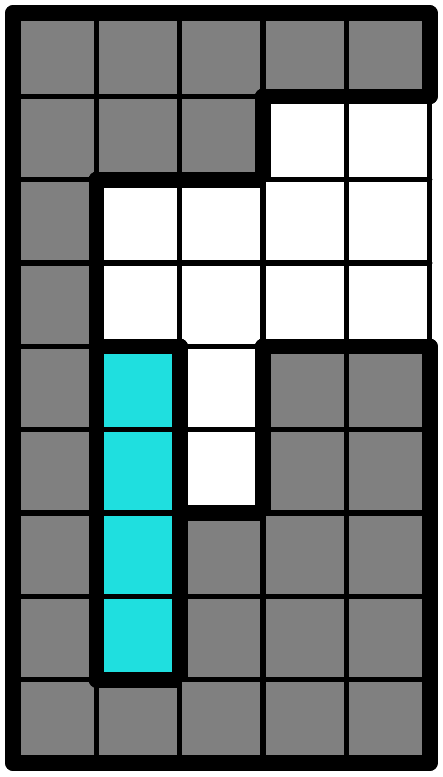}
    \caption{}
\end{subfigure}
\hspace{0.015\textwidth}
\begin{subfigure}[b]{0.1\textwidth}
    \includegraphics[width=\textwidth]{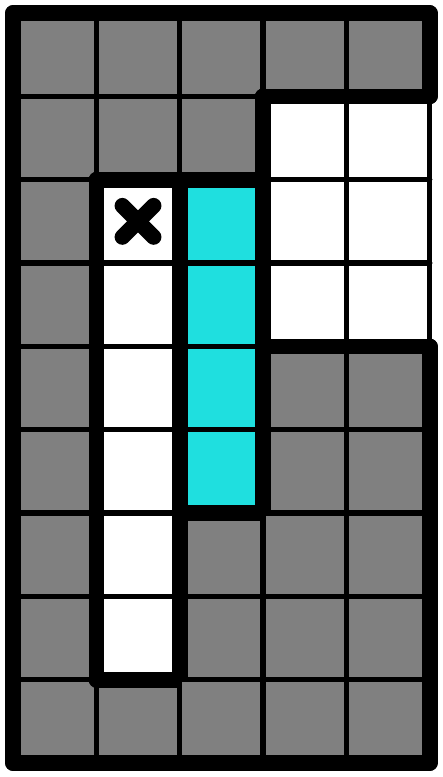}
    \caption{}
\end{subfigure}
\hspace{0.015\textwidth}
\begin{subfigure}[b]{0.1\textwidth}
    \includegraphics[width=\textwidth]{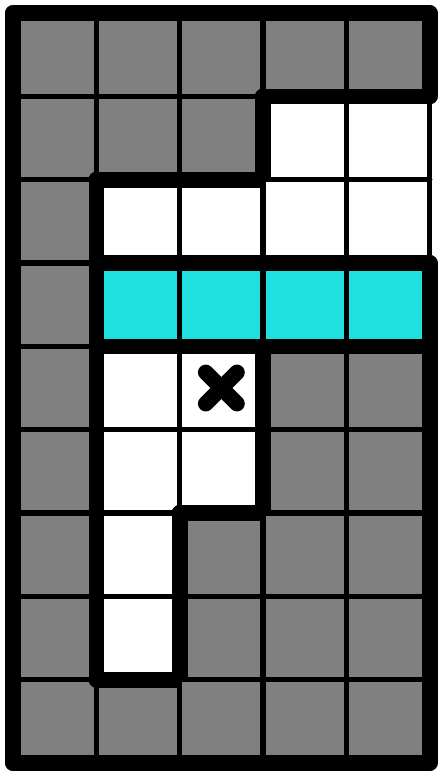}
    \caption{}
\end{subfigure}
\hspace{0.015\textwidth}
\begin{subfigure}[b]{0.1\textwidth}
    \includegraphics[width=\textwidth]{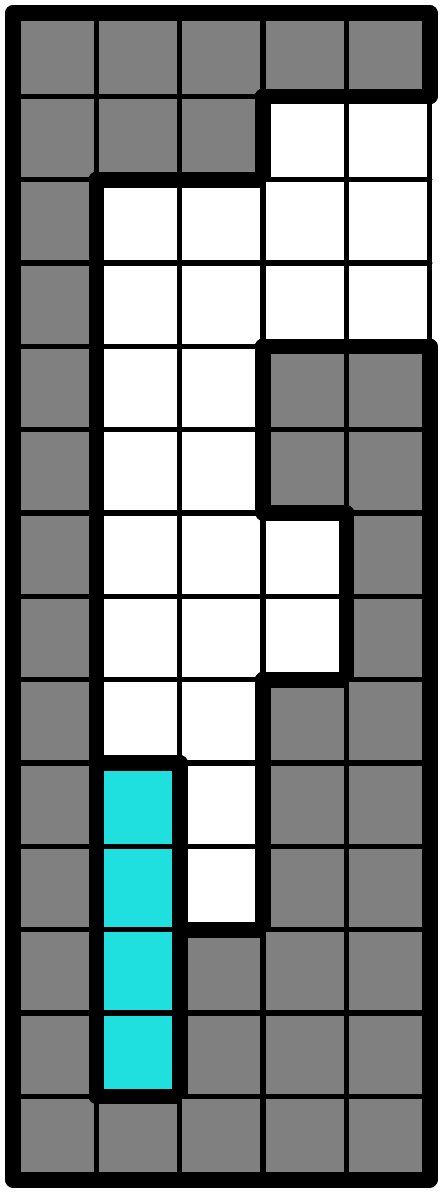}
    \caption{}
\end{subfigure}
\hspace{0.015\textwidth}
\begin{subfigure}[b]{0.1\textwidth}
    \includegraphics[width=\textwidth]{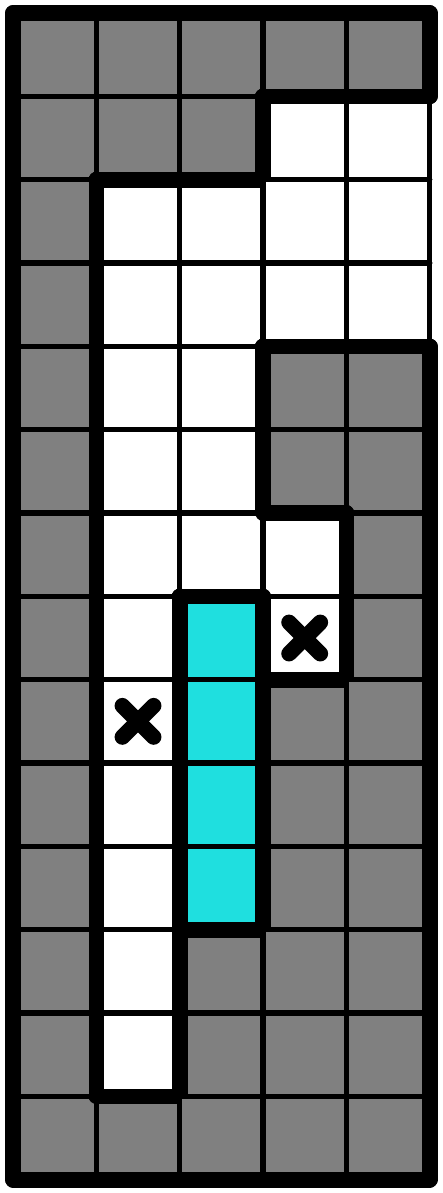}
    \caption{}
\end{subfigure}
\hspace{0.015\textwidth}
\begin{subfigure}[b]{0.1\textwidth}
    \includegraphics[width=\textwidth]{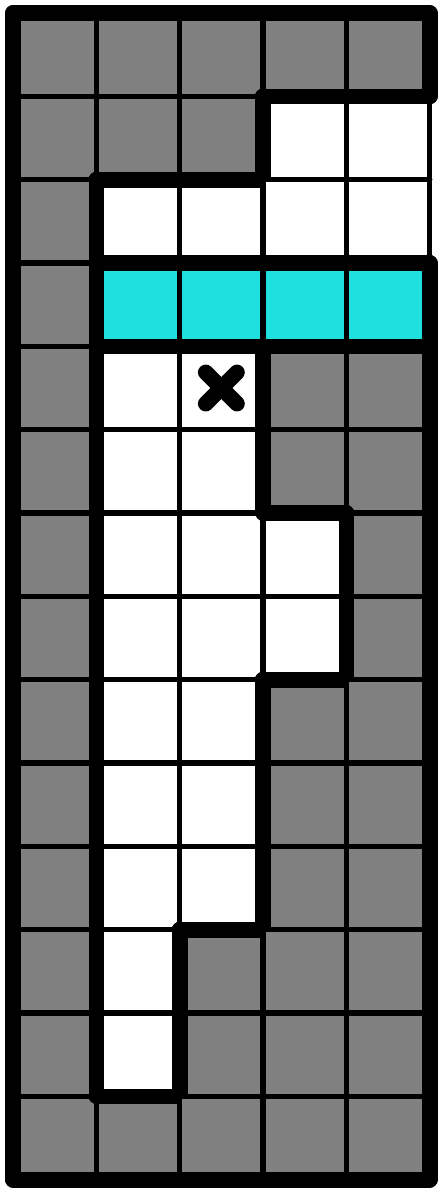}
    \caption{}
\end{subfigure}
\caption{Possibilities for placing the $\protect\II$. All but the first leave the puzzle unsolvable.}
\label{fig:Ifinal}
\end{figure}

\begin{claim} \label{struc5}
When the sequence $\langle \OO, \II \rangle$ is placed in a prepped bucket of height $h$, the bucket must end as an unprepped bucket of height $h - 1$. (We know $h \ge 1$ by Claim \ref{struc3}.)
\end{claim}
\begin{proof}
By the exact same cases as in Claim \ref{struc4}, the $\OO$ must be placed as in \ref{fig:O1inseq}(b).

Now, Figure~\ref{fig:Ifinal} shows all possible placements of the $\II$ afterward, where again crosses show unfillable cells. (In this case, it is necessary to split between the $h = 1$ and $h > 1$ cases.) Again, in Figure~\ref{fig:Ifinal}(e), there are two cells with crosses, which cannot both be filled since there is no $\SS$ in the sequence. In the only valid placements, Figures \ref{fig:Ifinal}(a) and \ref{fig:Ifinal}(d), the result is a prepped bucket of height $h-1$.
\end{proof}

The following corollary follows directly from Claims \ref{struc1}, \ref{struc2}, \ref{struc3}, \ref{struc4}, and \ref{struc5}:
\begin{corq} \label{strucmain-pre}
Suppose that before the $a_i$-sequence arrives, all buckets are unprepped.
If the $\LL$ starting the $a_i$-sequence is placed into a bucket, then the entire $a_i$-sequence must be placed into that bucket.
Furthermore, the height of that bucket decreases by $a_i + 1$; or if the height were to become negative, then these placements are impossible.
\end{corq}

\begin{figure}
\centering
\subcaptionbox{}{\includegraphics[scale=0.3]{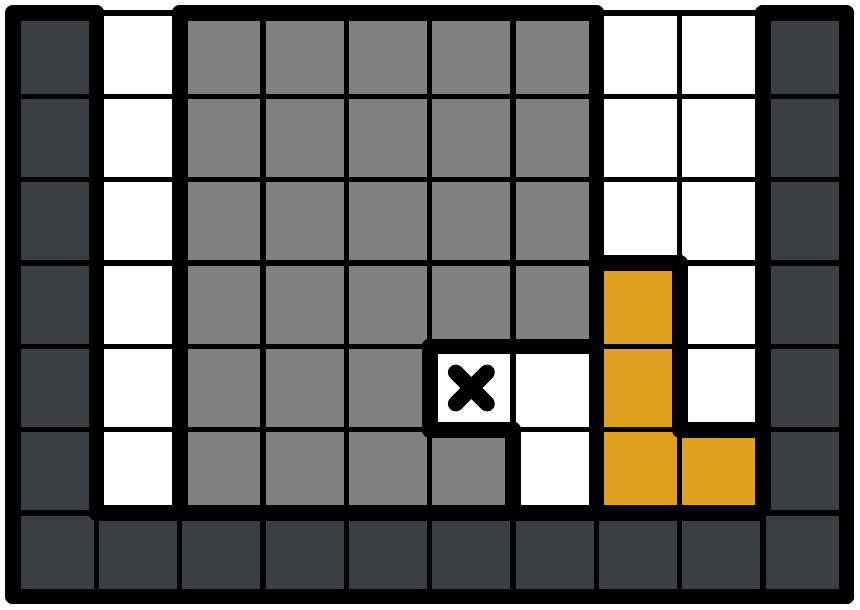}}
\subcaptionbox{}{\includegraphics[scale=0.3]{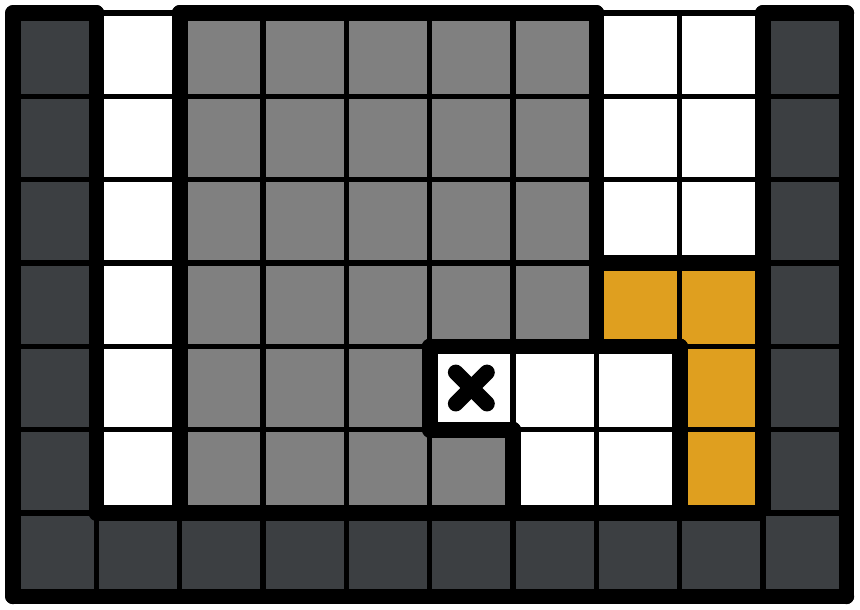}}
\subcaptionbox{}{\includegraphics[scale=0.3]{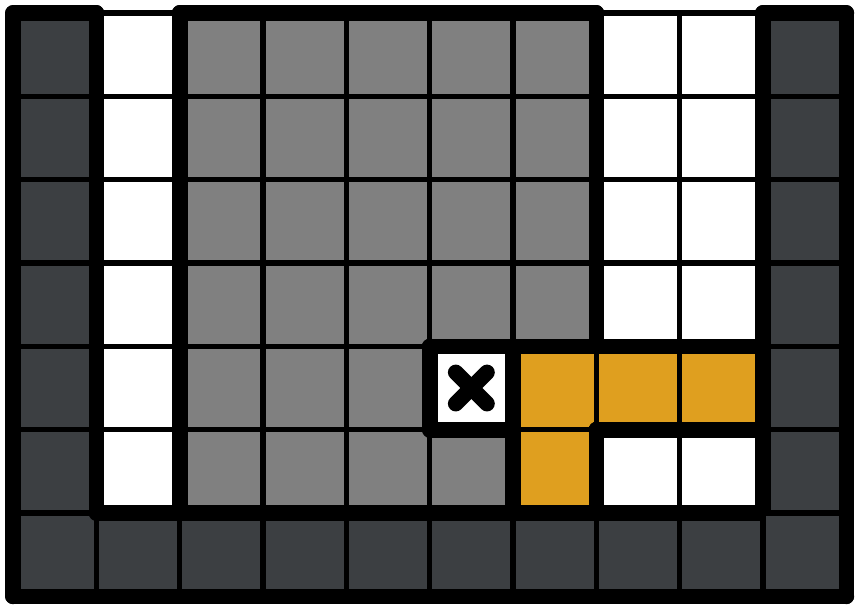}}
\subcaptionbox{}{\includegraphics[scale=0.3]{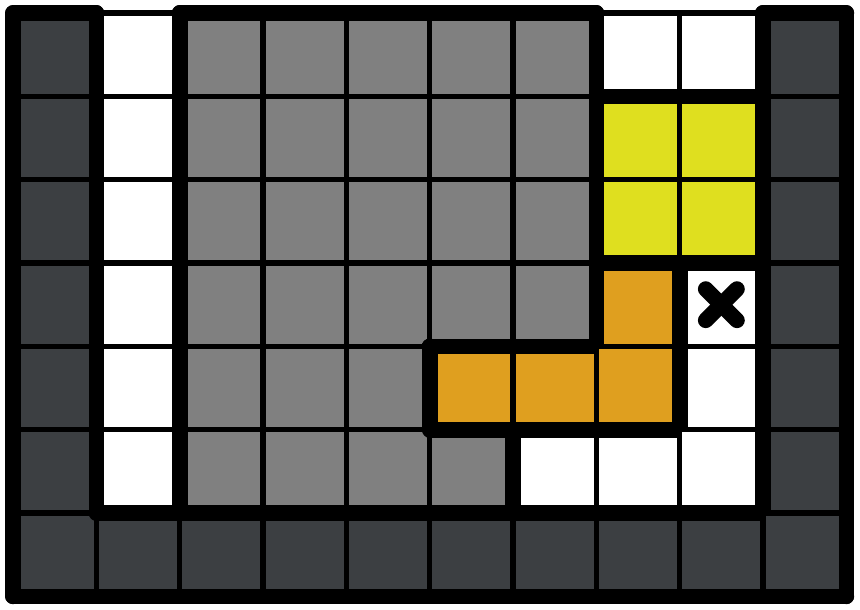}}
\subcaptionbox{}{\includegraphics[scale=0.3]{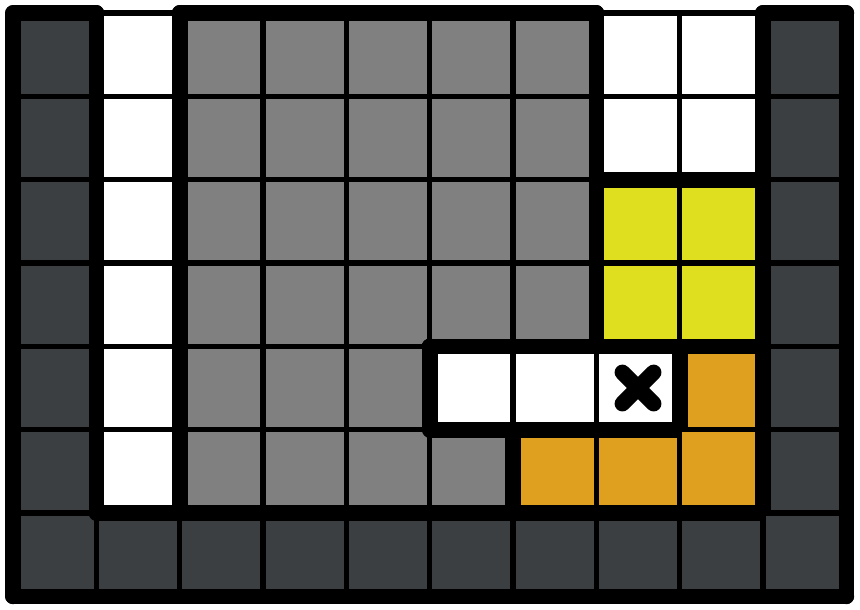}}
\caption{Possibilities for placing $\protect \LL$ into the corridor and then $\protect \OO$ into the corridor.}
\label{fig:no_o_after_l_in_corridor}
\end{figure}

We can now remove the assumption that the $\LL$ starting the $a_i$-sequence is placed into a bucket:

\begin{claim} \label{claim:a_i L in bucket}
  The $\LL$ starting each $a_i$-sequence must be placed into a bucket.
\end{claim}
\begin{proof}
  By the definition of the $a_i$-sequence, the $\LL$ is immediately followed by an~$\OO$.
  Consider for a contradiction the first $\LL$ from an $a_i$-sequence that is placed into the corridor. We claim that that the following $\OO$ cannot be placed.
  %creates an unfillable hole.
  By Claim~\ref{claim:no-corridor} and by induction, no piece other than the just-placed $\LL$ has been placed in the corridor.
  If the $\OO$ piece is placed in the corridor, the casework in Figure~\ref{fig:no_o_after_l_in_corridor} shows that it would create an unfillable hole.
  However, by Corollary~\ref{strucmain}, before each $a_i$-sequence, all buckets are unprepped,
  so by Claim~\ref{struc2}, the $\OO$ also cannot be placed in a bucket, a contradiction.
\end{proof}

Combining Corollary~\ref{strucmain-pre} with Claim~\ref{claim:a_i L in bucket}, we obtain the following:

\begin{corq} \label{strucmain}
Suppose that before the $a_i$-sequence arrives, all buckets are unprepped.
Then the entire $a_i$-sequence must be placed into that bucket.
Furthermore, the height of that bucket decreases by $a_i + 1$; or if the height were to become negative, then these placements are impossible.
\end{corq}

We are finally ready to prove the other direction of the bijection.

\begin{thm}
If $\mc G(\mc P)$ has a surviving trajectory sequence, then the {\partit} instance $\mc P$ has a solution.
\end{thm}
\begin{proof}
Numbering the buckets $1, 2, \dots, s$, let $S_b$ be the set of $i$ such that the $a_i$-sequence is placed in bucket~$b$.
By the first half of Corollary~\ref{strucmain}, the $S_b$'s form a partition of $\{a_1, a_2, \dots, a_{3s}\}$.
By the second half of Corollary~\ref{strucmain}, the sum $\sum_{i \in S_b} (a_i + 1)$ is at most the original height of each bucket, which is $T + 3$.
However, $\sum_{i=1}^{3s} (a_i + 1) = s(T + 3)$, so equality must hold.

Thus, we have $\sum_{i \in S_b} (a_i + 1) = T + 3$ for each bucket $b$. But the condition $T/4 < a_i < T/2$ means that this sum cannot have at most 2 or at least 4 terms, so it must have 3 terms, and thus $|S_b| = 3$. Then the condition simplifies to $\sum_{i \in S_b} a_i = T$, and thus the $S_b$ represent a valid 3-partition.
\end{proof}

%% file: tetris/nphard_row.tex
\section{\row-row Tetris is NP-hard}
\label{sec:4-row}

\begin{thm} \label{thm:4 row}
It is NP-complete to survive or clear the board in \rtet{$r$} for any $r \geq 4$.
\end{thm}

The proof by reduction from {\partit}. Given an instance $\mc P$ of {\partit} with elements $\{a_1, a_2, \dots, a_{3s}\}$ and target $T$, we create an instance $\mc G = \mc G(\mc P)$ of \rtet{$r$} which has a valid 3-partition if and only if there is a sequence of moves to survive, and an extension of such a surviving sequence to leave the Tetris board empty.

\begin{figure}
    \centering
    \includegraphics[width=\textwidth]{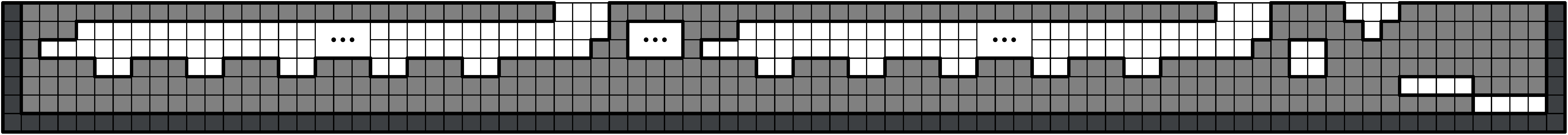}
    \caption{Shows the initial board for $r=6$}
\label{fig:row-figure-full}
\end{figure}

The initial board, illustrated in Figure~\ref{fig:row-figure-full} for $r=6$
(where filled cells are grey and the rest of the cells are unfilled), has $15sT+8s+8 + 4r$ columns and $r$ rows, containing the following unfilled cells:

\begin{itemize}
\item A number $s$ of \emph{buckets}, which branch off the corridor to its right. These are similar in shape to the buckets used for the proof of $\ctet{$c$}$ except for the first few and last few columns. Each bucket has a total width of $15T+6$, and contains $3T$ \emph{notches} (the pairs of adjacent empty cells in row 5, counting rows from the top). \emph{Buckets} are separated by $2$ columns. % formerly width 5T+21, T+3 notches
\item A \emph{T-lock} in the shape of a $\TT$ piece in the top two rows.
\item An \emph{O-lock} in the shape of a $\OO$ piece in the next two rows.
\item Right filler: in each row below the top four, there are exactly four empty spaces, to the right of any columns empty in any higher row.
\end{itemize}

The piece sequence is as follows.
First, for each $a_i$, we send the following \emph{$a_i$ sequence}
(see Figures~\ref{fig:bucketFilling}(i--m)):
$\langle \LL, \langle \OO, \JJ, \OO \rangle^{3a_i-1}, \OO, \II \rangle$.
%$\langle \LL, \langle \OO, \JJ, \OO \rangle^{a_i}, \OO, \II \rangle$.
After all these pieces, we send the following \emph{clearing sequence}
(see Figures~\ref{fig:bucketFilling}(n) and (b--h)):
$\langle \langle \LL, \OO, \JJ, \OO, \LL, \OO \rangle^{s}, \TT, \OO \rangle$.
% $\langle \langle \LL, \LL, \OO \rangle^{s}, \TT, \OO \rangle $.
Finally, if $r > 4$, we send $\langle \II \rangle^{r-4}$.

The total size of the board is $r(15sT+8s+8 + 4r)$ and the total number of pieces is
\[\sum_{i=1}^{3s} (3 + 3(3a_i-1)) + 2 + 6s\]
which are both polynomial in the size of the {\partit} instance (recalling that {\partit} is strongly NP-hard). 
\begin{lem}[\rtet{$r$} $\iff {\partit}$] \label{row-main}

  For a ``yes'' instance of {\partit}, there is a way to drop the pieces that survives and clears the entire board. Conversely, if the piece sequence can be survived, then the {\partit} instance has a solution.
\end{lem}

\begin{figure}
    \centering
    \begin{subfigure}[t]{\textwidth}
        \includegraphics[width=\textwidth]{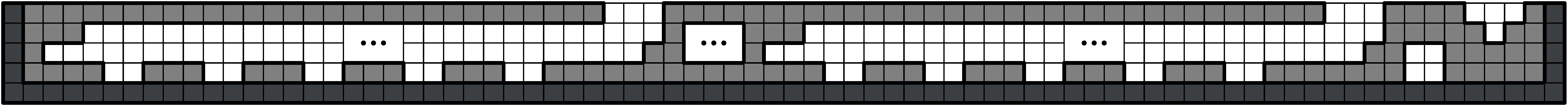}
        \caption{}
        \label{fig:row-init}
    \end{subfigure}

    \begin{subfigure}[t]{\textwidth}
        \includegraphics[width=\textwidth]{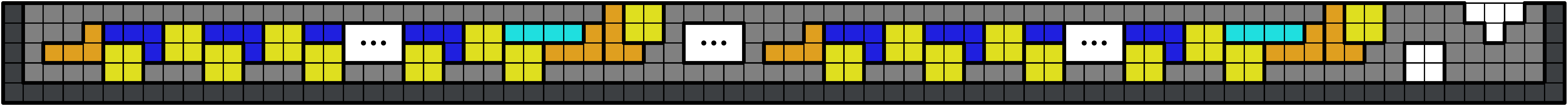}
        \caption{}
    \end{subfigure}

    \begin{subfigure}[t]{\textwidth}
        \includegraphics[width=\textwidth]{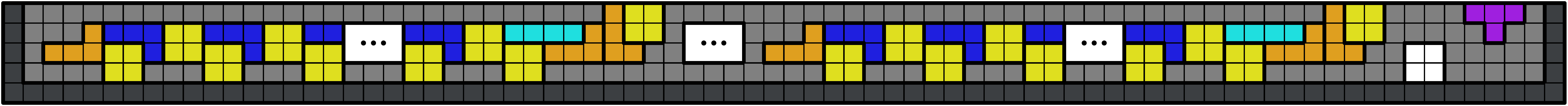}
        \caption{}
    \end{subfigure}

    \begin{subfigure}[t]{\textwidth}
        \includegraphics[width=\textwidth]{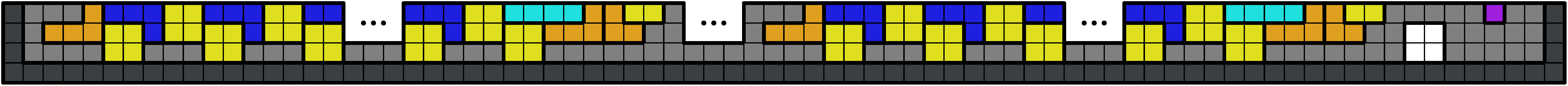}
        \caption{}
    \end{subfigure}

    \begin{subfigure}[t]{\textwidth}
        \includegraphics[width=\textwidth]{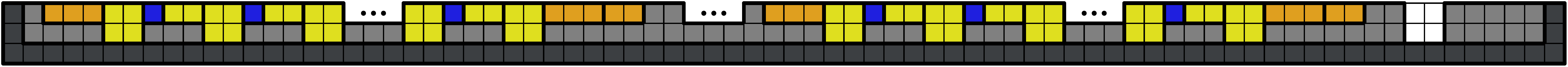}
        \caption{}
    \end{subfigure}

    \begin{subfigure}[t]{\textwidth}
        \includegraphics[width=\textwidth]{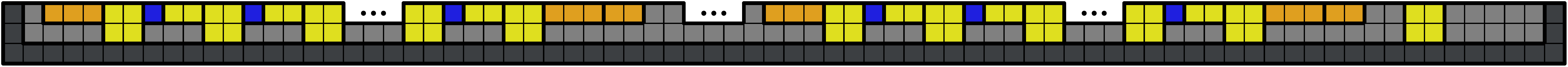}
        \caption{}
    \end{subfigure}

    \begin{subfigure}[t]{\textwidth}
        \includegraphics[width=\textwidth]{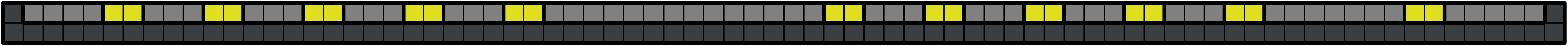}
        \caption{}
    \end{subfigure}
    
    \begin{subfigure}[t]{\textwidth}
        \includegraphics[width=\textwidth]{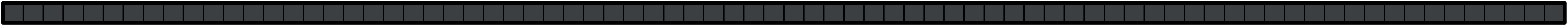}
        \caption{}
    \end{subfigure}
\caption{(a) shows the initial board for $r=4$. (b) shows a correctly filled board. After (c), there are many ways to survive; (c--h) show the clearing sequence.}
\label{fig:row-figure}
\end{figure}

Figure~\ref{fig:row-figure} illustrates that a solution to {\partit} survives and clears the Tetris board. 

To show that if $\mc G(\mc P)$ has a sequence of moves that survives, then the {\partit} instance $\mc P$ must also have a solution (i.e., a valid partition), we progressively constrain any {\tet} solution to a form that directly encodes a {\partit} solution.

\begin{claim}\label{row-tlock}
  Nothing may be placed in the T-lock except a $\TT$.
\end{claim}

\begin{proof}
  No row can be cleared until some cell of the T-lock is filled and we also note that there is only one $\TT$ in the complete piece sequence. However, the only piece that can fill any cell of the T-lock without filling a cell above the horizon (causing a loss by partial lock out) is a $\TT$, so the claim follows from Claim \ref{row-horiz}. 
\end{proof}

\begin{corq}\label{row-olock}
  Nothing may be placed in the O-lock except an $\OO$.
\end{corq}

\begin{proof}
  The first two rows cannot be cleared until T-lock is filled with an $\TT$ by Claim \ref{row-tlock}. This means that no piece may reach the O-lock since it is covered by the first two rows. The only piece that follows the $\TT$ is an $\OO$ which must go into the O-lock as desired.
\end{proof}

So we have the following corollary which follows directly from Claim \ref{row-olock}:

\begin{corq}
\label{row-noclear}
No row may be cleared until the first $\TT$ has arrived.
\end{corq}

We implicitly use Corollary \ref{row-noclear} throughout this paper, since it implies that the buckets must maintain their shape until after the $\TT$ arrives.

\begin{claim} \label{row-horiz}
No cell above the horizon may ever be filled. 
\end{claim}
\begin{proof}
The area of the pieces sent up to the first $\TT$ is exactly equal to 
\[4(9sT+6s+1),\] 
the area of the unfilled cells in the first four rows outside the $\OO$ lock is
\[s(36T+24) + 4= 4(9sT+6s+1),\] so when the first $\TT$ arrives, every cell in the first four rows outside the $\OO$ lock must be full for survival, and no cell can ever be placed in an empty row.
\end{proof}

% \begin{claim} \label{row-horiz}
% No cell above the horizon may ever be filled. 
% \end{claim}

% \begin{proof}
% The number of unfilled cells below the horizon is
% \[
% s(12T + 48) % buckets
% + 4 % O-Lock
% + 4 % T-Lock
% = 12sT + 48s + 8.
% \]
% Also, the total number of pieces to arrive is (recalling that $\sum_{i=1}^{3s} a_i = sT$)

% \[
% \sum_{i=1}^{3s} (3 + 3a_i) + 3s + 2 
% = 3sT + 12s + 2.
% \]
% Thus, the number of unfilled cells below the horizon is exactly four times the number of pieces to arrive. Each piece fills four cells, so the number of unfilled cells below the horizon is equal to the total number of cells in the pieces that will arrive. Thus, the number of rows that can be cleared is (by counting total filled cells) at most the number of rows below the horizon. Thus, in order to completely clear the board, no cell above the horizon can be filled.
% \end{proof}

If we \emph{survive} until the first $\TT$ by filling all cells in the first four rows except the $\OO$ lock, then the remaining $\OO$ and $\II$ pieces \emph{clear} the board. Hence it suffices to show that for $r=4$, the given Tetris game is clearable if and only if the instance of {\partit} is solvable. Henceforth we assume $r=4$.

% \begin{claim} \label{row-horiz}
% No cell above the horizon may ever be filled. 
% \end{claim}
% \begin{proof}
% We first note that we have shown above that the total number of unfilled cells below the horizon ($4((9sT+6s+2) + (r-4))$) is exactly equal to the area of the pieces that arrives since there are exactly $(9sT+6s+2) + (r-4)$ pieces. Thus, the number of rows that can be cleared is (by counting total filled cells) at most the number of rows below the horizon. Thus, in order to completely clear the board, no cell above the horizon can be filled.
% \end{proof}

\begin{figure}[t]
\centering
\begin{minipage}[b]{0.45\linewidth}
\centering
\begin{subfigure}[b]{0.365\textwidth}
    \includegraphics[scale=0.3]{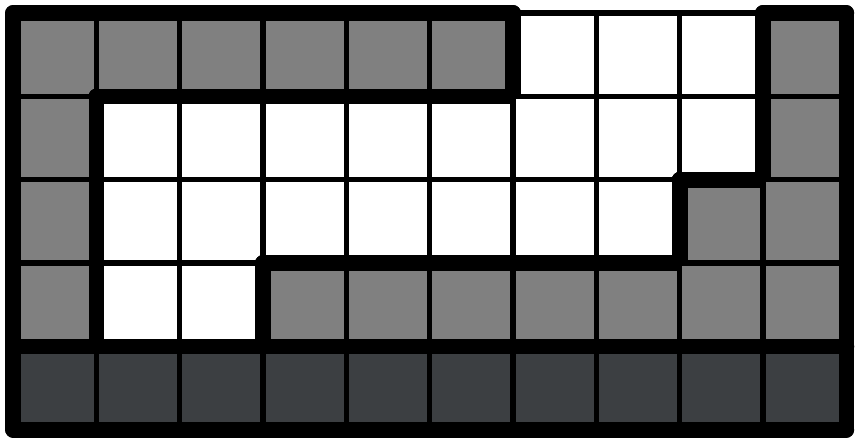}
    \caption{}
\end{subfigure}
\hspace{0.015\textwidth}
\begin{subfigure}[b]{0.225\textwidth}
    \includegraphics[scale=0.3]{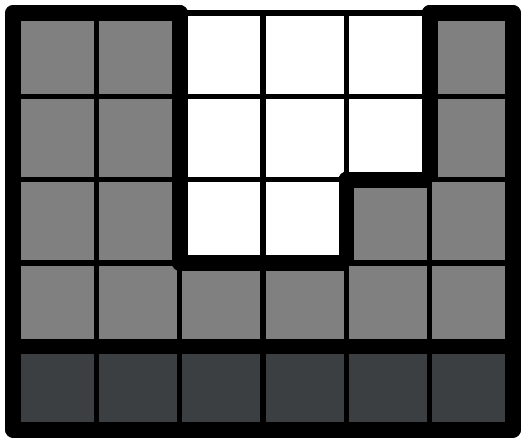}
    \caption{}
\end{subfigure}
\caption{Prepped buckets of heights 1 and 0, respectively}
\label{fig:row-unprep}
\end{minipage}\hfil\hfil
\begin{minipage}[b]{0.45\linewidth}
\centering
\begin{subfigure}[b]{0.5\textwidth}
    \includegraphics[scale=0.3]{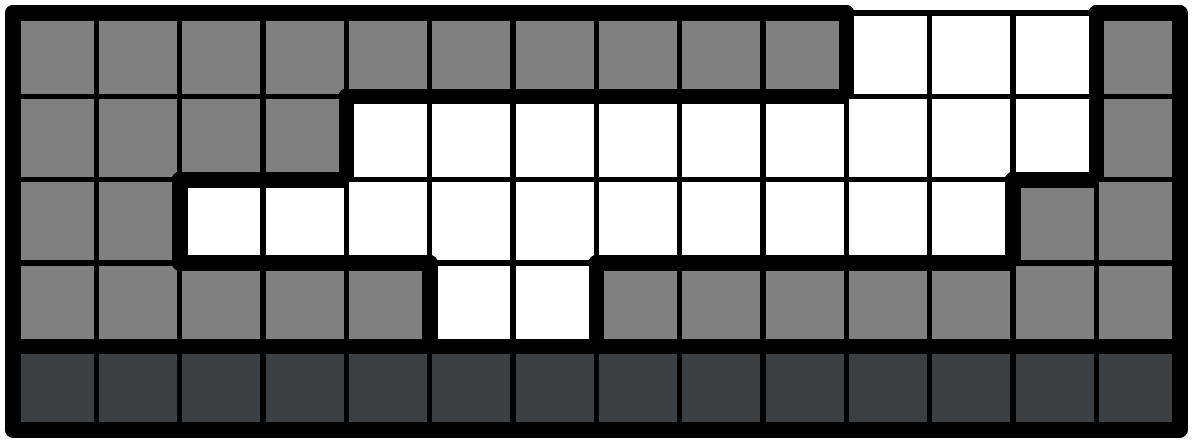}
    \caption{}
\end{subfigure}
\hspace{0.015\textwidth}
\begin{subfigure}[b]{0.33\textwidth}
    \includegraphics[scale=0.3]{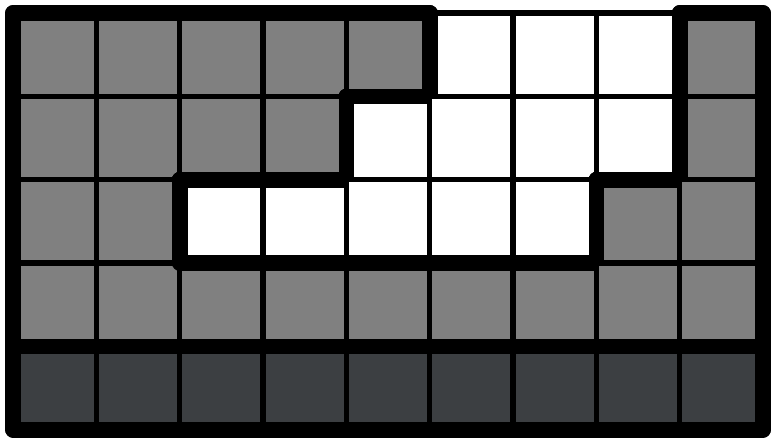}
    \caption{}
\end{subfigure}
\caption{Unprepped buckets of heights 1 and 0, respectively}
\label{fig:row-prep}
\end{minipage}
\end{figure}

Now, to complete the proof, we show that the buckets must be filled in the manner shown in Figure~\ref{fig:row-figure}, so some cell must be left empty if there is no {\partit} solution. Define \emph{prepped} (Figure~\ref{fig:row-prep}) and \emph{unprepped} (Figure~\ref{fig:row-unprep}) buckets as in Section \ref{soundness}. We define the height in the same manner by the number of notches.

\xxx{Are the height-0 figures necessary anymore? Maybe it's better to replace them with height-2 figures (for O(1) rows only, not columns)}

We can now prove an analogue of Claim~\ref{struc1}:
\begin{claim} \label{row-struc1}
If an $\LL$ is placed in an unprepped bucket of height at least 1, it must form a prepped bucket of the same height.
%If an $\LL$ is placed in an unprepped bucket, it must form a prepped bucket of the same height.
\end{claim}

\begin{figure}
    \centering
    \subcaptionbox{}{\includegraphics[scale=0.3]{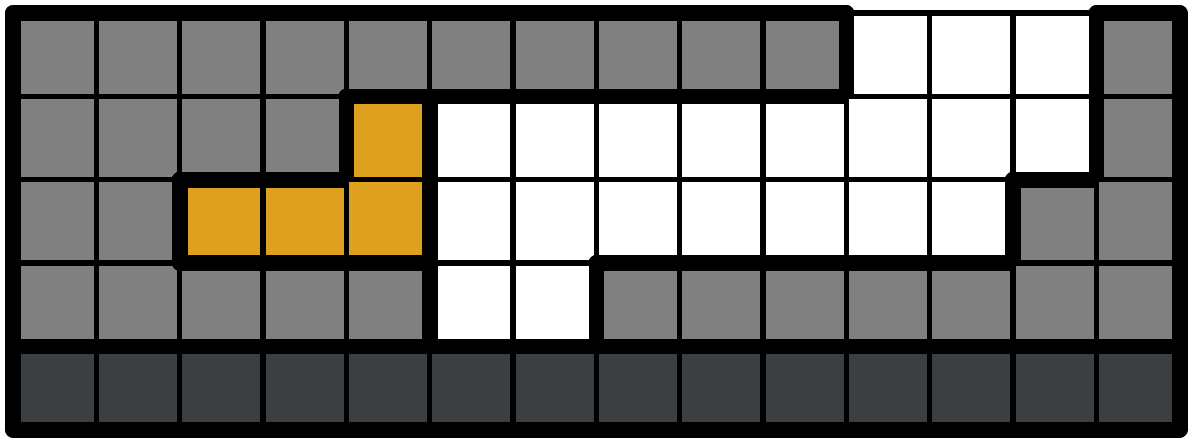}}
    % \subcaptionbox{}{\includegraphics[scale=0.3]{tetris/figures/rowalley_badconfigs/L_unprep_h1_L2.pdf}}
    % \subcaptionbox{}{\includegraphics[scale=0.3]{tetris/figures/rowalley_badconfigs/L_unprep_h1_L3.pdf}}
    % \subcaptionbox{}{\includegraphics[scale=0.3]{tetris/figures/rowalley_badconfigs/L_unprep_h1_L4.pdf}}
    % \subcaptionbox{}{\includegraphics[scale=0.3]{tetris/figures/rowalley_badconfigs/L_unprep_h1_L5.pdf}}
    % \subcaptionbox{}{\includegraphics[scale=0.3]{tetris/figures/rowalley_badconfigs/L_unprep_h1_L6.pdf}}
    % \subcaptionbox{}{\includegraphics[scale=0.3]{tetris/figures/rowalley_badconfigs/L_unprep_h1_L7.pdf}}
    \subcaptionbox{}{\includegraphics[scale=0.3]{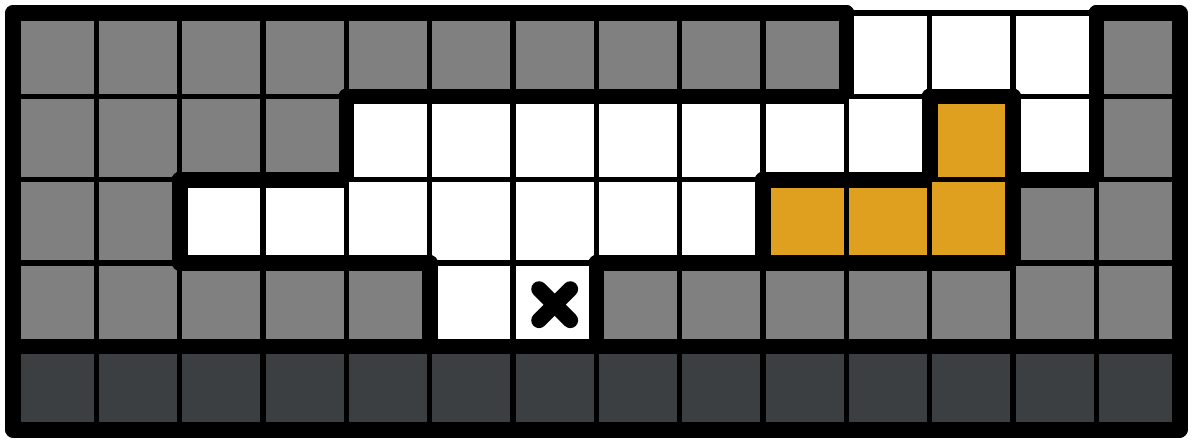}}
    % \subcaptionbox{}{\includegraphics[scale=0.3]{tetris/figures/rowalley_badconfigs/L_unprep_h1_L9.pdf}}
    % \subcaptionbox{}{\includegraphics[scale=0.3]{tetris/figures/rowalley_badconfigs/L_unprep_h1_L10.pdf}}
    \subcaptionbox{}{\includegraphics[scale=0.3]{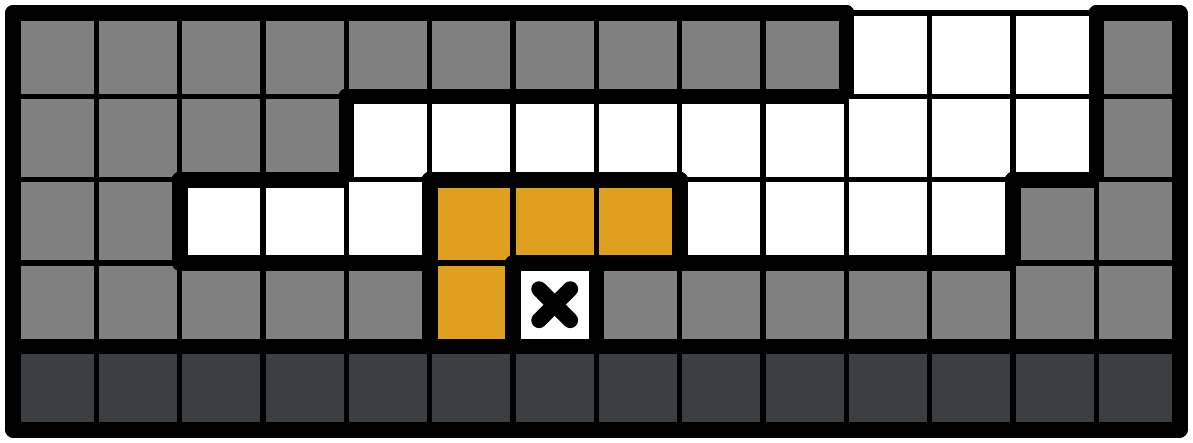}}
    \subcaptionbox{}{\includegraphics[scale=0.3]{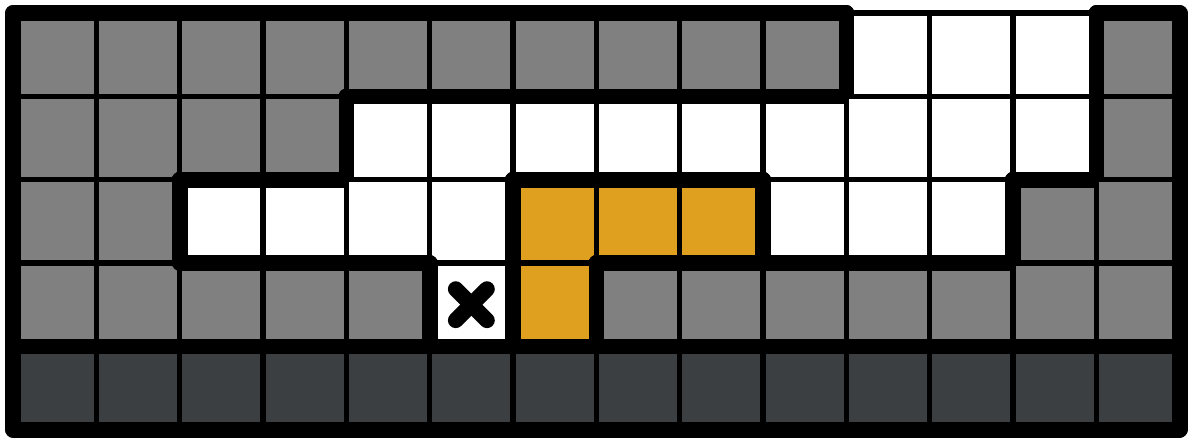}}
    % \subcaptionbox{}{\includegraphics[scale=0.3]{tetris/figures/rowalley_badconfigs/L_unprep_h1_L13.pdf}}
    % \subcaptionbox{}{\includegraphics[scale=0.3]{tetris/figures/rowalley_badconfigs/L_unprep_h1_L14.pdf}}
    % \subcaptionbox{}{\includegraphics[scale=0.3]{tetris/figures/rowalley_badconfigs/L_unprep_h1_L15.pdf}}
    % \subcaptionbox{}{\includegraphics[scale=0.3]{tetris/figures/rowalley_badconfigs/L_unprep_h1_L16.pdf}}
    % \subcaptionbox{}{\includegraphics[scale=0.3]{tetris/figures/rowalley_badconfigs/L_unprep_h1_L17.pdf}}
    % \subcaptionbox{}{\includegraphics[scale=0.3]{tetris/figures/rowalley_badconfigs/L_unprep_h1_L18.pdf}}
    % \subcaptionbox{}{\includegraphics[scale=0.3]{tetris/figures/rowalley_badconfigs/L_unprep_h1_L19.pdf}}
    % \subcaptionbox{}{\includegraphics[scale=0.3]{tetris/figures/rowalley_badconfigs/L_unprep_h1_L20.pdf}}
\caption{Ways to place an $\protect\LL$ into an unprepped bucket of positive height. Cases in which the leftmost empty cell is neither filled nor connected by a path of empty cells to the outside are not shown.}
\label{fig:row-L-unprep-2}
\end{figure}

\begin{proof}
% In Figure~\ref{fig:row-L-unprep}, we show all possible cases. 
In Figure~\ref{fig:row-L-unprep-2}, we show all cases that don't leave the leftmost cell disconnected from the outside. The first one works; in the others, we attempt to place an $\LL$ incorrectly, and mark some cell that can never thereafter be filled.
\end{proof}

% \begin{figure}
%     \centering
%     \subcaptionbox{}{\includegraphics[scale=0.3]{tetris/figures/rowalley_badconfigs/L_unprep_h0_L1.pdf}}
%     \subcaptionbox{}{\includegraphics[scale=0.3]{tetris/figures/rowalley_badconfigs/L_unprep/L_unprep_h0_L2.pdf}}
%     \subcaptionbox{}{\includegraphics[scale=0.3]{tetris/figures/rowalley_badconfigs/L_unprep/L_unprep_h0_L3.pdf}}
%     \subcaptionbox{}{\includegraphics[scale=0.3]{tetris/figures/rowalley_badconfigs/L_unprep_h0_L4.pdf}}

%     \subcaptionbox{}{\includegraphics[scale=0.3]{tetris/figures/rowalley_badconfigs/L_unprep_h0_L5.pdf}}
%     \subcaptionbox{}{\includegraphics[scale=0.3]{tetris/figures/rowalley_badconfigs/L_unprep/L_unprep_h0_L6.pdf}}
%     \subcaptionbox{}{\includegraphics[scale=0.3]{tetris/figures/rowalley_badconfigs/L_unprep/L_unprep_h0_L7.pdf}}
%     \subcaptionbox{}{\includegraphics[scale=0.3]{tetris/figures/rowalley_badconfigs/L_unprep_h0_L9.pdf}}
% \caption{Ways to place an $\LL$ into an unprepped bucket of height 0.}
% \label{fig:row-L-unprep}
% \end{figure}

\begin{claim} \label{row-struc2}
%None of $\OO, \JJ, \II$ may be placed in an unprepped bucket.
None of $\OO, \JJ, \II$ may be placed in an unprepped bucket of height at least $1$.
\end{claim}

\begin{figure}
    \centering
    \subcaptionbox{}{\includegraphics[scale=0.3]{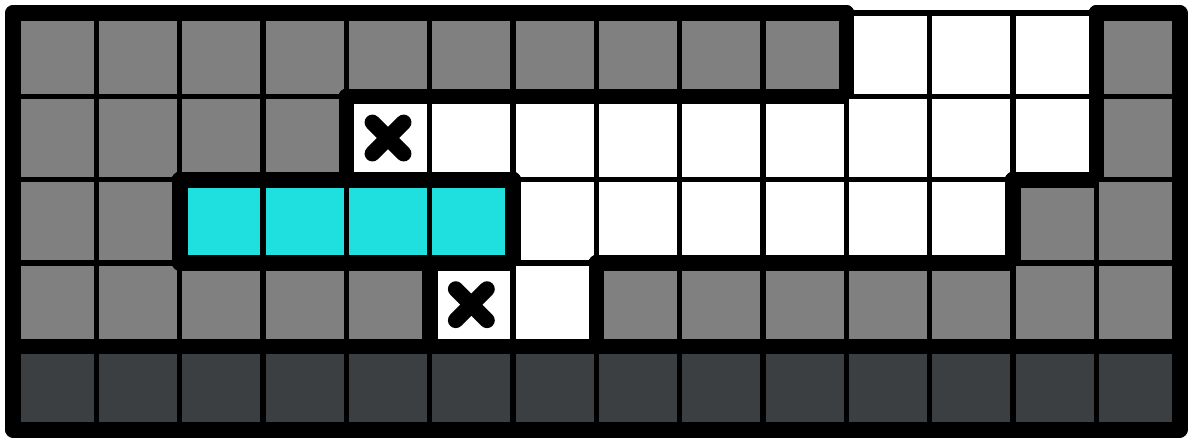}}
    \subcaptionbox{}{\includegraphics[scale=0.3]{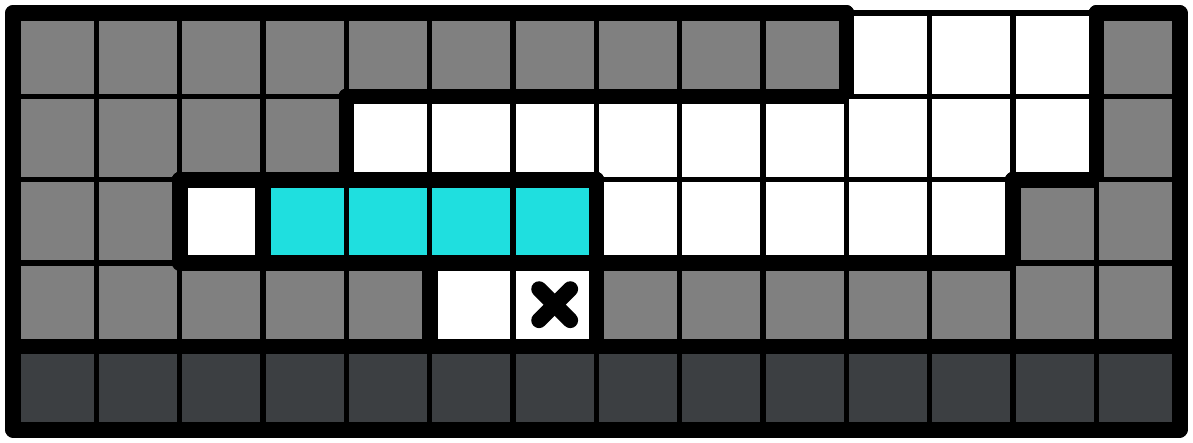}}
    \subcaptionbox{}{\includegraphics[scale=0.3]{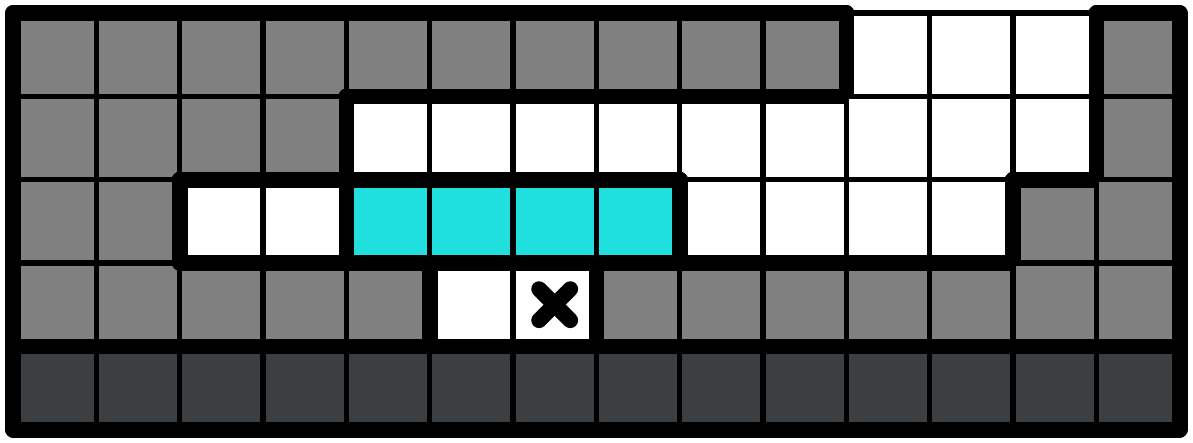}}
    \subcaptionbox{}{\includegraphics[scale=0.3]{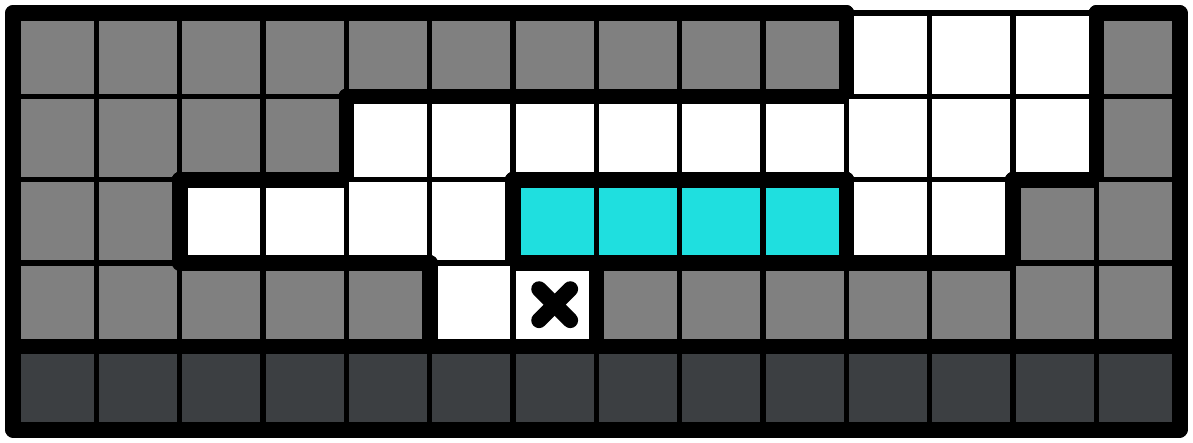}}
    \subcaptionbox{}{\includegraphics[scale=0.3]{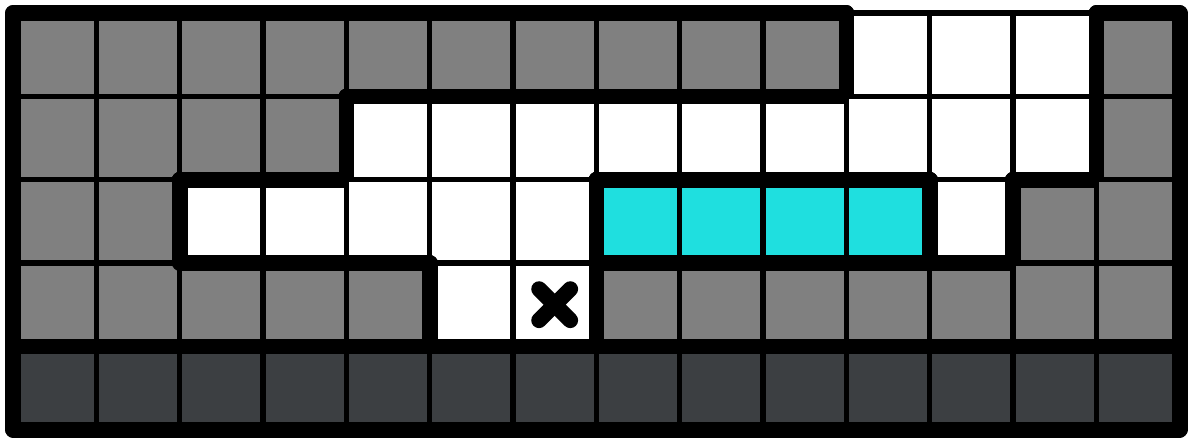}}
    \subcaptionbox{}{\includegraphics[scale=0.3]{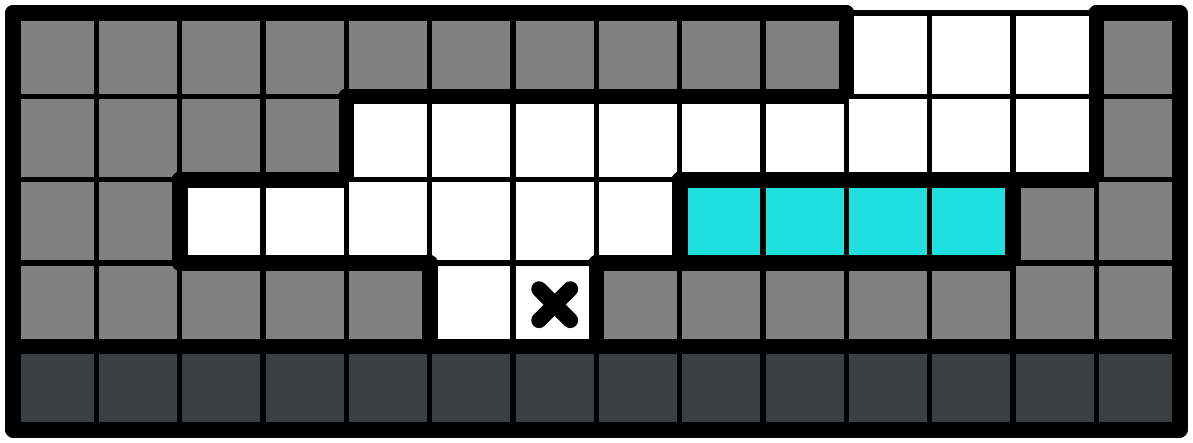}}
    \subcaptionbox{}{\includegraphics[scale=0.3]{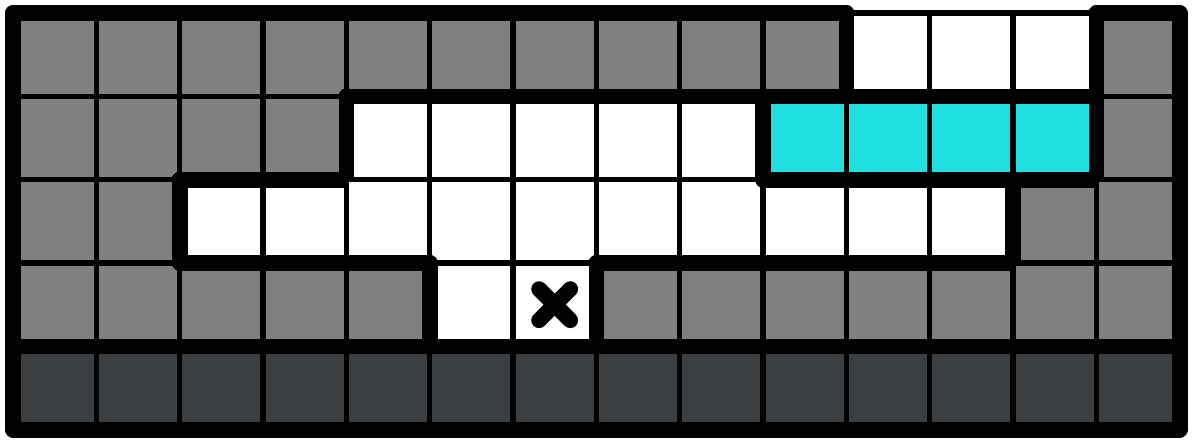}}
    \subcaptionbox{}{\includegraphics[scale=0.3]{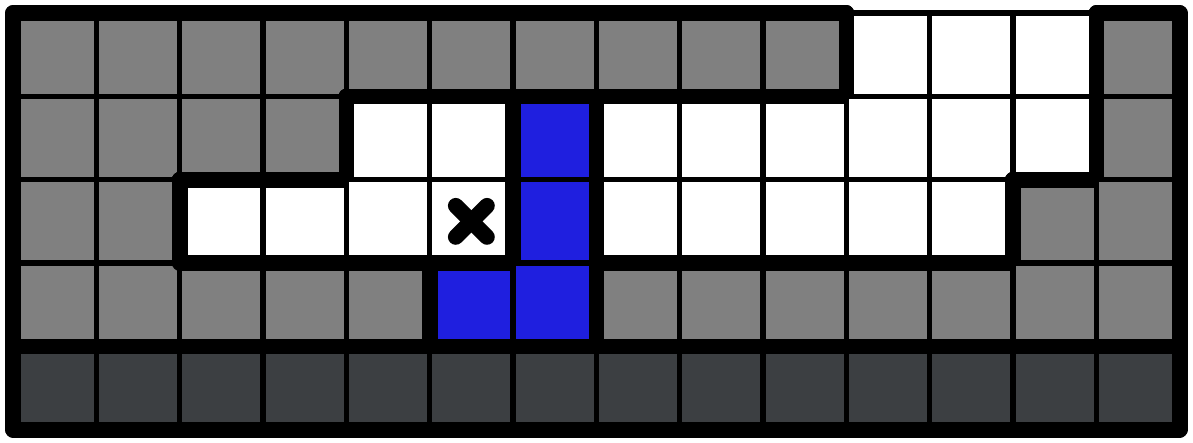}}
    \subcaptionbox{}{\includegraphics[scale=0.3]{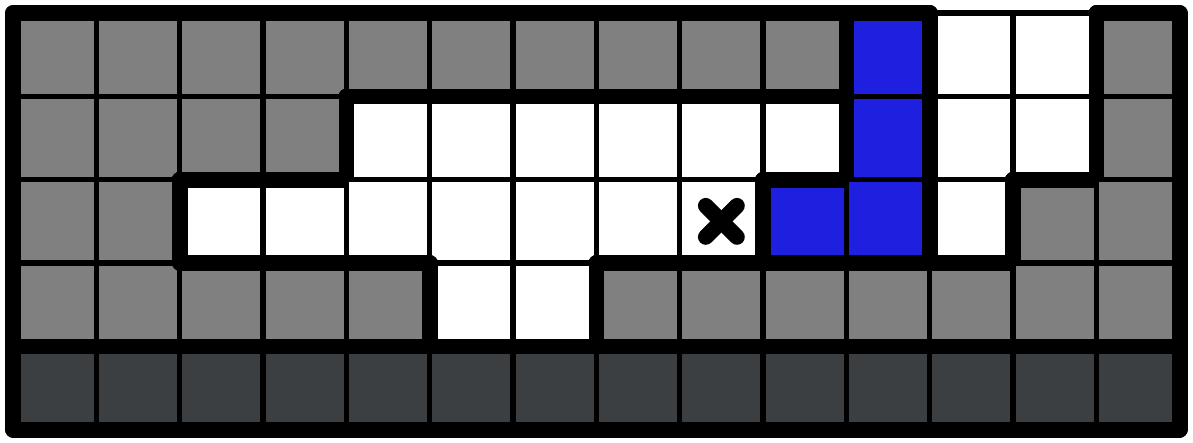}}
    \subcaptionbox{}{\includegraphics[scale=0.3]{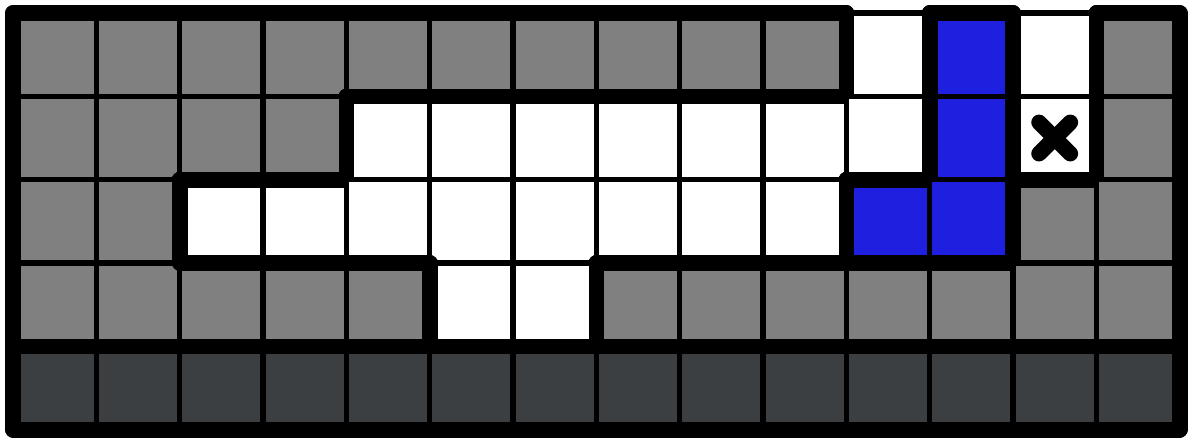}}
    \subcaptionbox{}{\includegraphics[scale=0.3]{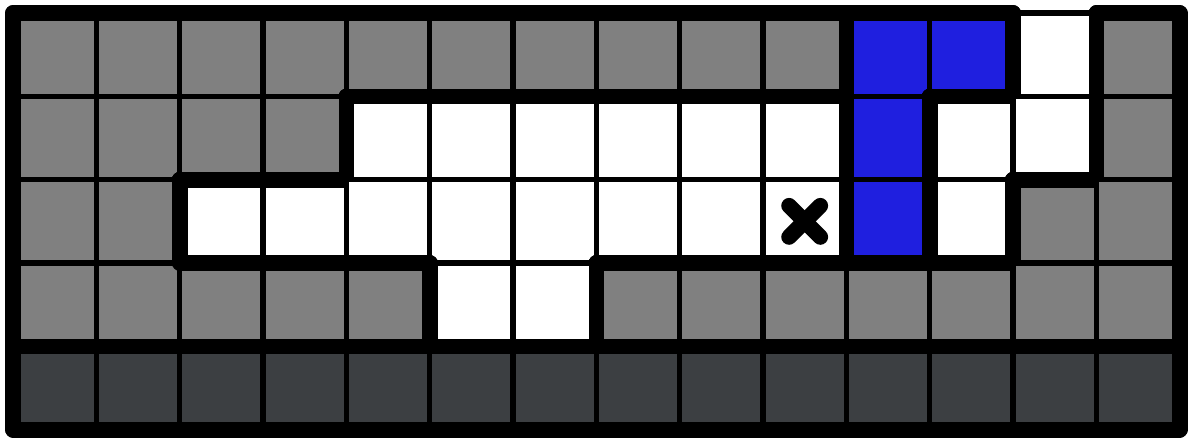}}
    \subcaptionbox{}{\includegraphics[scale=0.3]{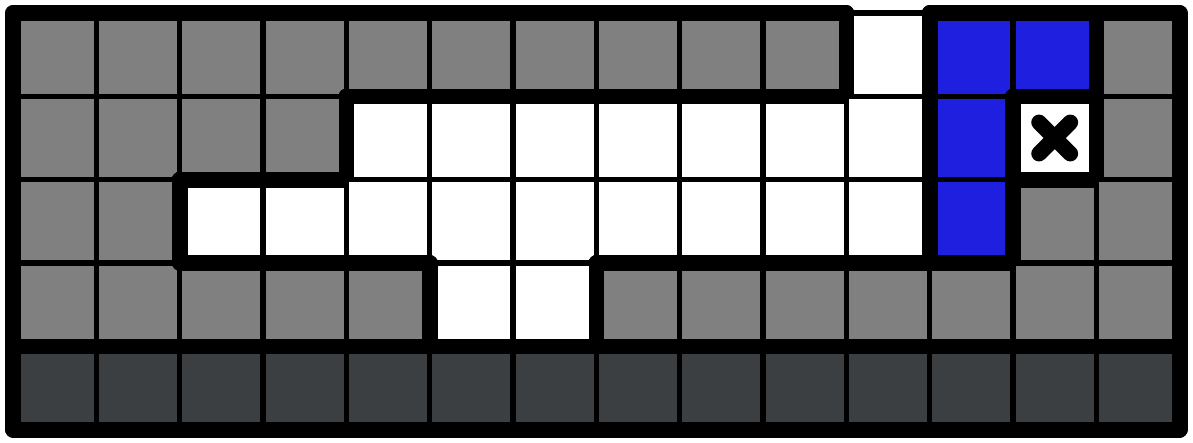}}
    \subcaptionbox{}{\includegraphics[scale=0.3]{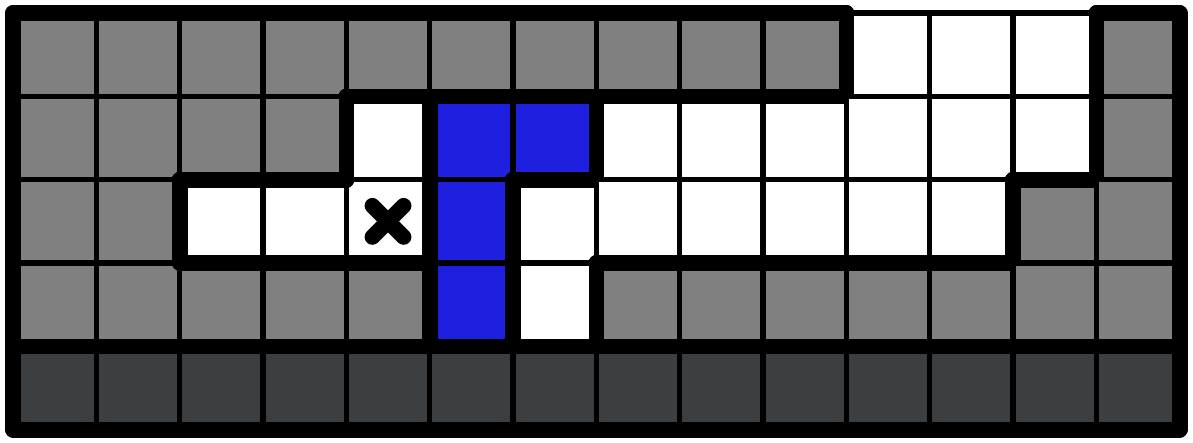}}
    \subcaptionbox{}{\includegraphics[scale=0.3]{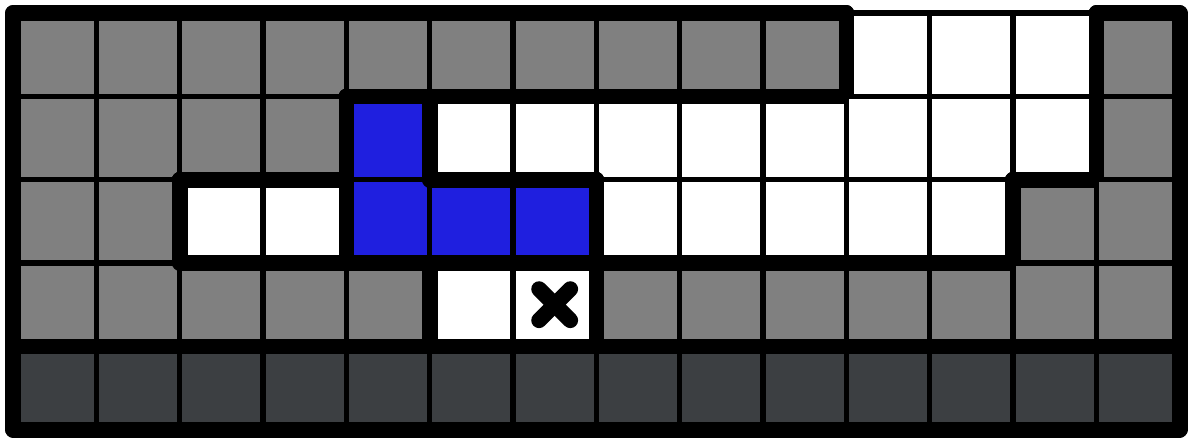}}
    \subcaptionbox{}{\includegraphics[scale=0.3]{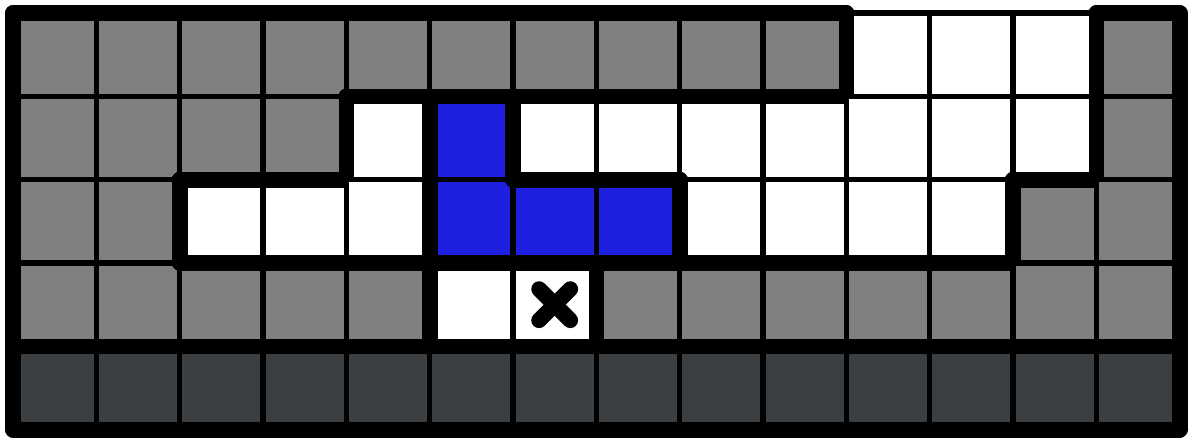}}
    \subcaptionbox{}{\includegraphics[scale=0.3]{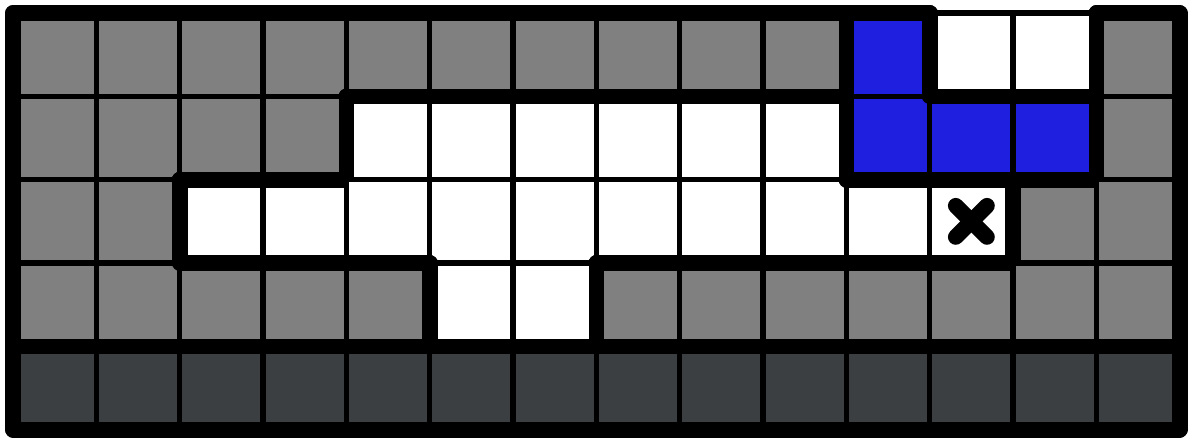}}
    \subcaptionbox{}{\includegraphics[scale=0.3]{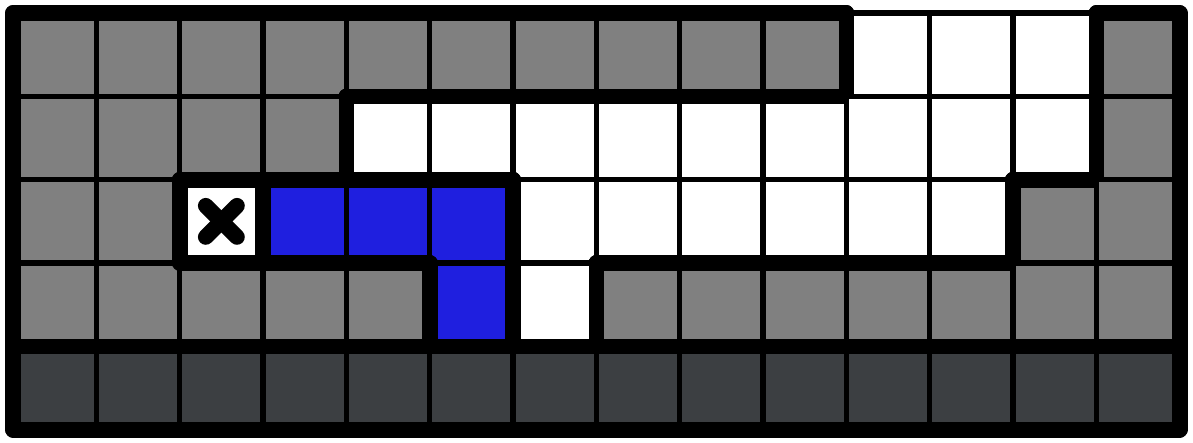}}
    \subcaptionbox{}{\includegraphics[scale=0.3]{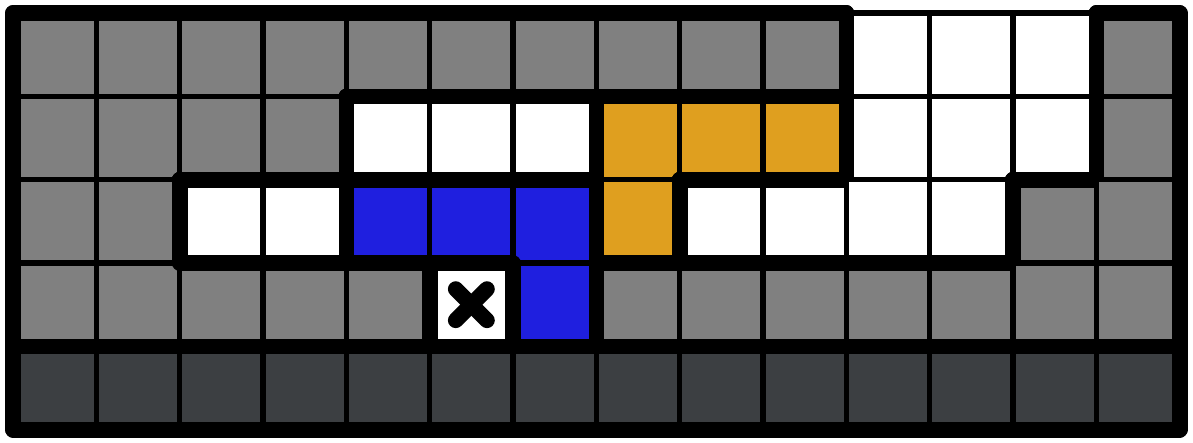}}
    \subcaptionbox{}{\includegraphics[scale=0.3]{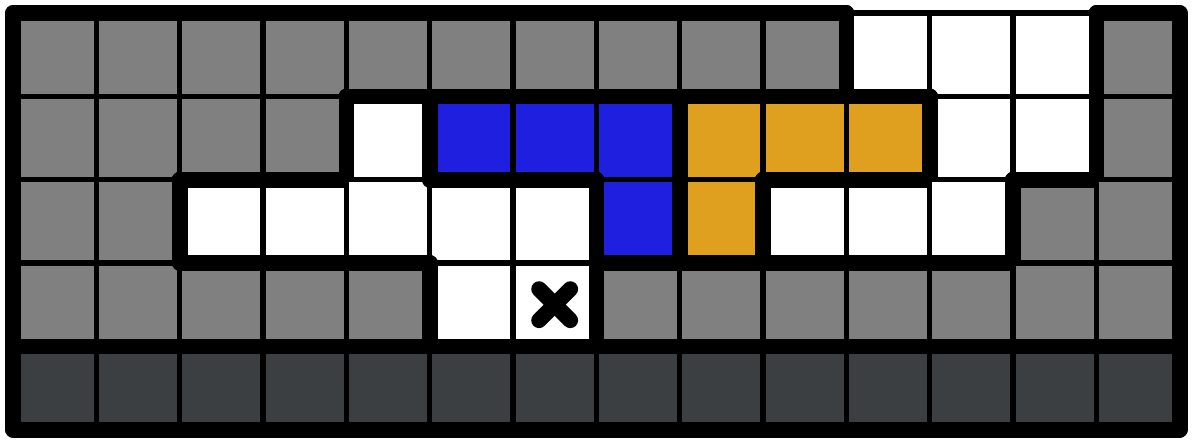}}
    \subcaptionbox{}{\includegraphics[scale=0.3]{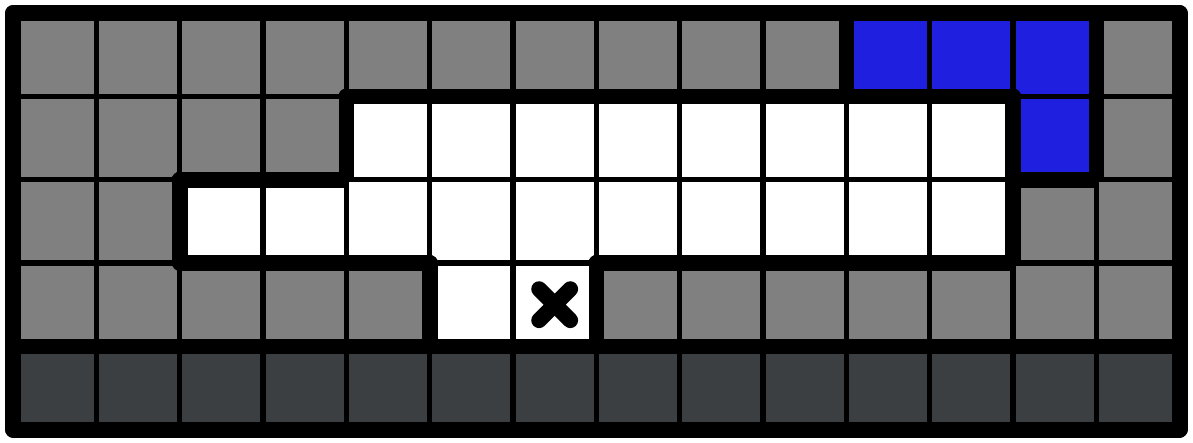}}
    \subcaptionbox{}{\includegraphics[scale=0.3]{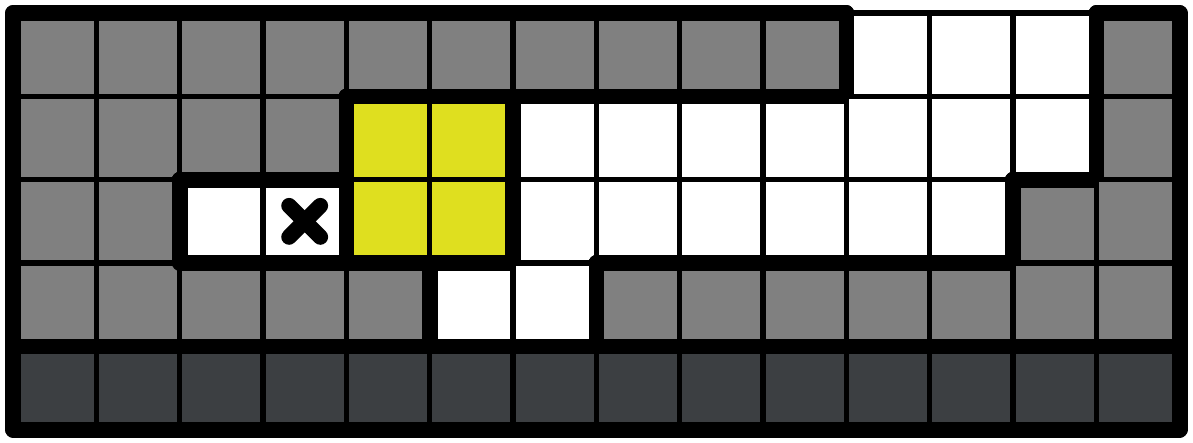}}
    \subcaptionbox{}{\includegraphics[scale=0.3]{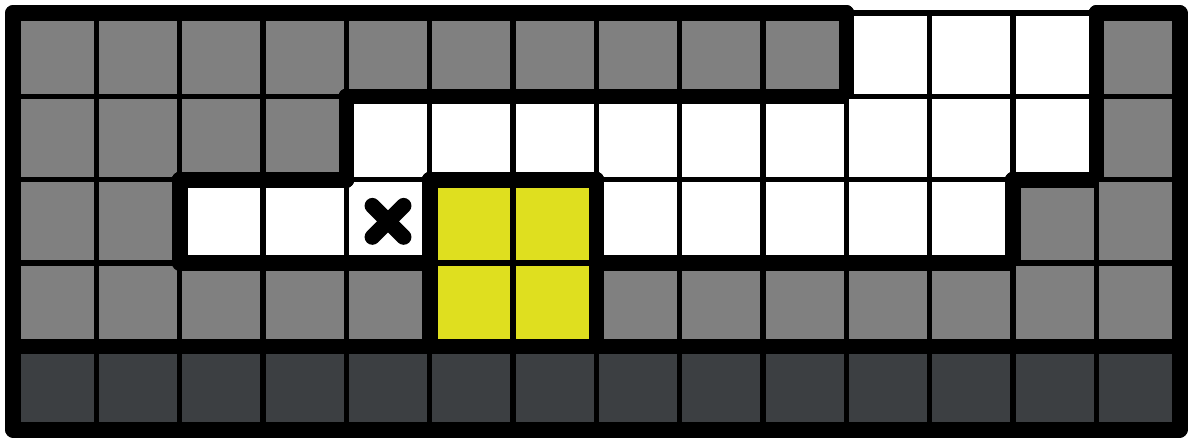}}

\caption{Ways to place an $\protect\II$, $\protect\JJ$, or $\protect\OO$ into an unprepped bucket.}
\label{fig:row-OJI-unprep}
\end{figure}

\begin{proof}
In Figure~\ref{fig:row-OJI-unprep}, we show all possible cases. In every one, we attempt to place an $\LL$ incorrectly, and mark some set of cells (usually exactly one) that can never thereafter be simultaneously filled. For instance, in Figure~\ref{fig:row-OJI-unprep}(p), the only piece (of the ones we ever use) that can fill the bottom marked cell is $\JJ$, but once a $\JJ$ is placed, no piece can fill the top marked cell. 
\end{proof}

% \begin{claim} \label{row-struc3}
% The sequence $\langle \OO, \JJ, \OO \rangle$ cannot be placed in a prepped bucket of height 0.
% \end{claim}
% \xxx{I think this claim is unnecessary in the current version of the proof --Mihir}

\begin{proof}
There are only eight empty cells in an unprepped bucket of height 0, less than the 12 needed.
\end{proof}

\begin{claim} \label{row-struc4}
When the sequence $\langle \OO, \JJ, \OO \rangle$ is placed in a prepped bucket of height $h \ge 2$, the bucket must end up as a prepped bucket of height $h-1$. %(We know $h \ge 1$ by Claim \ref{row-struc3}, and we know that the following piece in our sequence is an $\OO$.)
\end{claim}

\begin{figure}
    \centering
    \subcaptionbox{}{\includegraphics[scale=0.3]{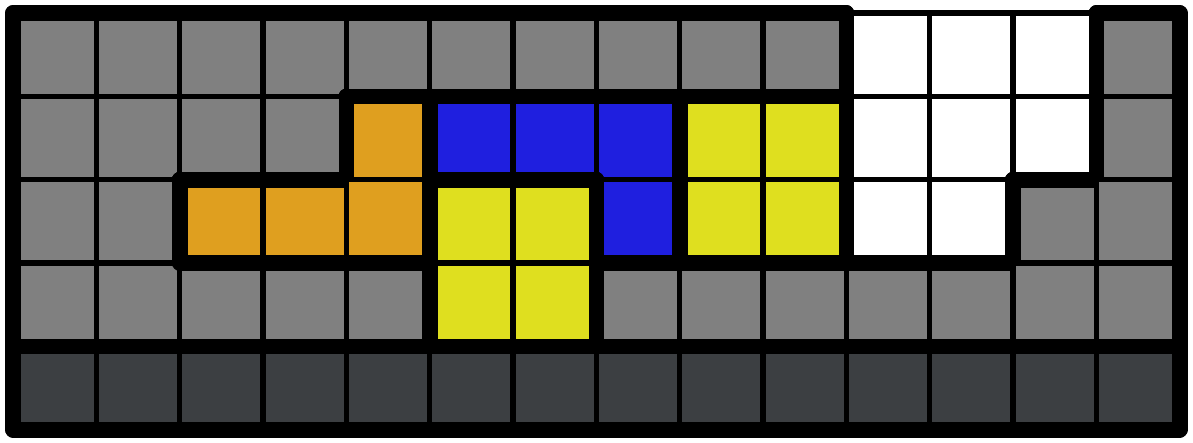}}
    \subcaptionbox{}{\includegraphics[scale=0.3]{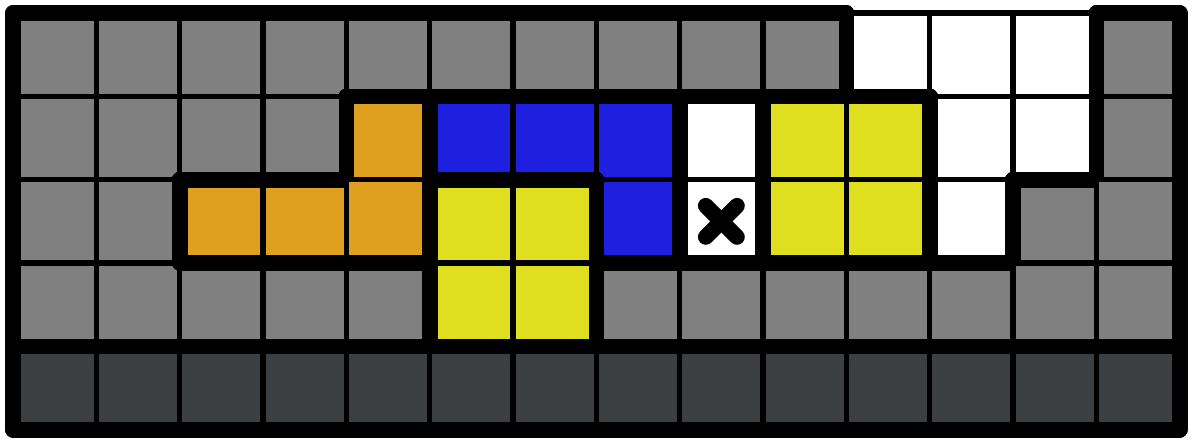}}
    \subcaptionbox{}{\includegraphics[scale=0.3]{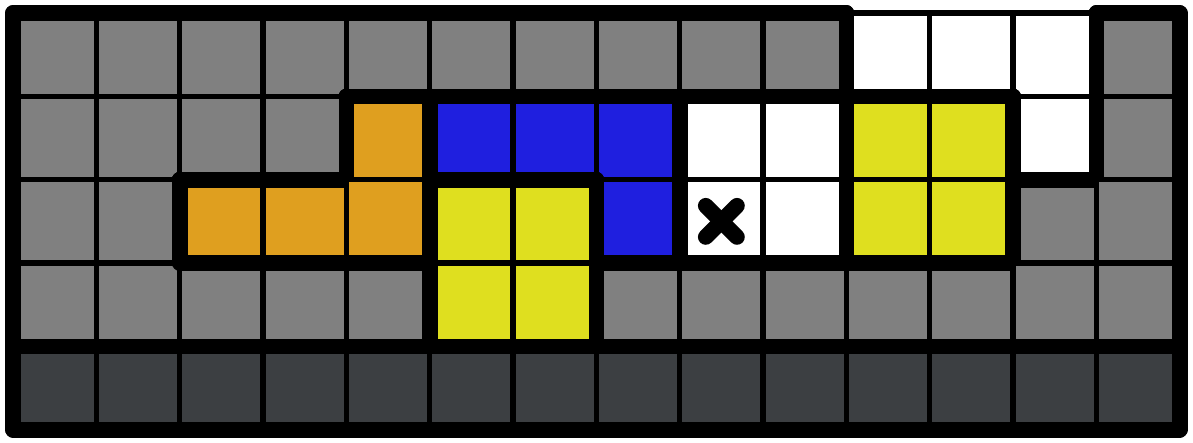}}
    \subcaptionbox{}{\includegraphics[scale=0.3]{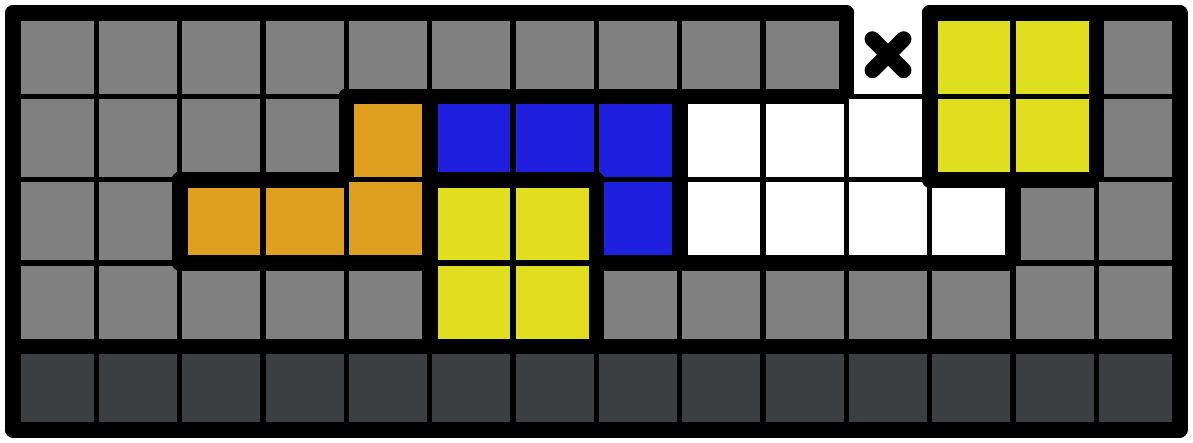}}
    \subcaptionbox{}{\includegraphics[scale=0.3]{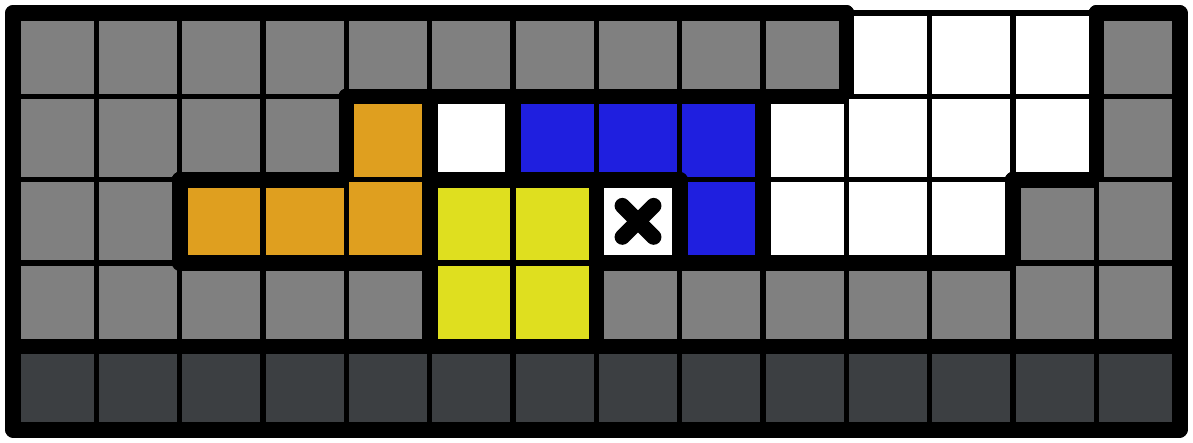}}
    \subcaptionbox{}{\includegraphics[scale=0.3]{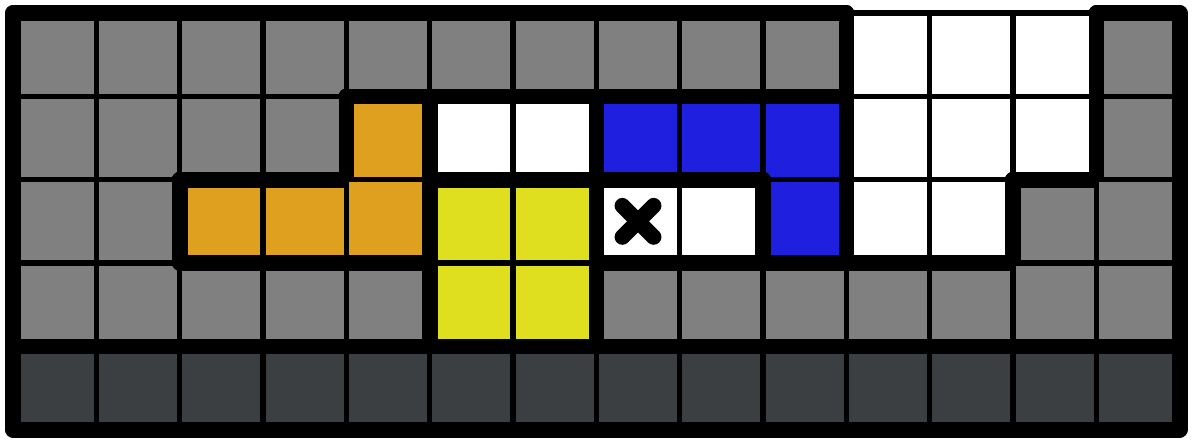}}
    \subcaptionbox{}{\includegraphics[scale=0.3]{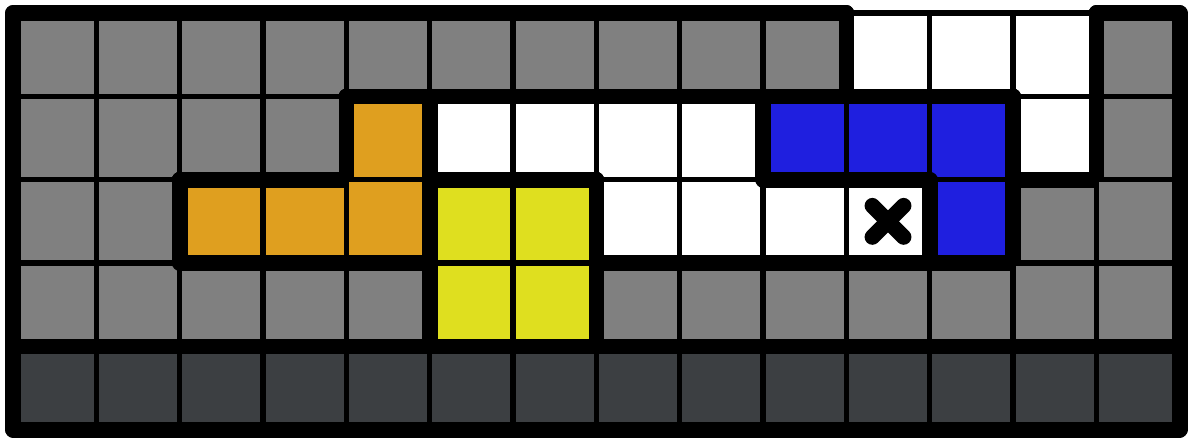}}
    \subcaptionbox{}{\includegraphics[scale=0.3]{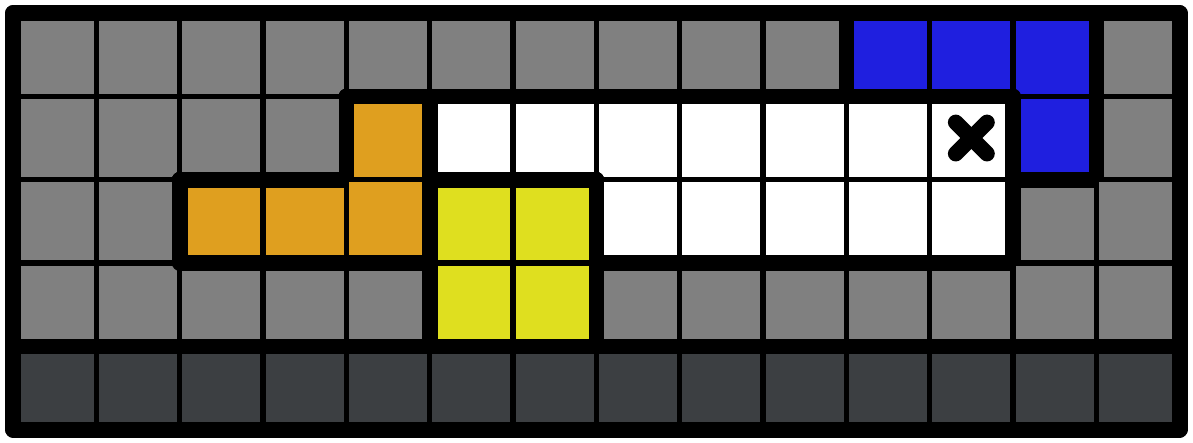}}
    \subcaptionbox{}{\includegraphics[scale=0.3]{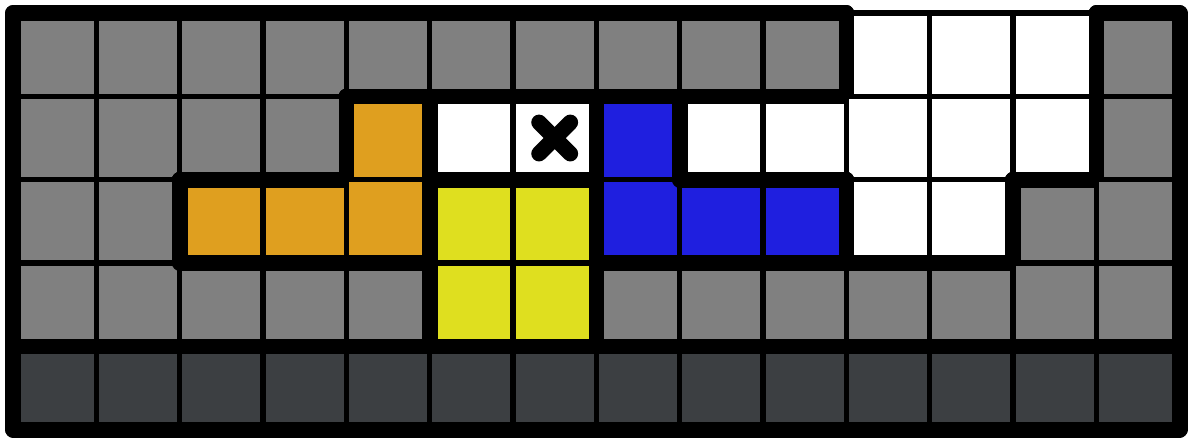}}
    \subcaptionbox{}{\includegraphics[scale=0.3]{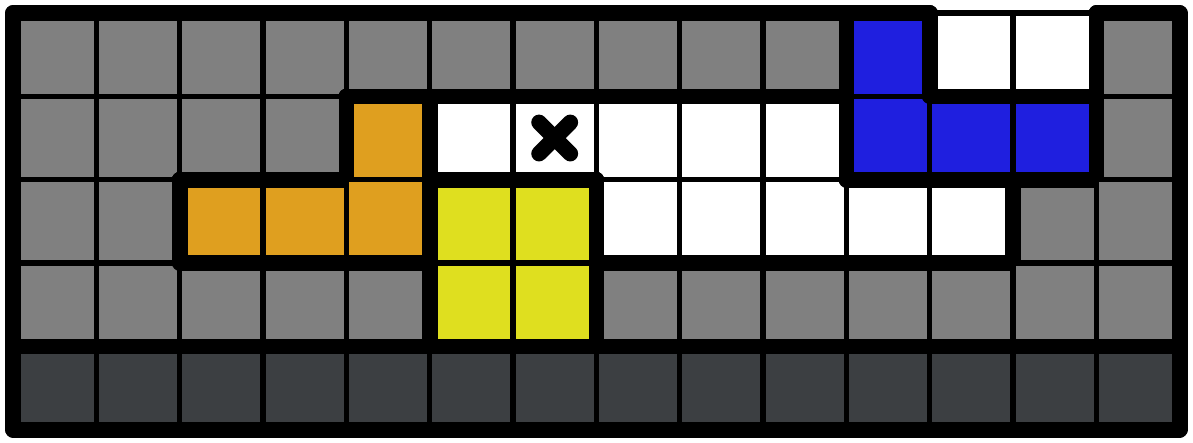}}
    \subcaptionbox{}{\includegraphics[scale=0.3]{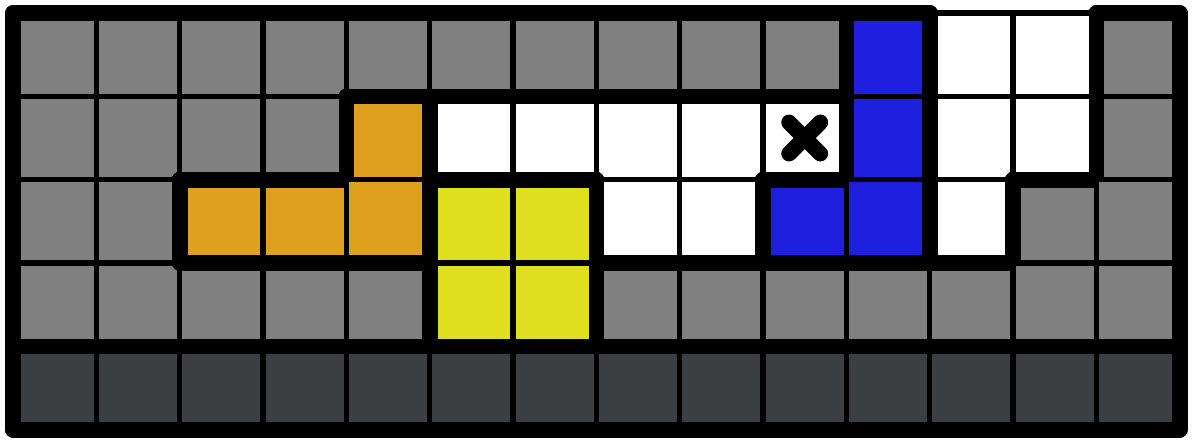}}
    \subcaptionbox{}{\includegraphics[scale=0.3]{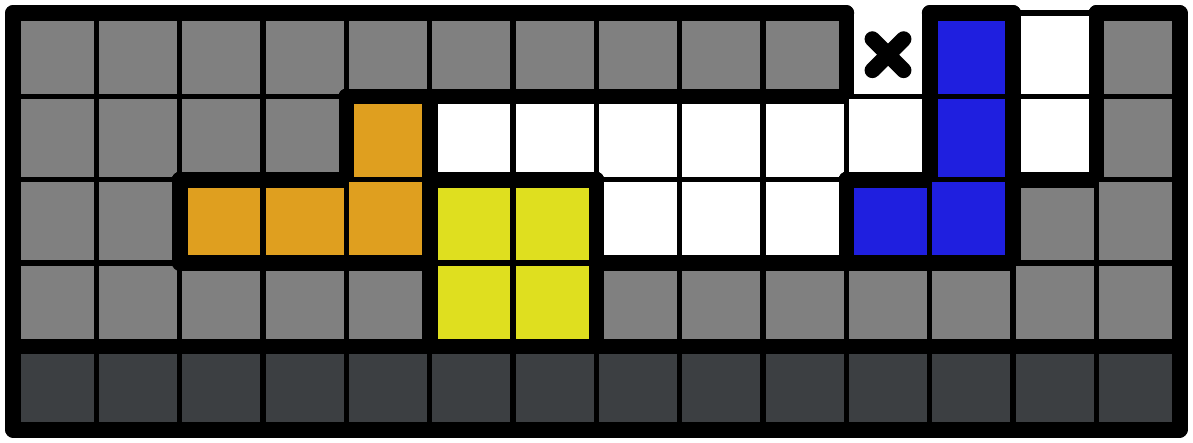}}
    \subcaptionbox{}{\includegraphics[scale=0.3]{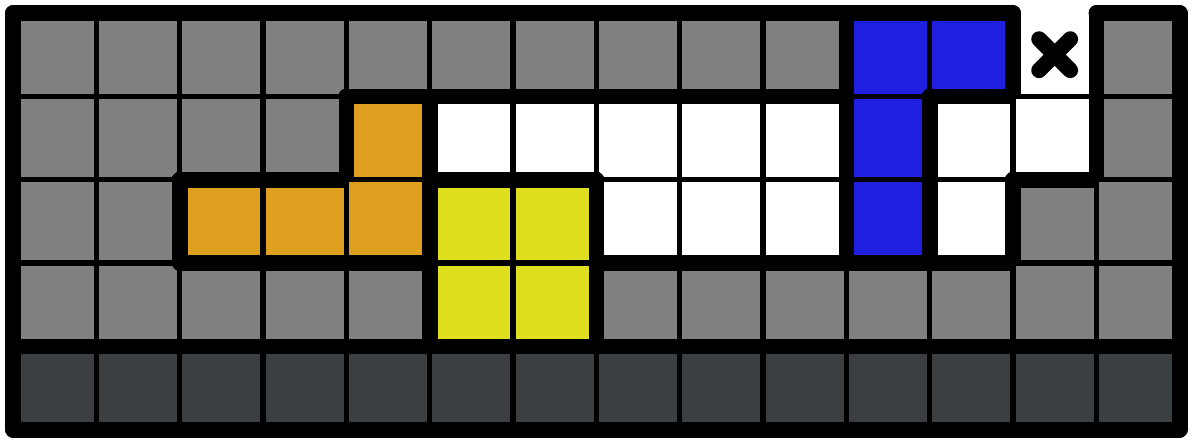}}
    \subcaptionbox{}{\includegraphics[scale=0.3]{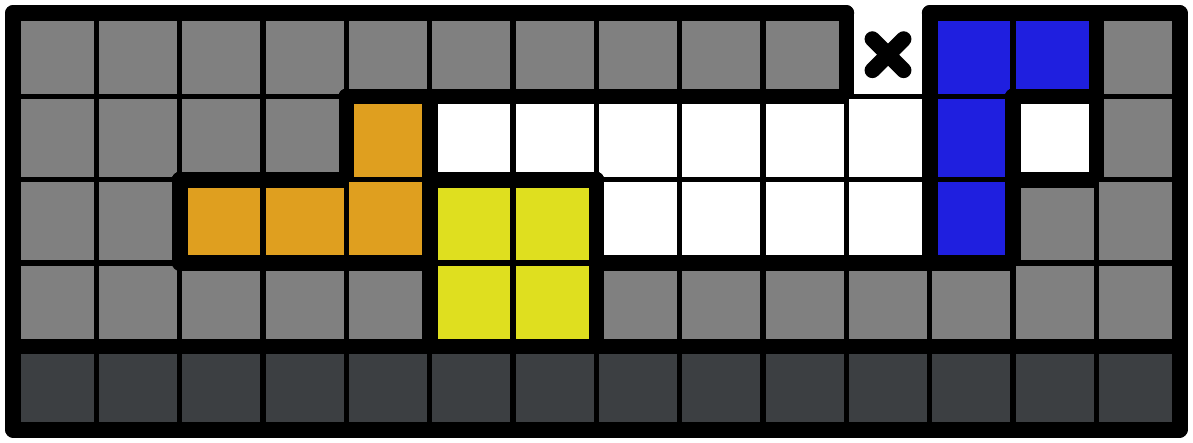}}
    \subcaptionbox{}{\includegraphics[scale=0.3]{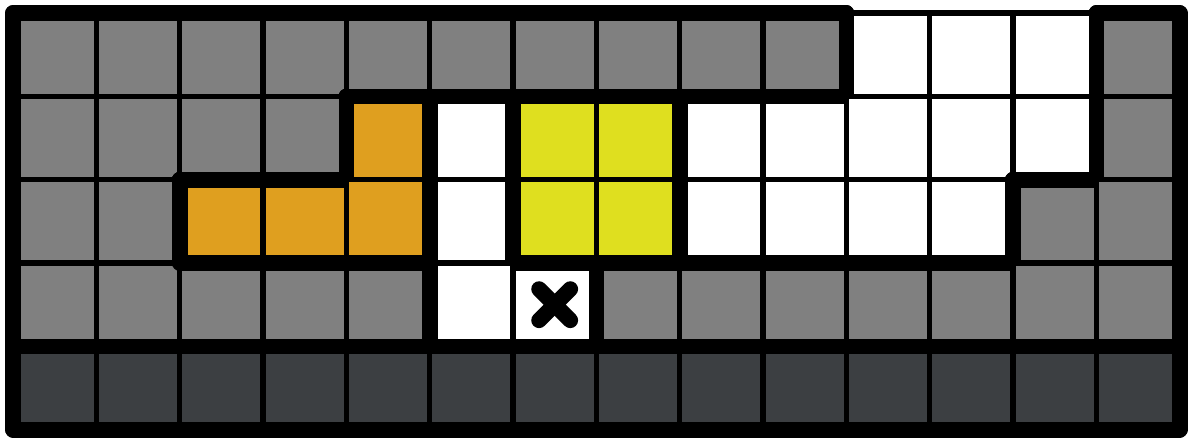}}
    \subcaptionbox{}{\includegraphics[scale=0.3]{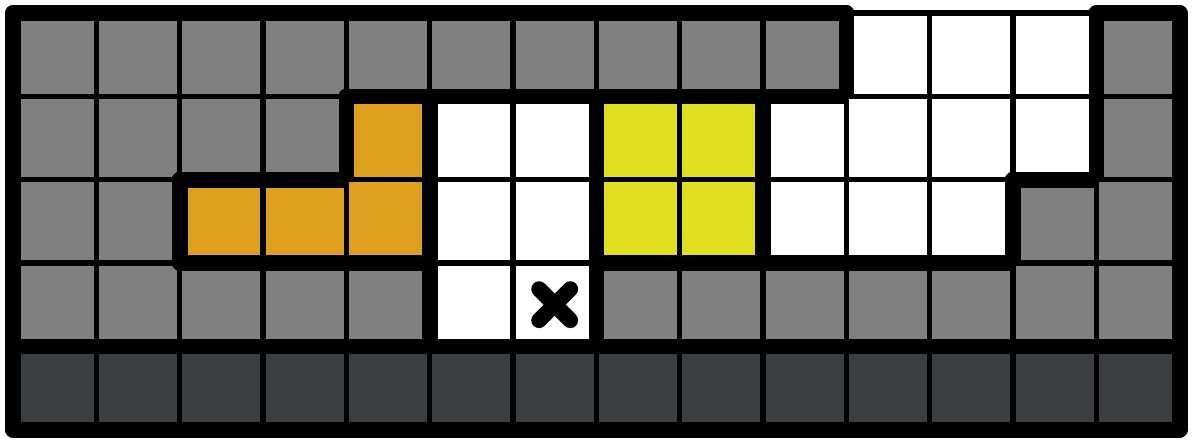}}
    \subcaptionbox{}{\includegraphics[scale=0.3]{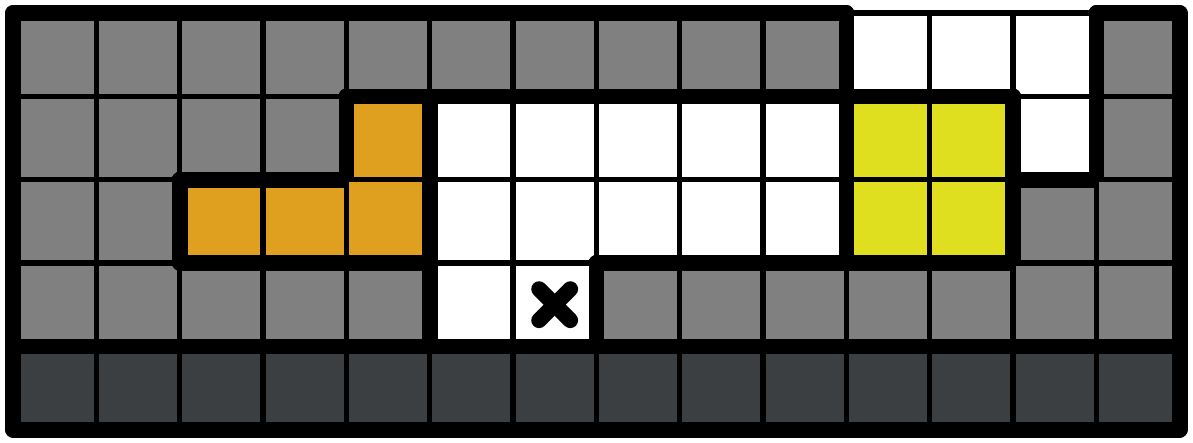}}
    \subcaptionbox{}{\includegraphics[scale=0.3]{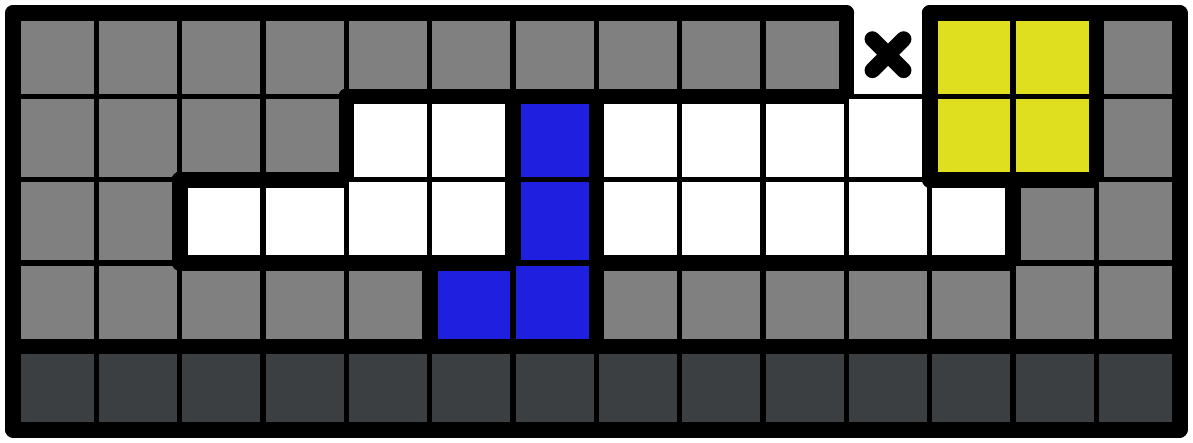}}
\caption{Ways to place an $\protect\OO$, then a $\protect\JJ$, then an $\protect\OO$ into a prepped bucket, where the following piece is an $\protect\OO$.}
\label{fig:row-OJO-prep}
\end{figure}

\begin{proof}
In Figure~\ref{fig:row-OJO-prep}, we show all possible cases. In every one, we attempt to place a $\OO, \JJ, \OO$. The first one works. For every other one, we show some cell that cannot be filled by the next piece. %noting that the piece after $\OO, \JJ, \OO$ is another $\OO$, and only $\II$ can fit through bottlenecks of size 1.
\end{proof}

\xxx{replace OJO-prep with height-2 version}

\begin{claim} \label{row-struc5}
When the sequence $\langle \OO, \II \rangle$ is placed in a prepped bucket of height $h \ge 2$, the bucket must end as an unprepped bucket of height $h - 1$. %(We know $h \ge 1$ by Claim \ref{row-struc3}, and we know that the following piece in our sequence is an $\OO$.)
\end{claim}

\begin{figure}
    \centering
    \subcaptionbox{}{\includegraphics[scale=0.3]{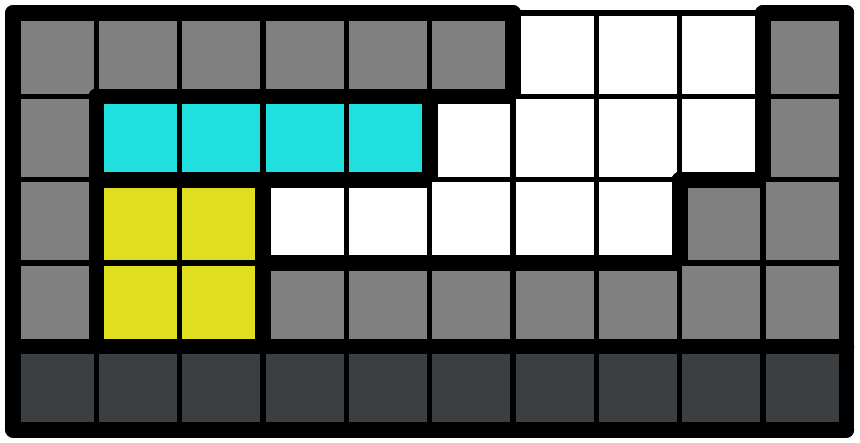}}
    \subcaptionbox{}{\includegraphics[scale=0.3]{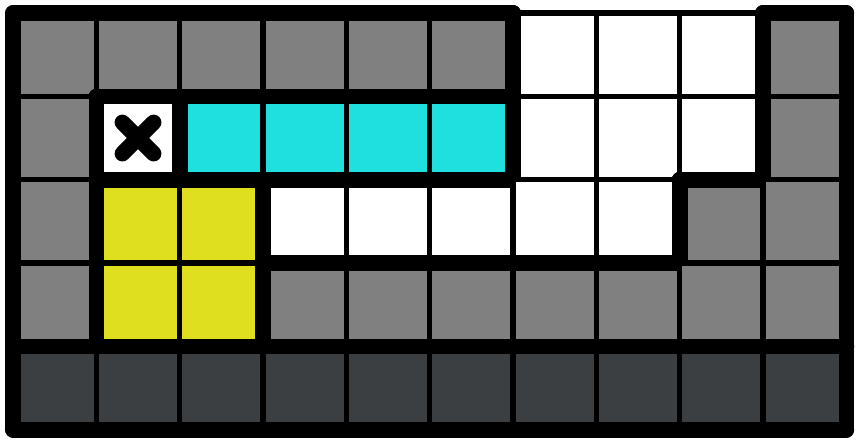}}
    \subcaptionbox{}{\includegraphics[scale=0.3]{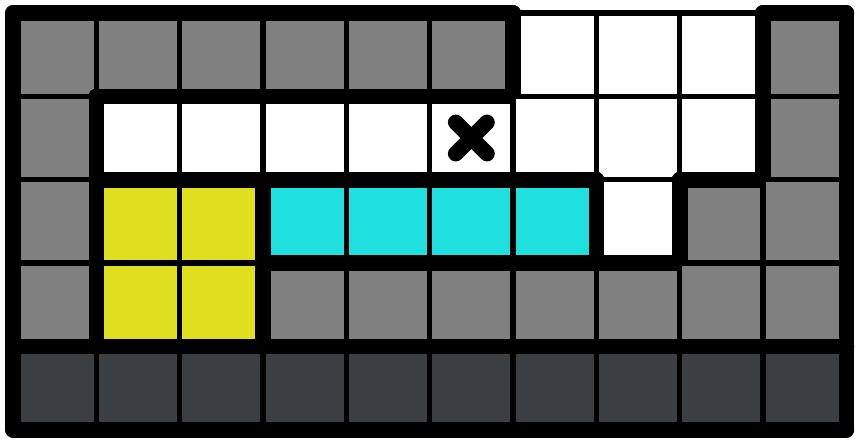}}
    \subcaptionbox{}{\includegraphics[scale=0.3]{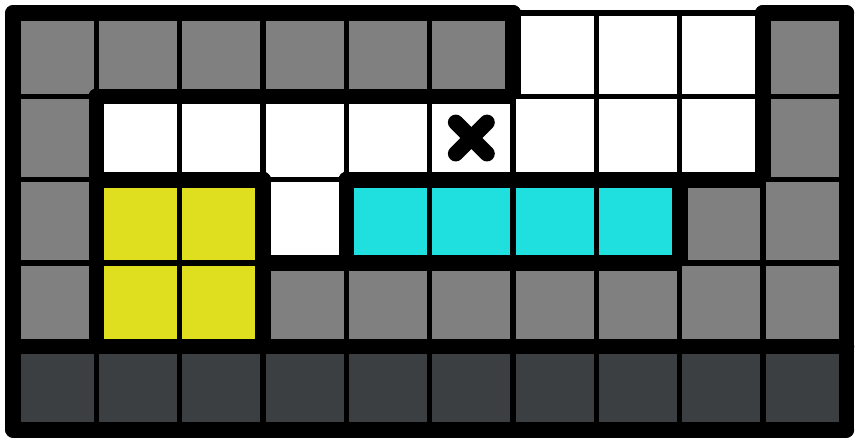}}
    \subcaptionbox{}{\includegraphics[scale=0.3]{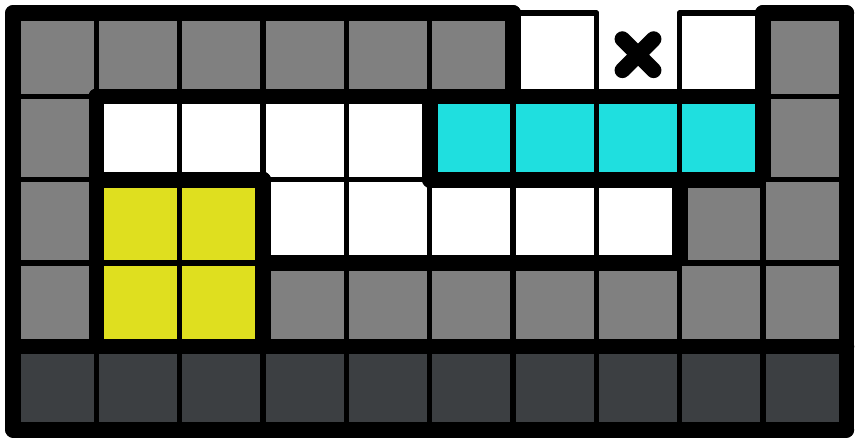}}
    \subcaptionbox{}{\includegraphics[scale=0.3]{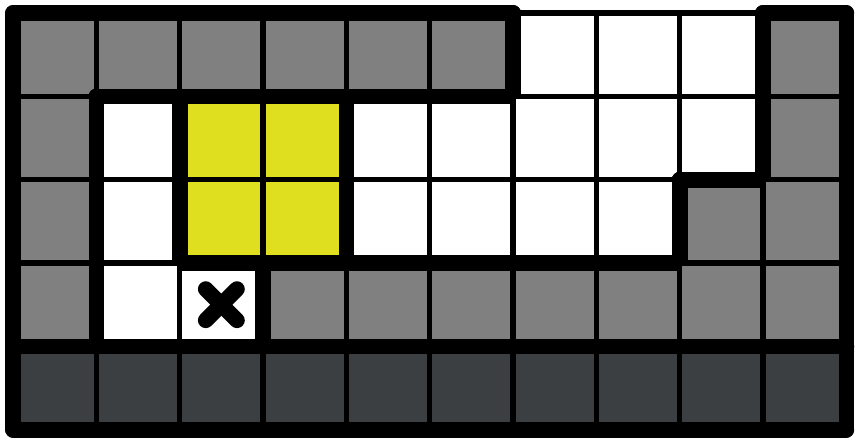}}
    \subcaptionbox{}{\includegraphics[scale=0.3]{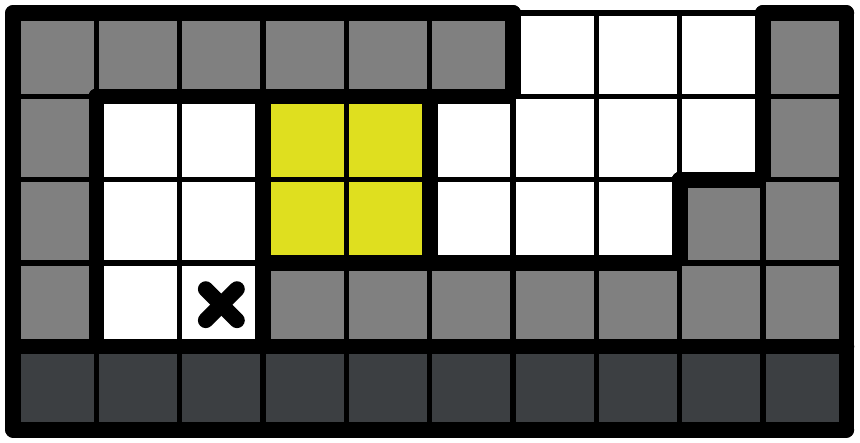}}
    \subcaptionbox{}{\includegraphics[scale=0.3]{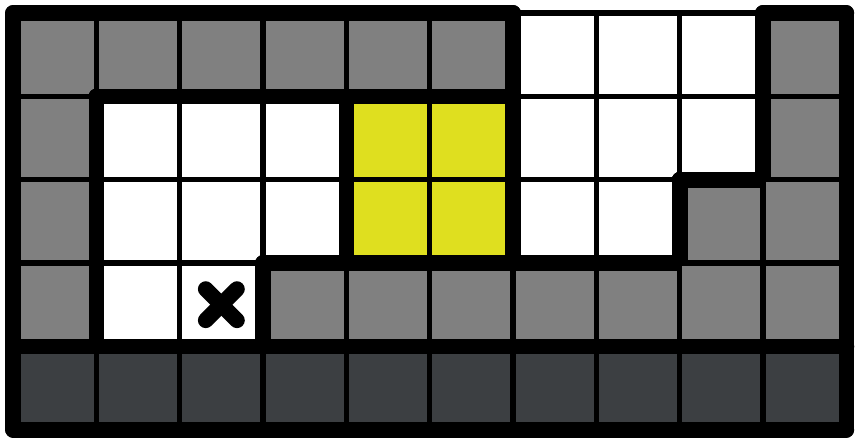}}
    \subcaptionbox{}{\includegraphics[scale=0.3]{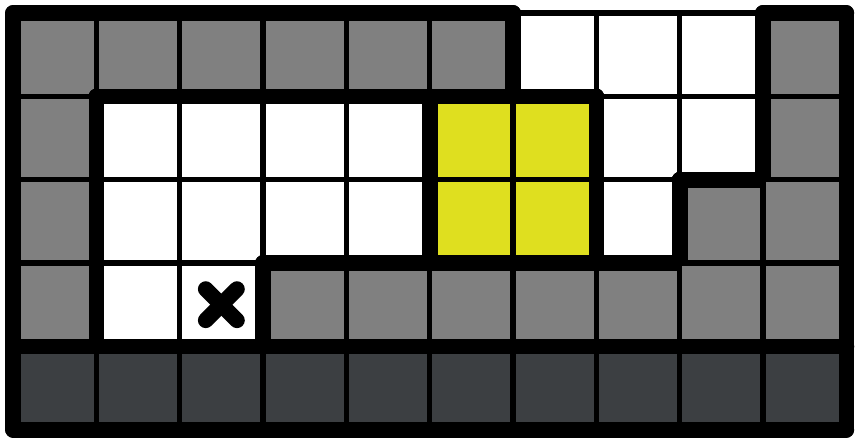}}
    \subcaptionbox{}{\includegraphics[scale=0.3]{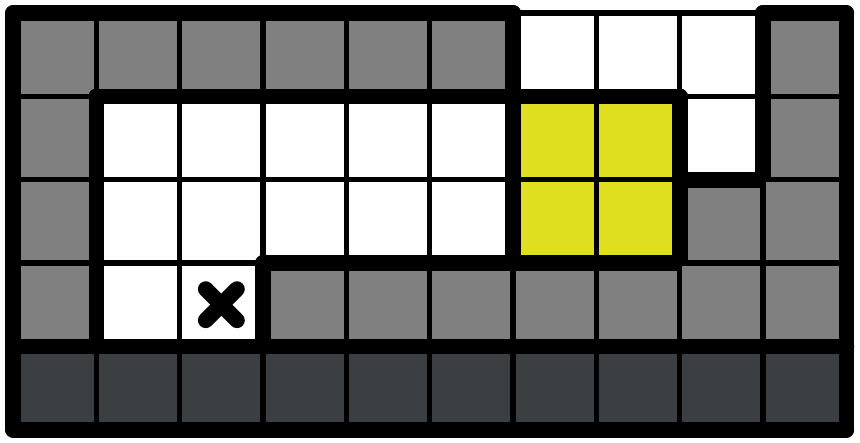}}
    \subcaptionbox{}{\includegraphics[scale=0.3]{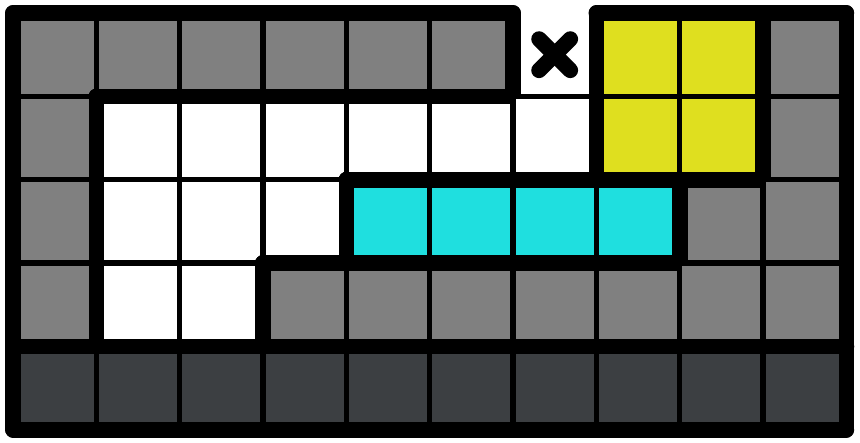}}
    \subcaptionbox{}{\includegraphics[scale=0.3]{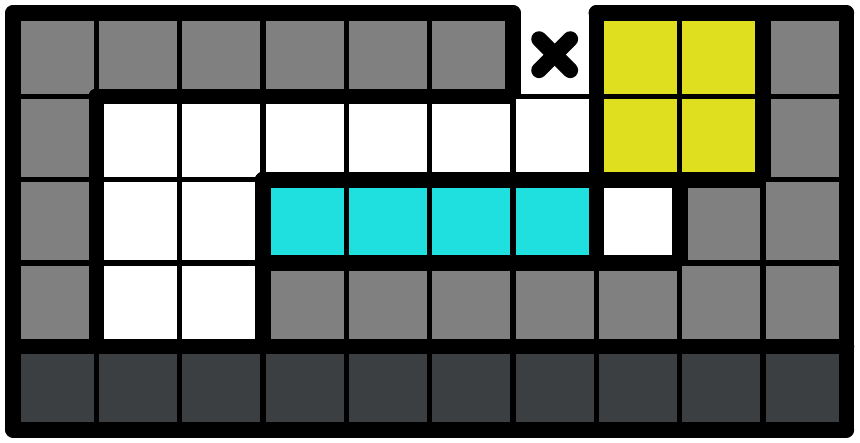}}
    \subcaptionbox{}{\includegraphics[scale=0.3]{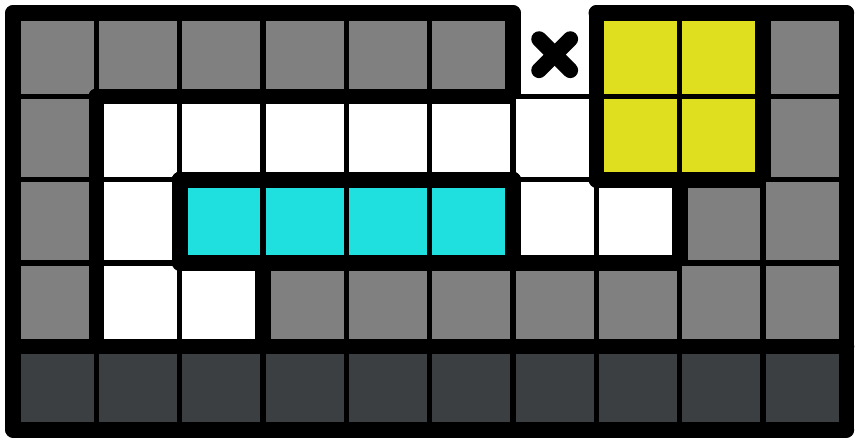}}
    \subcaptionbox{}{\includegraphics[scale=0.3]{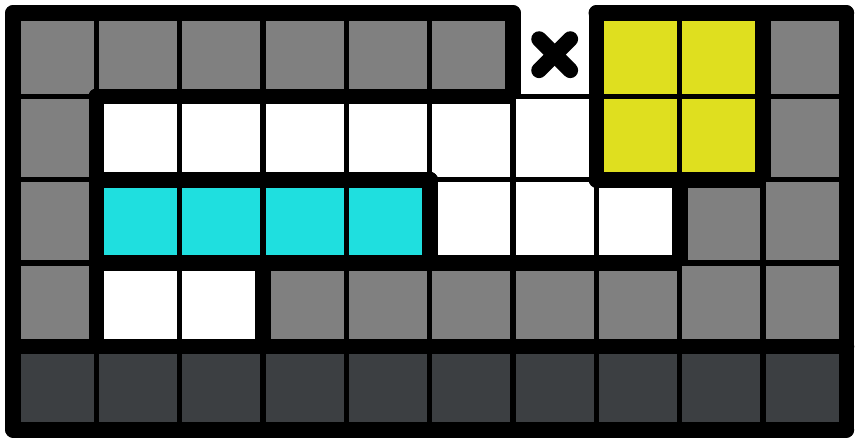}}
\caption{Ways to place an $\protect\OO$ and then an $\protect\II$ into a prepped bucket, where the following piece is an~$\protect\OO$.}
\label{fig:row-OI-prep}
\end{figure}

\begin{proof}
In Figure~\ref{fig:row-OI-prep}, we show all possible cases. In every one, we attempt to place a $\OO, \II$. The first one works. For every other one, we show some cell that cannot be filled by the next piece. %noting that the piece after $\OO, \II$ is another $\OO$.
\end{proof}

\xxx{replace OI-prep with height-2 version}

The following corollary follows from Claims \ref{row-struc1}, \ref{row-struc2}, \ref{row-struc4}, and \ref{row-struc5}:
% \xxx{didn't use Claim \ref{row-struc3} in the proof}
\begin{cor} \label{row-strucmain}
Suppose that before the $a_i$-sequence arrives, all buckets are unprepped and have height $1 \pmod 3$. Then, the entire $a_i$-sequence must be placed in one bucket, and the height of that bucket decreases by $3a_i$ (the height cannot go below 0) and is unprepped at the end.
\end{cor}
\begin{proof}
The initial $\LL$ of the $a_i$-sequence must go in some bucket; say it has height $3h + 1$. By Claim \ref{row-struc1}, the bucket is now prepped with height $3h + 1$. By Claim \ref{row-struc2}, all pieces of the $a_i$-sequence must go into this bucket. Now, we have that the total area of pieces remaining in the $a_i$-sequence is $36a_i - 4$, while the total area remaining in the bucket is $36h + 20$. Thus, since the total area of the pieces cannot exceed the area of the bucket, we must have $36a_i - 4 \le 36h + 20$, and therefore (since $a_i, h$ are integers), $a_i \le h$. Now, by Claim \ref{row-struc4}, each $\langle \OO, \JJ, \OO \rangle$ sequence must decrease the height of the bucket by 1, so after all of these the bucket is now prepped and has height $3h - 3a_i + 2$ (note that at each step the bucket had height at least 2). Now, the height of the bucket is still at least 2, so we can apply Claim \ref{row-struc5}. Thus, after the final $\langle \OO, \II \rangle$, the bucket must become an unprepped bucket of height $3h - 3a_i + 1$, as desired.
\end{proof}

\begin{thm}
If $\mc G(\mc P)$ has a clearing trajectory sequence, then the {\partit} instance $\mc P$ has a solution.
\end{thm}
\begin{proof}
Numbering the buckets $1, 2, \dots, s$, let $S_b$ be the set of $i$ such that the $a_i$-sequence is placed in bucket $b$, so the $S_b$ form a partition of $\{a_1, a_2, \dots, a_{3s}\}$. By Corollary \ref{row-strucmain}, the sum $\sum_{i \in S_b} 3a_i$ is at most the original width of each bucket, which is $3T + 1$. But note that the sum is a multiple of 3, so it must in fact be at most $3T$. However, $\sum_{i=1}^{3s} 3a_i = 3sT$, so equality must hold for each individual sum.

Thus, we have $\sum_{i \in S_b} 3a_i = 3T$ for each $b$. Dividing out by 3, $\sum_{i \in S_b} a_i = T$, and thus the $S_b$ represent a valid 3-partition.

% Numbering the buckets $1, 2, \dots, s$, let $S_b$ be the set of $i$ such that the $a_i$-sequence is placed in bucket $b$, so the $S_b$ form a partition of $\{a_1, a_2, \dots a_{3s}\}$. By Corollary \ref{row-strucmain}, the sum $\sum_{i \in S_b} (a_i + 1)$ is at most the original width of each bucket, which is $T + 3$. However, $\sum_{i=1}^{3s} (a_i + 1) = s(T + 3)$, so equality must hold.

% Thus, we have $\sum_{i \in S_b} (a_i + 1) = T + 3$ for each $b$. But the condition $T/4 < a_i < T/2$ means that this sum can't have at most 2 or at least 4 terms, so it must have 3 terms, and thus $|S_b| = 3$. Then the condition simplifies to $\sum_{i \in S_b} a_i = T$, and thus the $S_b$ represent a valid 3-partition.
\end{proof}

%\clearpage

\subsection{Linear-size pieces, 1 row}

As a transition to the next topic, which considers \ktet{$\leq k$}, we point
out that even just one row is trivially NP-hard when we allow $k \gg 4$:

\begin{prop} \label{prop:1 row big}
  \rktet{1}{$\leq k$} is strongly NP-hard.
\end{prop}

\begin{proof}
  We reduce from {\partit}.
  The board is $s (T+1)$ units wide,
  with an initially filled square every $T+1$ spaces,
  leaving $s$ gaps of length exactly~$T$.
  The first $3 s$ pieces are $1 \times a_i$ for $i = 1, 2, \dots, 3 s$.
  The line can clear if and only if {\partit} has a solution.
  A final $1 \times s (T+1)$ piece forces a loss otherwise.
\end{proof}

%% file: tetris/empty.tex
\section{Starting from an empty board is NP-hard}
\label{sec:empty-start}

\subsection{$O(1)$-size pieces, 8 columns}

%--- No longer true:
%The following result assumes that all lines clear (and lines above descend)
%before the game checks whether the player lost by breaching the ceiling.

\begin{thm} \label{thm:65}
\cektet{$c$}{$(\leq c^2+1)$} is NP-complete for any $c \geq 8$.
In particular, \cektet{$8$}{$(\leq 65)$} is NP-complete.
\end{thm}

\begin{proof}
We force the player to build the board initial configuration $B$ from Theorem~\ref{thm:8 col}'s proof, starting from an empty board, using pieces of size at most $c^2+1$.
We build $B$ from the bottom up using pieces of height $c+1$ (so they cannot be rotated) and width $c$ (so they also cannot be translated).
Specifically, if we want to add a cell in column $i$ of the top row that already has a cell, we send a $c \times c$ square with an extra cell in column $i$ below it (in the bottom $(c+1)$st row); the $c \times c$ square lands above the top existing row and clears, leaving just a cell in column $i$ in the previous top row.
To start a new row with a cell in column $i$, we send a $c \times c$ square with an extra cell in column $i$ above it (in the top $(c+1)$st row); the $c \times c$ square lands above the top existing row and clears, leaving just a cell in column $i$ in its own row.

These two operations suffice to create any legal board configuration,
until we get near the top of the board in which case the partial lock out
would cause the player to lose.
When we fill the last pixel in the $(c+1)$st row of the board, we send a
different piece: instead of putting a $c \times c$ square above that pixel,
we put the desired board configuration for the top $c$ rows.
By modifying the construction of Theorem~\ref{thm:8 col} to have a lot of
rows after the first two where just the first, seventh, and eighth columns
are empty, we can guarantee the additional property that
this piece shape is a connected polyomino.

Then we proceed as in the reduction of Theorem~\ref{thm:8 col}.
\end{proof}

\subsection{Linear-size pieces, 3 columns}

\begin{thm} \label{thm:3 columns}
\cektet{3}{$O(n)$} is NP-complete.
\end{thm}

\begin{proof}
The problem is in NP because checking whether a sequence of Tetris piece placements clears the board can be done in polynomial time, so it suffices to prove that 3-column Tetris with polynomially sized pieces is NP-hard.

As in the previous section, we reduce from {\partit} to perfect-information Tetris.

Given an instance of {\partit} with target sum $T$ and integers $\{a_1, \dots, a_{3s}\}$, we construct an instance of 3-column Tetris as follows, and as pictured in Figure~\ref{fig:3-column}:
\begin{enumerate}
    \item The board is 3 columns wide and $(3t+1)s$ rows tall.
    \item In the left column, the cells whose row is a multiple of $3t+1$ are filled (starting with the bottom one, row $0$).
    \item The middle column is empty.
    \item In the right column, the cells whose row is not a multiple of $3t+1$ are filled.
    \item The sequence of pieces is an \emph{$a_i$ sequence}, a \emph{right filling sequence}, and a \emph{clearing sequence}:
    \begin{enumerate}
        \item The $a_i$ sequence is, for each $a_i$, a $3a_i \times 1$ rectangle.
        \item The right filling sequence is $s$ $1 \times 1$ squares.
        \item The clearing sequence is a single $(3T+1)s \times 1$ rectangle.
    \end{enumerate}
\end{enumerate}

\begin{figure}
    \centering
    \includegraphics[scale=0.3]{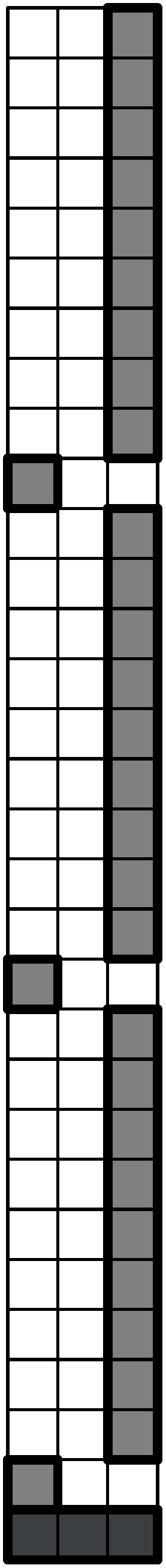}
    \caption{The initial board when $T=3$ and $s=3$.}
\label{fig:3-column}
\end{figure}

This initial position can be reached by normal Tetris play, by placing $\LL$ pieces, $\JJ$ pieces, and one $1 \times 1$ piece at the top. If, as in Section~\ref{sec:empty-start}, all lines clear (and lines above descend), before the game checks whether the player lost by breaching the ceiling, then we can force this position from an empty board, by sending pieces consisting of a $3\times 3$ square and one extra cell, which can't rotate and therefore add a pixel.

If the {\partit} instance has a solution, then the constructed Tetris problem has a solution: for each triple $\{a_i, a_j, a_k\}$ with sum $T$ in the ${\partit}$ solution, we'll fill one of the empty blocks of size $3t$ in the left column by moving the rectangles of size $3a_i \times 1$, $3a_j \times 1$, and $3a_k \times 1$ down the empty middle column and then left. Then use the $s$ squares of the right filling sequence to fill the $s$ empty cells in the right column, which are again accessible by the middle column. Finally, place the $(3T+1)s \times 1$ rectangle in the middle column, which clears the puzzle.

If the Tetris problem has a solution, we can construct a solution to the ${\partit}$ instance. 

First, there are $2(3T+1)s$ empty cells in the starting board, so at least $2(3T+1)s$ cells from the given pieces must fill those empty cells. The total number of cells in the given sequence of pieces is $2(3T+1)s$: $(3T)s$ from the $a_i$ sequence, $s$ from the right filling sequence, and $(3T+1)s$ from the clearing sequence, so every cell from the given pieces must fill one of the empty cells in the $(3T+1)s$ rows that are initially nonempty. The final $(3T+1)s \times 1$ rectangle puts pieces in $(3T+1)s$ rows, so those must be exactly the initially nonempty rows; that is, no rows can be cleared before the final piece, and no pieces can be placed in the center column before the final piece.

The $T$ empty cells in the right column can be filled only by $1 \times 1$ rectangles, so the $T$ pieces of the right filling sequence must be placed there.

Finally, the pieces of the $a_i$ sequence can only go in the left column and fill the $s$ empty spaces of size $1 \times 3T$, so the assignment of $a_i$ blocks to those spaces gives a solution to the {\partit} problem.
\end{proof}

If we relax the constraint that the initial position can be reached by normal Tetris play, then essentially the same proof shows that Tetris is hard even with only 2 columns: just delete the right column and the right filling sequence.
More interestingly, we can apply this idea to the regular game with two rows:

\begin{thm} \label{thm:2-row empty}
\rektet{2}{$O(n)$} is NP-complete.
\end{thm}

\begin{proof}
  We reduce from {\partit}. If the instance of {\partit} has $s$ triples and target sum $t$, the Tetris board has 2 rows and $s(4t+1)+2$ columns. We first send a piece of width $s(4t+1)+1$ columns; on the bottom row, with every cell present in on the bottom row and only the multiples of $4t+1$ (starting with 0, for a total of $s+1$ of them) present on the top row. This piece can only be placed in two positions, and either of them leaves $s$ buckets of size $4t$ on the top row and a single $1 \times 2$ hole at the side. We then send, for each $a_i$, a $1 \times 4a_i$ rectangle; these can't fill the $1 \times 2$ hole without losing, so they must be placed in the buckets on the top row. If there's a solution to the {\partit} instance, the buckets can be filled by the $1 \times 4a_i$ rectangles exactly by filling the buckets according to such a solution; if there is no solution, the Tetris game is lost. Finally, we send a $1 \times 2$ piece, which can go in the hole at the side to clear the Tetris board and survive.
\end{proof}

This result is tight:

\begin{prop} \label{prop:1 row empty}
\rektet{1}{$O(n)$} can be solved in linear time.
\end{prop}

\begin{proof}
  In \rektet{1}{$\leq k$}, any piece that is not $1 \times k$
  results in an immediate loss for the player
  (as in the proof of Proposition~\ref{prop:1 col 1 row}).
  If there are only $1 \times k$ pieces in the sequence,
  we proceed in rounds between line clears.
  For each round except the last, we compute the number $m$ of pieces
  before the next round by finding the prefix of remaining pieces
  whose total area is exactly the board width.
  If such an $m$ exists, we can place those $m$ pieces greedily
  from left to right in the initially empty row, and the row clears.
  If no such $m$ exists, then we attempt to place the remaining pieces
  greedily from left to right; if we run out of space, then no strategy
  could have survived.
\end{proof}

%% file: tetris/font.tex
\section{Font}
\label{sec:font}

\begin{figure}
  \centering
  % http://erikdemaine.org/fonts/tetris/?text=abcdefg%0Ahijklmn%0Aopqrst%0Auvwxyz&anim=0
  \includegraphics[width=0.75\linewidth]{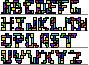}
  \caption{Tetris font: each 8-row letter can be made by stacking each of the
    seven tetrominoes exactly once in some order.}
  \label{font solved}
\end{figure}

\begin{figure}
  \centering
  % http://erikdemaine.org/fonts/tetris/?text=abcdefg%0Ahijklmn%0Aopqrst%0Auvwxyz&anim=0&puzzle=1
  \includegraphics[width=0.63\linewidth]{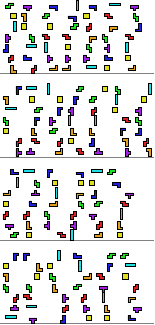}
  \caption{Tetris puzzle font: if each piece falls vertically, the result
    is Figure~\protect\ref{font solved}.}
  \label{font puzzle}
\end{figure}

\begin{figure}
  \centering
  % http://erikdemaine.org/fonts/tetris/?text=tetris&anim=0
  \includegraphics[width=0.75\linewidth]{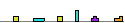}

  \bigskip

  \includegraphics[width=0.75\linewidth]{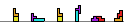}

  \bigskip

  \includegraphics[width=0.75\linewidth]{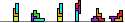}

  \bigskip

  \includegraphics[width=0.75\linewidth]{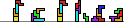}

  \bigskip

  \includegraphics[width=0.75\linewidth]{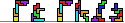}

  \bigskip

  \includegraphics[width=0.75\linewidth]{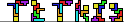}

  \bigskip

  \includegraphics[width=0.75\linewidth]{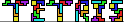}
  \caption{Tetris animated font.
    See \protect\url{http://erikdemaine.org/fonts/tetris/?text=tetris}
    for animation with falling, sliding, and rotation.}
  \label{font anim}
\end{figure}

To demonstrate the versatility of Tetris constructions, we designed an
8-row font where each letter of the alphabet is constructed as a stacking
of exactly one copy of each tetromino (treating reflections as distinct,
as in Tetris).  Figure~\ref{font solved} shows the fully assembled font.
Crucially, these letters can actually be constructed in Tetris by stacking
the pieces one at a time in some order (dependent on the letter), while being
supported by the previously stacked pieces according to Tetris physics.
Figure~\ref{font puzzle} illustrates the stacking order in a puzzle version
of the font, where the pieces are spread out vertically according to their
fall order, but placed correctly horizontally;
letting the pieces fall straight down reveals the letters in
Figure~\ref{font solved}.
A companion web app%
\footnote{\url{http://erikdemaine.org/fonts/tetris/}}
allows you to type a custom message, and animate the stacking.
Figure~\ref{font anim} shows a sample animation.

